\RequirePackage{fix-cm}
\documentclass[twocolumn,epjc3]{svjour3}  
\smartqed  
\usepackage{amsmath,amssymb}
\usepackage[dvips]{graphicx,color}
\usepackage{color}
\usepackage{comment}
\usepackage[abc]{overpic}
\usepackage{cite}
\journalname{Eur. Phys. J. B}
\begin{document}

\title{Relations between anomalous diffusion and fluctuation scaling: The case of ultraslow diffusion and time-scale-independent fluctuation scaling in language}
\subtitle{Relations between anomalous diffusion and fluctuation scaling}


\author{Hayafumi Watanabe\thanksref{e1,addr1,addr2,addr3,addr4}       
}

\thankstext{e1}{e-mail: hayafumi.watanabe@gmail.com}


\institute{Faculty of Economics, Seijo University, 6-1-20 Seijo, Setagaya, Tokyo 157-8511, Japan \label{addr1}
           \and
           The Institute of Statistical Mathematics, 10-3 Midori-cho, Tachikawa, Tokyo 190-8562, Japan \label{addr2}
           \and
           College of Science and Engineering, Kanazawa University, Kanazawa, Ishikawa 920-1192, Japan \label{addr3}
           \and
           Hottolink, Inc., 6 Yonbancho Chiyoda-ku, Tokyo 102-0081, Japan \label{addr4}
}

\date{Received: date / Accepted: date}

\abstractdc{
Fluctuation scaling (FS) and anomalous diffusion have been discussed in different contexts, even though both are often observed in complex systems.
To clarify the relationship between these concepts, we investigated approximately \textcolor{black}{three billion Japanese blog articles} over a period of six years and analyzed the corresponding Poisson process driven by a random walk model with power-law forgetting, which reproduces both the anomalous diffusion and the FS.
From the analysis of the model, we have identified the relationship between the time-scale dependence of FS and characteristics of anomalous diffusion and 
 showed that the time-scale-independent FS corresponds to essentially a logarithmic diffusion (i.e., a kind of ultraslow diffusion).
In addition, we confirmed that this relationship is also valid for the actual data.
This finding may contribute to the discovery of actual examples of ultraslow diffusion,  which have been nearly unobserved in spite of many mathematical theories, because we can detect the time-scale-independent FS  more easily and more distinctly than through direct detection of the logarithmic diffusion based on the mean squared displacement. \\\\
}

\maketitle

%
\noindent
{\bf Program Summary and Specifications}\\
\begin{small}
\noindent
{Program title:}\\
{Licensing provisions:}\\
{Programming language:}\\
{Repository and DOI:}\\
{Description of problem:}\\
{Method of solution:}\\
{Additional comments:}\\
\end{small}
\section{Introduction}
\textcolor{black}{Anomalous diffusion is observed in various complex systems.
It is basically characterized by the mean squared displacement (MSD), }
	\textcolor{black}{
	\begin{equation}
	\left< x^2(t) \right> \propto t^{\alpha}.
	\end{equation}
	}
	In the case of $\alpha=1$, the diffusion corresponds to normal diffusion, such as the diffusion of particles in water, which is modeled using a random walk.  
	In other cases, it is known as anomalous diffusion, being termed subdiffusion for $0<\alpha<1$ and superdiffusion for $\alpha>1$. 
	Many complex systems have been shown to exhibit this power-law type of anomalous diffusion in diverse areas, such as physics, chemistry, geophysics, biology, and economy \cite{metzler2000random, da2014ultraslow}. 
	In theoretical studies,
	\textcolor{black}{anomalous diffusion is explained using the correlation of random noise (e.g., a random walk in disordered media) \cite{bouchaud1990anomalous}, a finite-variance (e.g., a Levy flight) \cite{bouchaud1990anomalous, metzler2000random}, a power-law wait time (e.g., a continuous random walk) \cite{bouchaud1990anomalous, metzler2000random, burov2011single}, and a long memory (e.g., a fractional random walk) \cite{lowen2005fractal, burov2011single}.}
	\par
	%
	%
	Another class of anomalous diffusion predicted by theories entails  logarithmic MSD growth:  
	\begin{equation}
	\left<x^2(t) \right> \propto \log(t)^{\alpha}. \label{eq_usd}
	\end{equation}
	This type of diffusion is known as ``ultraslow diffusion.''
	One of the best-known examples that was first discovered is the diffusion in a disordered medium (known as Sinai diffusion for $\alpha=4$) \cite{sinai1983limiting}. 
	Thereafter, other types of models that explain ultraslow diffusion have also been proposed; these include a continuous random walk (CTRW) with a waiting time generated by the logarithmic-form probability density function (PDF) \cite{godec2014localisation}, a CTRW with a waiting time generated by the power-form PDF and the excluded volume effect \cite{sanders2014severe}, temporal change of diffusion coefficients \cite{bodrova2015ultraslow}, spatial changes \cite{cherstvy2013population}, \textcolor{black}{two-dimensional comb-like structure\cite{sandev2016comb}} and \textcolor{black}{fractional dynamics \cite{eab2011fractional}.} 
	\textcolor{black}{Ref. \cite{liang2019survey} is a very informative review that systematically summarizes the various generative models of ultraslow diffusion, including related empirical observations.}
	 \par
	\textcolor{black}{
	Although many theoretical studies of ultraslow diffusion have been reported, few empirical examples are available. 
	\textcolor{black}{Rare real example of ultraslow-like diffusion in the real world can be observed in the aging colloidal glass at high density, where the heterogeneity can be tuned to move between normal,
abnormal, and ultra-slow diffusion \cite{liang2019survey,liang2018non}.
	Another example of ultraslow-like diffusion in the real world can be observed in the time series of word counts of already popular words in three different nationwide \textcolor{black}{word counts} databases: (i) newspaper articles (in Japanese), (ii) blog articles (in Japanese), and (iii) page views of Wikipedia (in English, French, Chinese, and Japanese) \cite{watanabe2018empirical}.}} 
	The observed logarithmic diffusion, namely, very slow changes, is not in conflict with our intuition that languages are basically stable but change constantly. 
	\textcolor{black}{Note that diffusion related to the logarithmic function, which is similar to but different from the ``ultraslow diffusion'' defined by Eq. \ref{eq_usd} (i.e., $\propto \log(t)^{\alpha}$), is observed in the mobility of humans ($\log(\log(t))^{\alpha}$ or becomes saturated) \cite{song2010modelling} and the mobility of monkeys \cite{boyer2011non}. 
	 In addition, logarithmic ``relaxation'' phenomena, which are known as ``aging'', are also observed in many systems such as paper crumpling \cite{matan2002crumpling} and grain compaction \cite{richard2005slow}. 	
	}  
	\par
\textcolor{black}{One possible reason for there being so few real-world examples of ultraslow diffusion is the technical difficulty in making observations.  
The difficulty is attributed to the logarithmic changes, namely, very slow changes, which make it difficult to distinguish the diffusion signal from stationary noise or can easily hide it by responses  to irregular outer shocks.
One of the purposes of this study is to overcome this technical difficulty. } \par
\textcolor{black}{
To alleviate the difficulty in making observations, we adopt the concept of fluctuation scaling (FS).
FS, which is also known as ``Taylor's law'' \cite{taylor1961aggregation} in ecology, is a power-law relation between the system size (e.g., a mean) and the magnitude of fluctuation (e.g., a standard deviation).
FS is observed in various complex systems, such as a random work on a complex network \cite{PhysRevLett.100.208701}, internet traffic \cite{argollo2004separating}, river flows \cite{argollo2004separating}, animal populations \cite{xu2015taylor}, insect numbers \cite{xu2015taylor, eisler2008fluctuation}, cell numbers \cite{eisler2008fluctuation}, foreign exchange markets \cite{sato2010fluctuation}, the download numbers of Facebook applications \cite{onnela2010spontaneous}, word counts of Wikipedia \cite{gerlach2014scaling}, academic papers \cite{gerlach2014scaling}, old books \cite{gerlach2014scaling}, crimes \cite{10.1371/journal.pone.0109004}, and Japanese blogs \cite{sano2010macroscopic,RD_base}.}  Even though FS and anomalous diffusion are often observed in complex systems, the two concepts have been discussed in different contexts previously.
 \par
 A certain type of FS can be explained through the random diffusion (RD) model \cite{PhysRevLett.100.208701}.
The RD model, which has been introduced as a mean-field approximation for a random walk on a complex network, is described by a Poisson process 
with a random variable Poisson parameter.
It can be demonstrated that the fluctuation of the RD model obeys FS with an exponent of $0.5$ for a small system size (i.e., a small mean) or $1.0$ for a large system size (i.e., a large mean). 
Because this model is based only on a Poisson process, it is applicable not only to random walks on complex networks but also to a wide variety of phenomena related to random processes. For instance, this model can reproduce a type of FS regarding the appearance of words in Japanese blogs on a daily basis \cite{sano2009, PhysRevE.87.012805}.  \par
    \textcolor{black}{Note that physicists have studied linguistic phenomena using concepts of complex systems \cite{link1} such as competitive dynamics \cite{abrams2003linguistics}, statistical laws \cite{altmann2015statistical}, and complex networks \cite{cong2014approaching}. Our study can also be positioned within this context, that is, we study properties of the time series of word counts in nationwide blogs (a linguistic phenomenon) using FS and anomalous diffusion.}
\textcolor{black}{Pioneering works on word count time series using anomalous diffusion and the scaling of fluctuation are available in Refs. \cite{gao2012culturomics} and \cite{petersen2012languages}.
In Ref. \cite{gao2012culturomics}, using annual n-gram time series data (similar to word count data) for books written in the past two centuries, authors reported that the  the anomalous diffusion are observed in the n-gram time series and their power-law exponents are different between natural and social phenomena. 
In Ref. \cite{petersen2012languages}, the authors also suggested that new words contribute more to the increase in the number of words used in the corpus than basic words. In addition, they found that the fluctuation of the growth rate decreases as the power-law function of the amount of usage of a focused word. } \par
\textcolor{black}{
In this study, we attempt to clarify the relationship between anomalous diffusion and FS through investigating the word count time series in nationwide Japanese blogs.}
In addition, we aim to apply the relation to observe ultraslow diffusion indirectly 
to alleviate the technical difficulty of direct observations based on the MSD. \par
In this study, first, we investigate FS of the word count time series for various time scales using five billion Japanese blog articles from 2007 and we clarify the empirical FS accompanying ultraslow diffusion.
 Second, we employ the Poisson process driven by the random walk model with the power law forgetting discussed in Ref. \cite{watanabe2018empirical},  which can reproduce the ultraslow diffusion of word count time series. 
 Third, we show that the model can also reproduce the time-scale dependence of empirical FS as well as ultraslow diffusion. 
\textcolor{black}{In addition, we derive the relation between the exponent of anomalous diffusion and the coefficient of FS. Particularly, we show that ultraslow diffusion is linked to the fact that the coefficient is independent of the time scale $L$.  }
Finally, we discuss the application possibility of FS to the detection of ultraslow diffusion. \par 
\begin{figure*}
\begin{minipage}{0.48\hsize}
\centering
\begin{overpic}[width=7cm,clip,height=4cm]{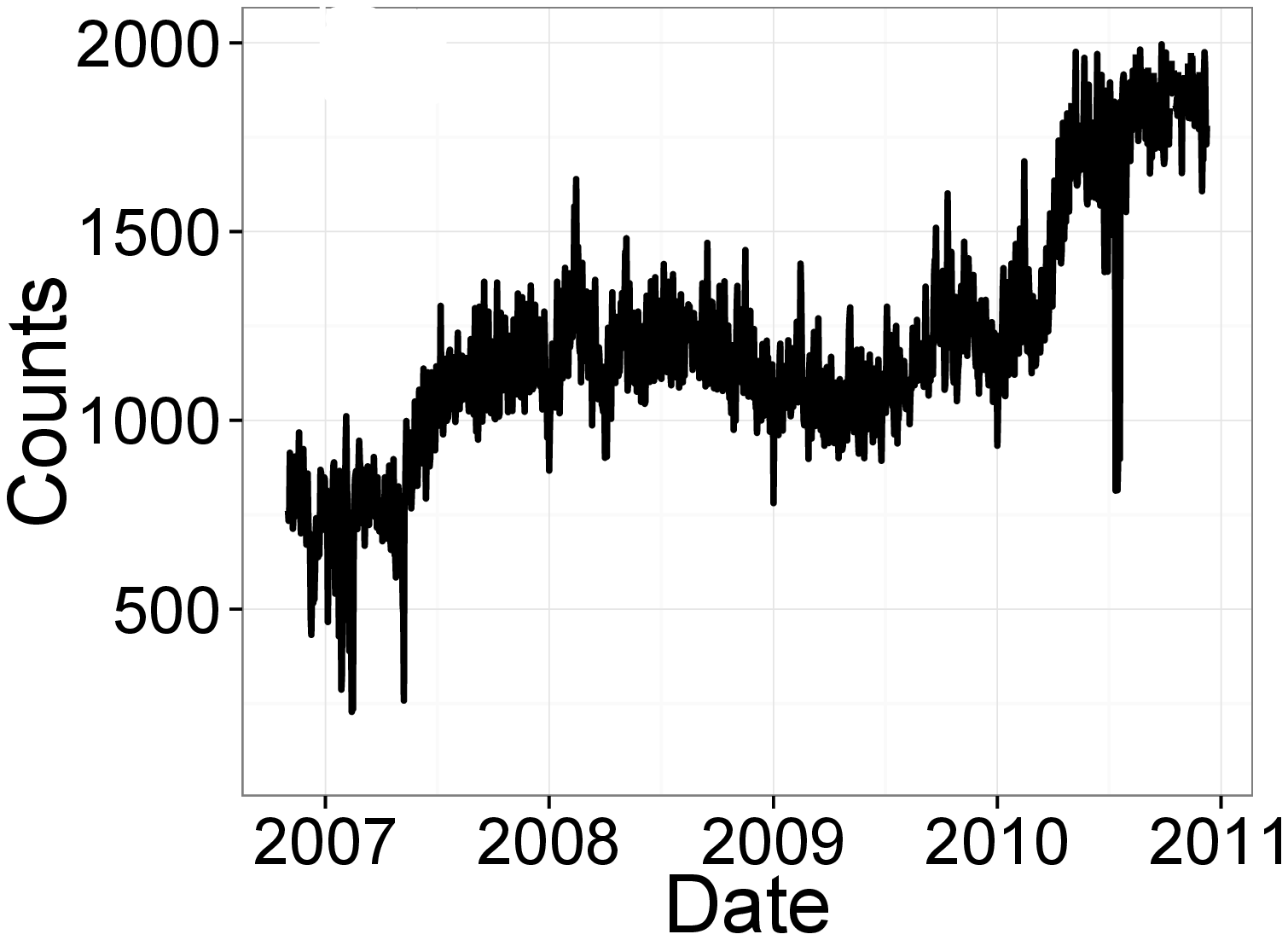}
\put(23,47){\Large{(a)}}
\end{overpic}
\end{minipage}
\begin{minipage}{0.48\hsize}
\centering
\begin{overpic}[width=7cm,clip,height=4cm]{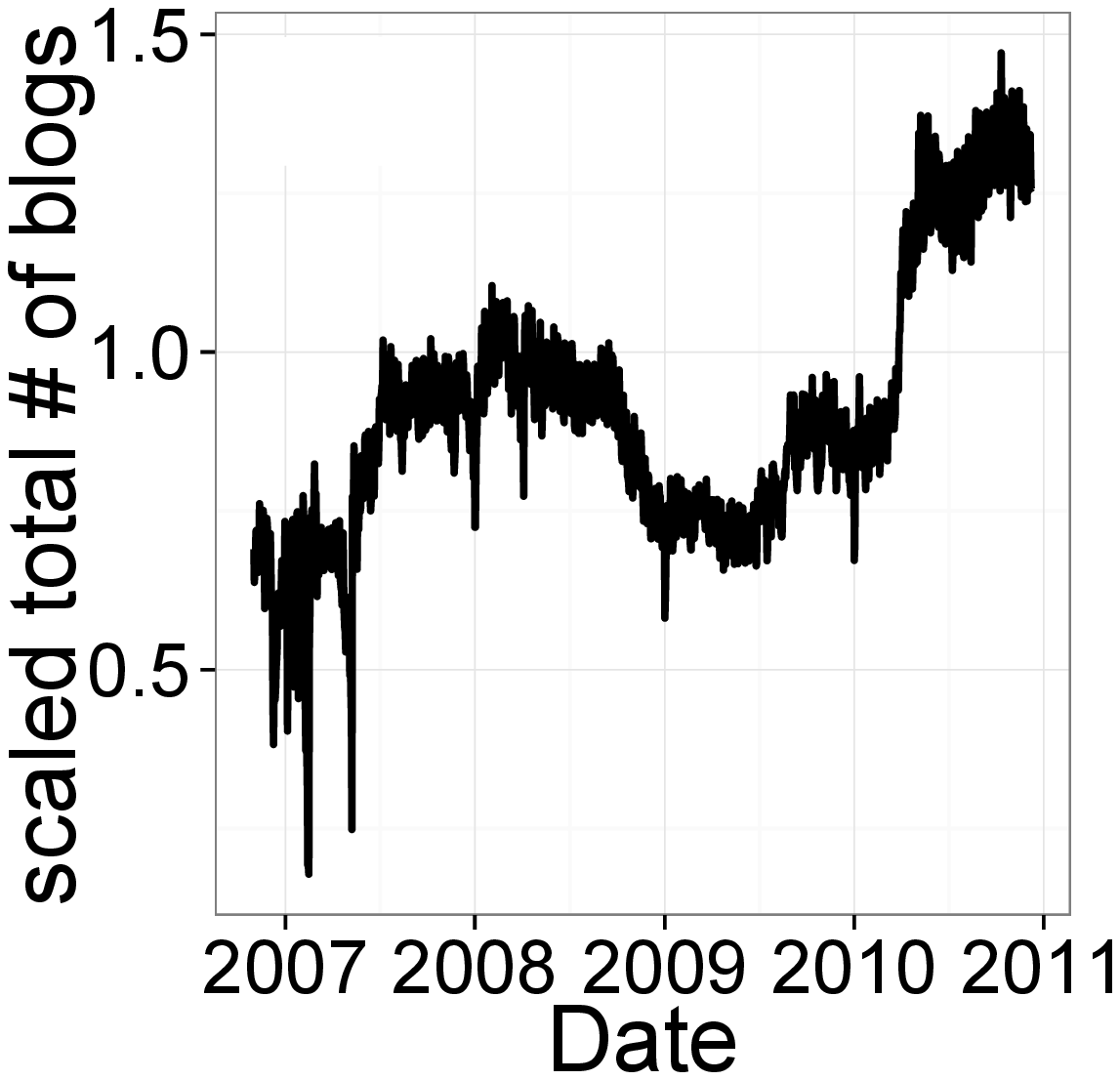}
\put(23,47){\Large(b)}
\end{overpic}
\end{minipage}
\begin{minipage}{0.48\hsize}
\centering
\includegraphics[width=7cm,height=4cm]{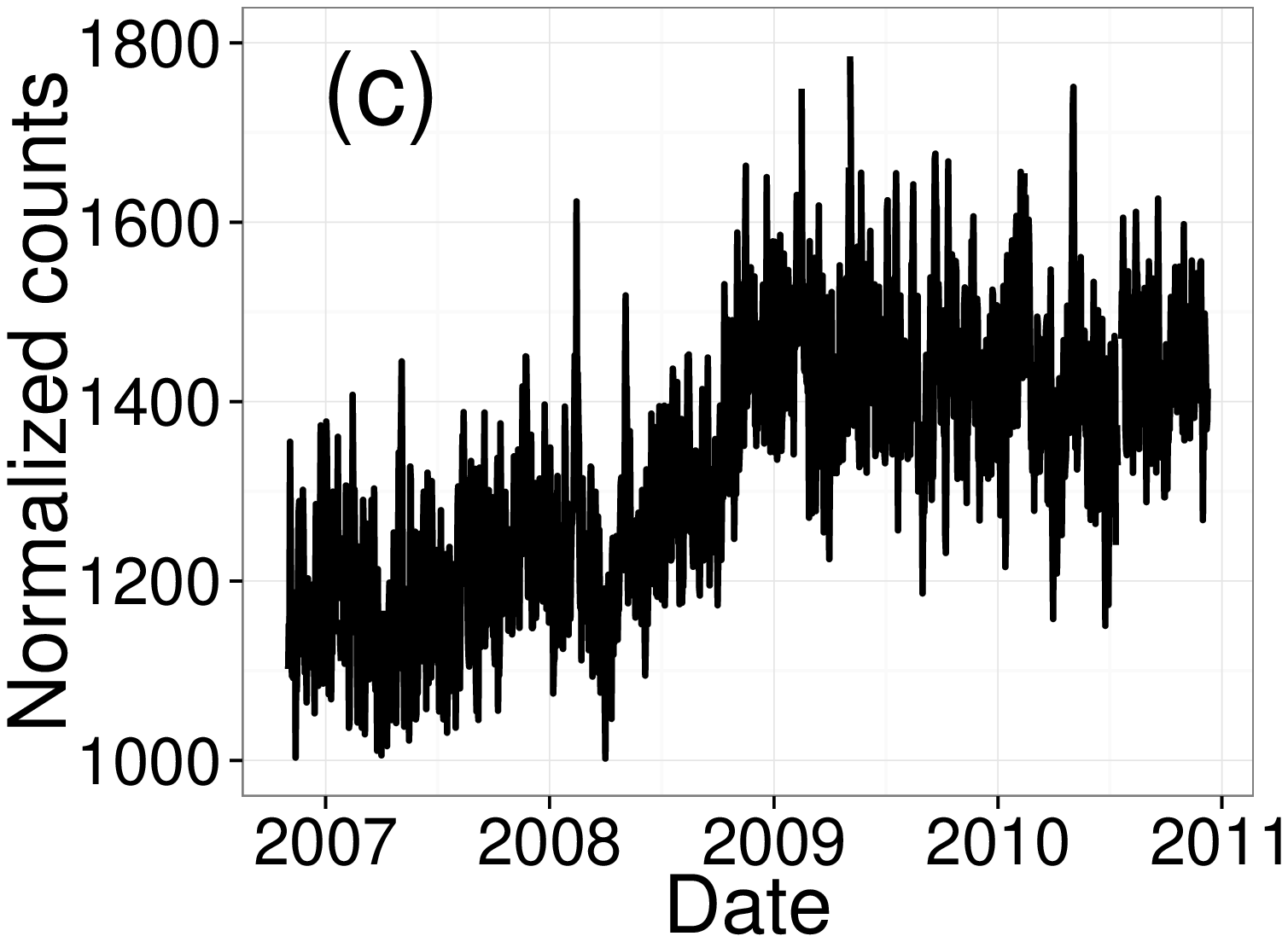}
\end{minipage}
\begin{minipage}{0.48\hsize}
\centering
\includegraphics[width=7cm,height=4cm]{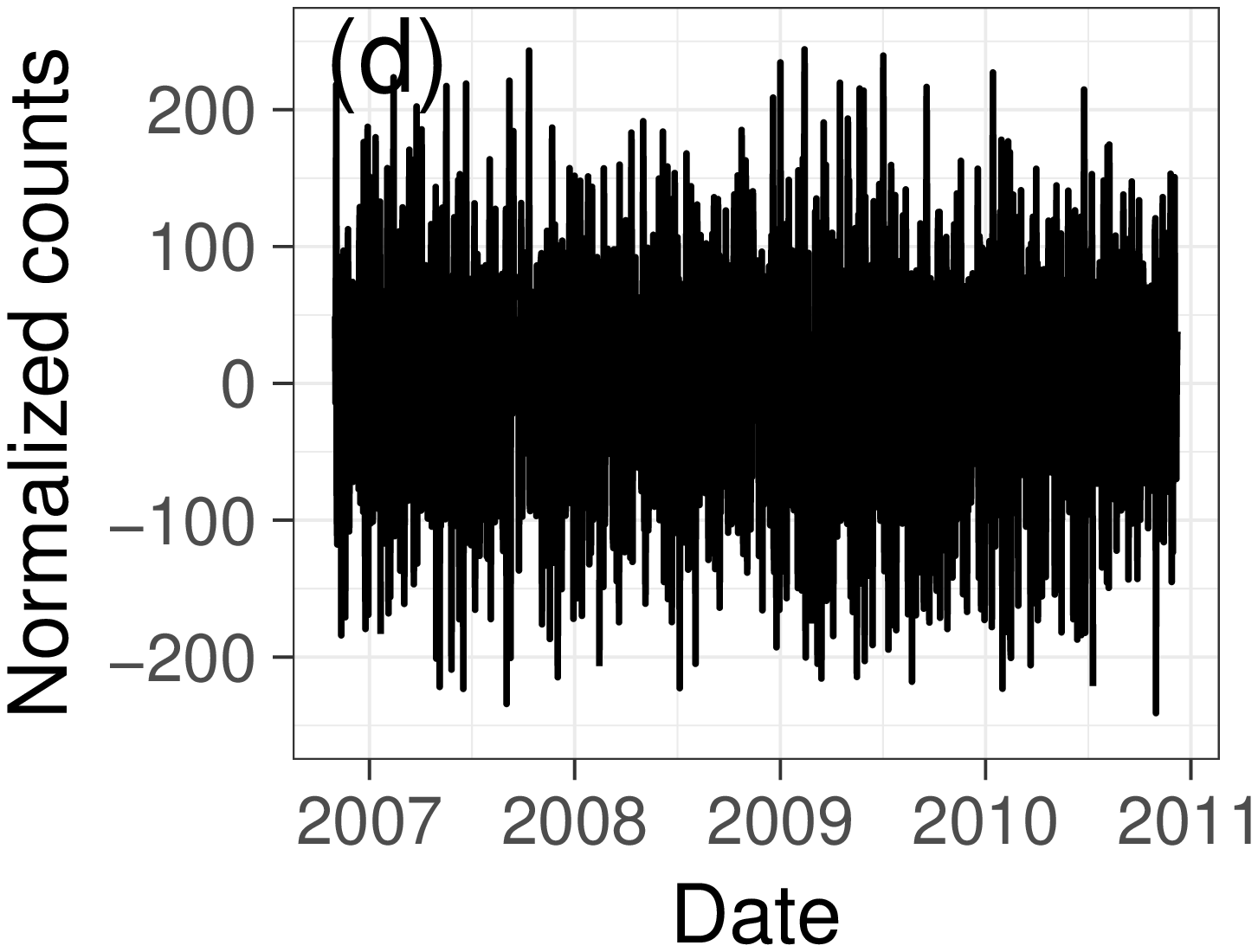}
\end{minipage}
\begin{minipage}{0.48\hsize}
\centering
\includegraphics[width=7cm,height=4cm]{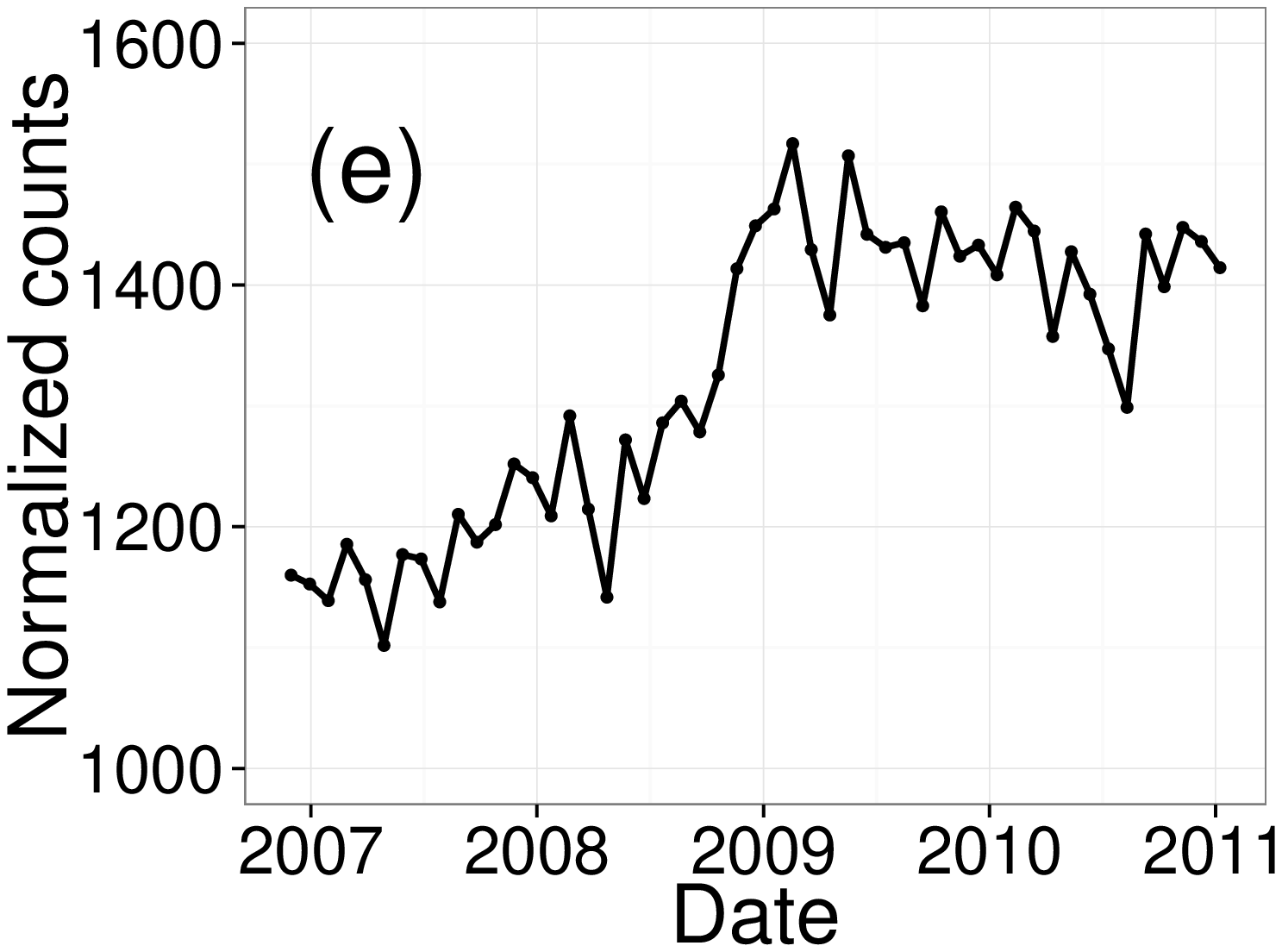}
\end{minipage}
\begin{minipage}{0.48\hsize}
\centering
\includegraphics[width=7cm,height=4cm]{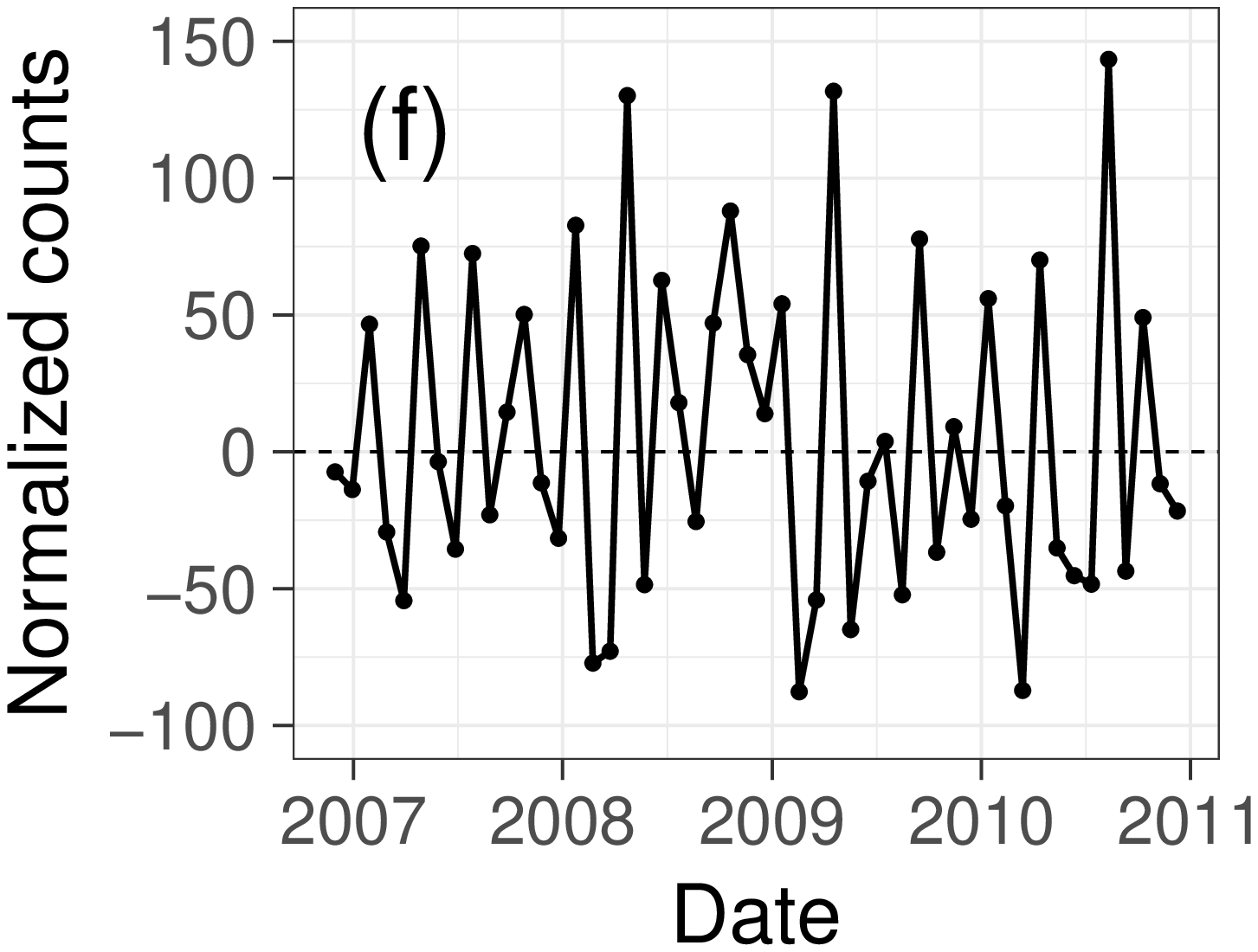}
\end{minipage}
\caption{
(a) Example of a daily time series of raw word appearances for ``oishii'' (delicious in English), $g_j(t)$. 
(b) Daily time series of the normalized total number of blogs, $m(t)$ (see \ref{app_m}).  
We can confirm that the time variation of raw word appearances $g_j(t)$ shown in panel (a) is almost the same as that of the total number of blogs, $m(t)$, shown in panel (b).
(c) Daily time series of word appearances scaled by the normalized total number of blogs for ``oishii,'' $f_j(t)=g_j(t)/m(t)$. 
(d) Differential time series of the word appearances scaled by the normalized total number of blogs for ``oishii,'' $\delta f_j(t) \equiv f_j(t)-f_j(t)$.
(e) Time series of the box means of the word appearances scaled by the normalized total number of blogs for ``oishii,'' $F_j^{(L)}(\tau^{(L)})$, given by Eq. \ref{eq_F} for $L=30$.
(f) Differential time series of the box means of the word appearances scaled by the normalized total number of blogs for ``oishii'', $\delta F(\tau^{(L)})=F_j^{(L)}(\tau^{(L)})-F_j^{(L)}(\tau^{(L)}-1)$, for $L=30$.
}
\label{tseries}
\end{figure*}

\section{Dataset}
In the data analysis, we analyzed the word frequency time series of Japanese blogs, that is, the time series of the numbers of word occurrences in Japanese blogs per day.
To obtain these time series, we used a large database of Japanese blogs (Kuchikomi@kakaricho), which was provided by Hottolink, Inc. This database contains three billion Japanese blog articles, which covers 90$\%$ of Japanese blogs from 1 November 2006 to 31 December 2012. We used 1,771 basic adjectives as keywords \cite{RD_base}. 
\textcolor{black}{These data are a typical example of a word count time series.  The data have properties in common with Japanese newspaper data and Wikipedia page view data (in Japanese, French, and English) in terms of ultraslow diffusion \cite{watanabe2018empirical}.} 
Note that if an article contains more than two focused keywords (e.g., key word ``dog'': ``There is a dog. The dog is big.''), the system counts it as one article.  \par
\subsection{Normalized time series of word appearances}
Here, we define the notation of the time series of the word appearances $g_j(t)$ and $f_j(t)$ as follows:
\begin{itemize}  
\item $g_j(t)$ $(t=1,2,\cdots, T)$ $(j=1,2,3,W)$ is a raw daily count of the appearances of the $j$th word within the dataset (see Fig. \ref{tseries}(a)).  Specifically, $g_j(t)$ corresponds to the daily number of articles containing the $j$th keyword in the blog database.
\item $f_j(t)=g_j(t)/m(t)$ is the time series of daily count normalized by the total number of blogs, $m(t)$ (see Fig. \ref{tseries}(c)).
\end{itemize}
Here, $m(t)$ is the normalized total number of blogs obtained by assuming that $\sum_{t=1}^{T}m(t)/T=1$ for normalization (see Fig. \ref{tseries}(b)), 
where $m(t)$ is estimated by the ensemble median of the number of words at time $t$, as described in \ref{app_m}.  
Note that $f_j(t)$ corresponds to the original time deviation of the $j$th word separated from the effects of deviations in the total number of blogs, $m(t)$ (see Figs. \ref{tseries}(a)--(c)).  
\begin{figure*}
\begin{minipage}{0.48\hsize}
\centering
\includegraphics[width=7cm]{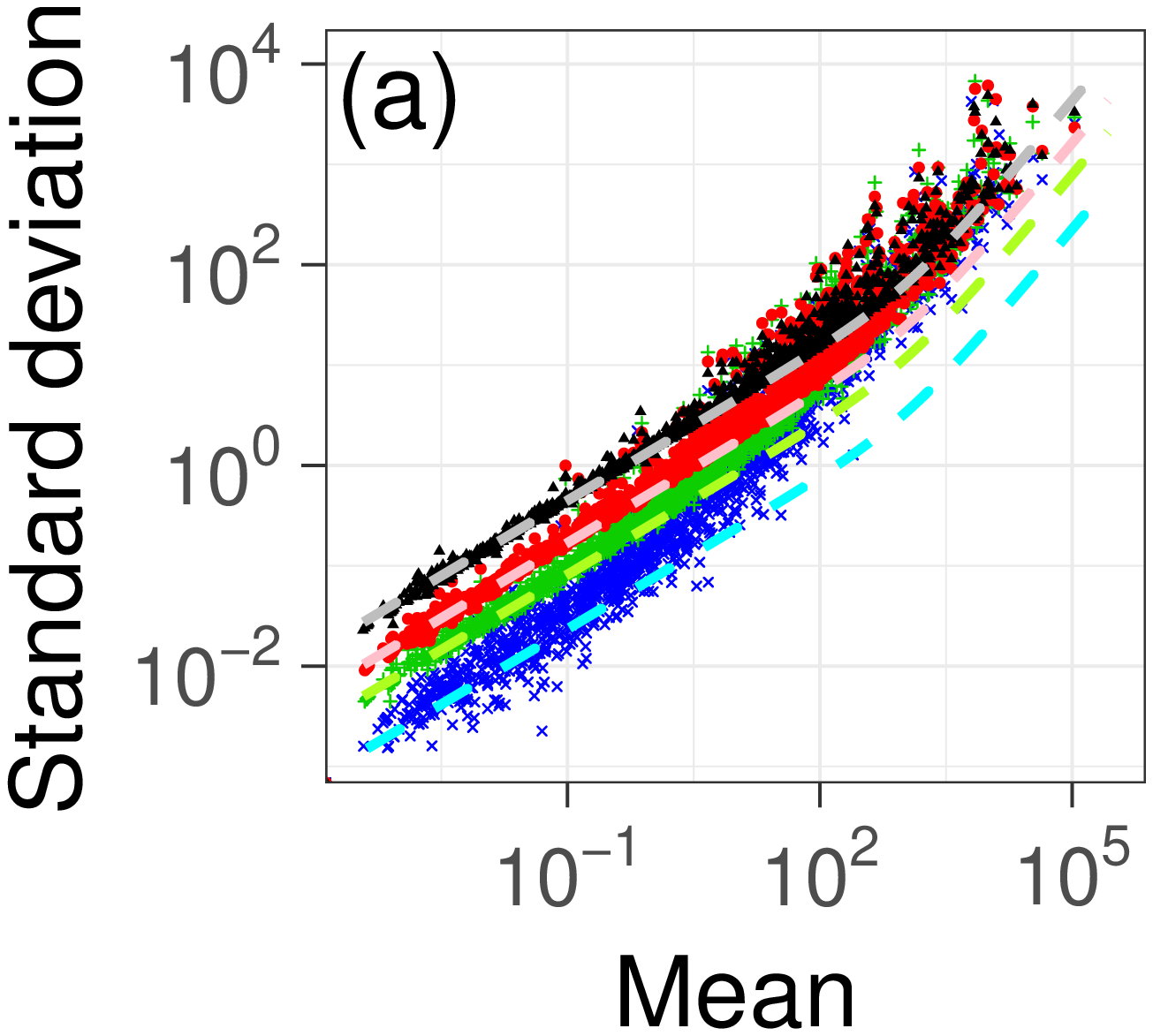}
\end{minipage}
\begin{minipage}{0.48\hsize}
\centering
\includegraphics[width=7cm]{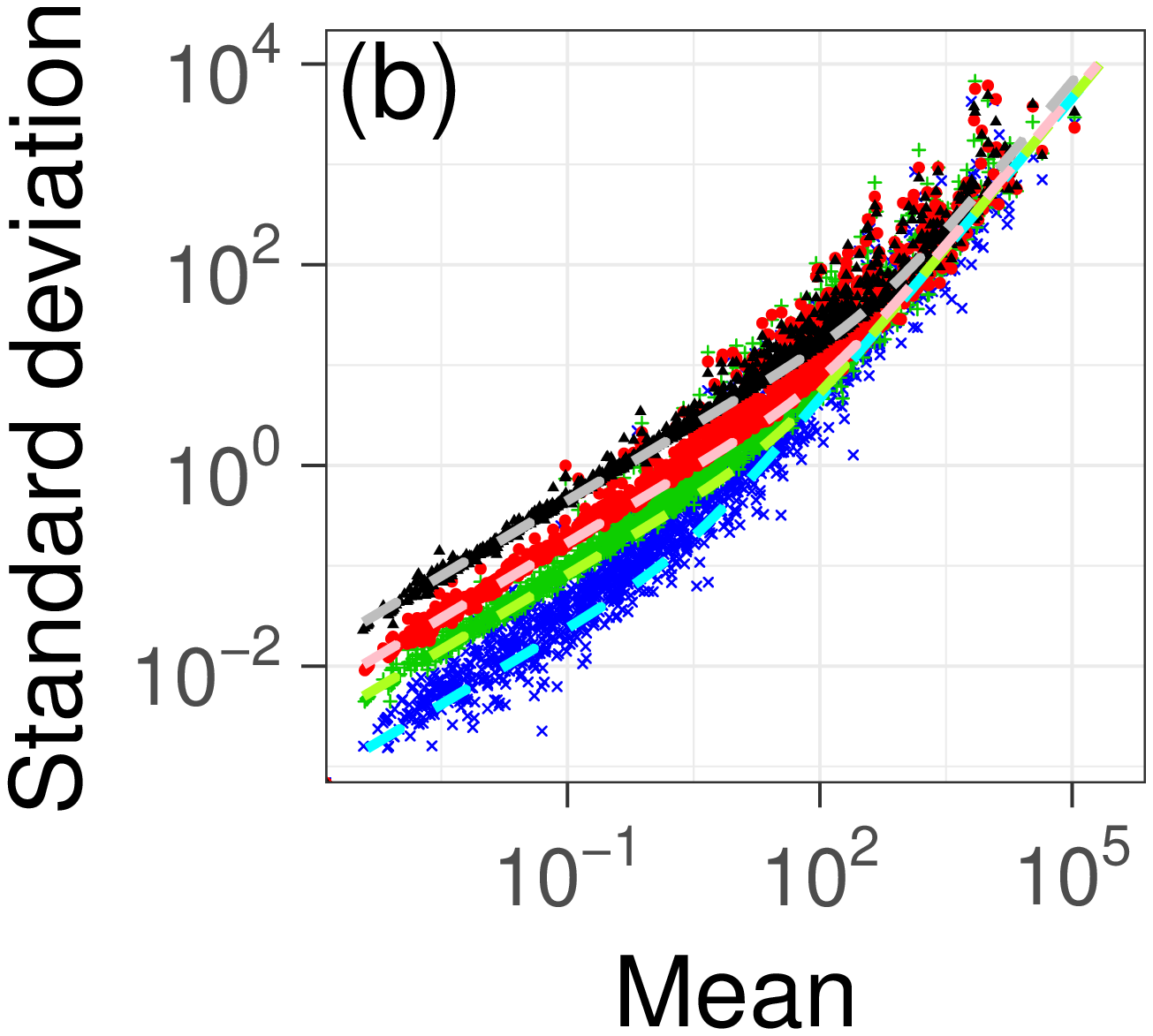}
\end{minipage}
\caption{(a) FS of the box mean time series, mean $E[F^{(L)}_j]$, versus standard deviation $V[\delta F^{(L)}_j]^{1/2}$. The data shown are empirical results of 1,771 adjectives for $L=1$ (black triangles),
$L=7$ (red circles), $L=30$(green plus signs), $L=365$ (blue crosses), and the corresponding theoretical curves given in Eq. \ref{V_df_emp} under the conditions that $a(L)=a_0/L$, $b(L)=b_0/L$, $a_0=E[1/m] \approx 1$, and $b_0=\check{\delta}_0^2=0.030^2$, which are deduced from the assumption that $\{f_j(t)\}$ is sampled from independent random variables, for $L=1$ (gray dashed line), $L=7$ (pink dashed line), $L=30$ (mint green dashed line), and $L=365$ (cyan dashed line) from top to bottom. We can see that the empirical data are in good agreement with the theoretical lower bounds for the small mean $V[\delta E^{(L)}_j]$ (the left part of figure) and are not in agreement for the large mean $V[\delta E^{(L)}_j]$  (the right part of the figure).  
(b) Theoretical lower bound calculated by using the random diffusion model based on the power-law forgetting process, in which $b(L)$ is given by Eqs. \ref{u_ans} and \ref{b_LL2}  ($\check{\eta}_{j}=0.029$, $\check{\delta}_{j}=0.030$). From panel (b), we can confirm that the empirical lower bound is in agreement with the theoretical curve at all scales.}
\label{fig_TFS}
\end{figure*}

\begin{figure*}
\begin{minipage}{0.48\hsize}
\includegraphics[width=7cm]{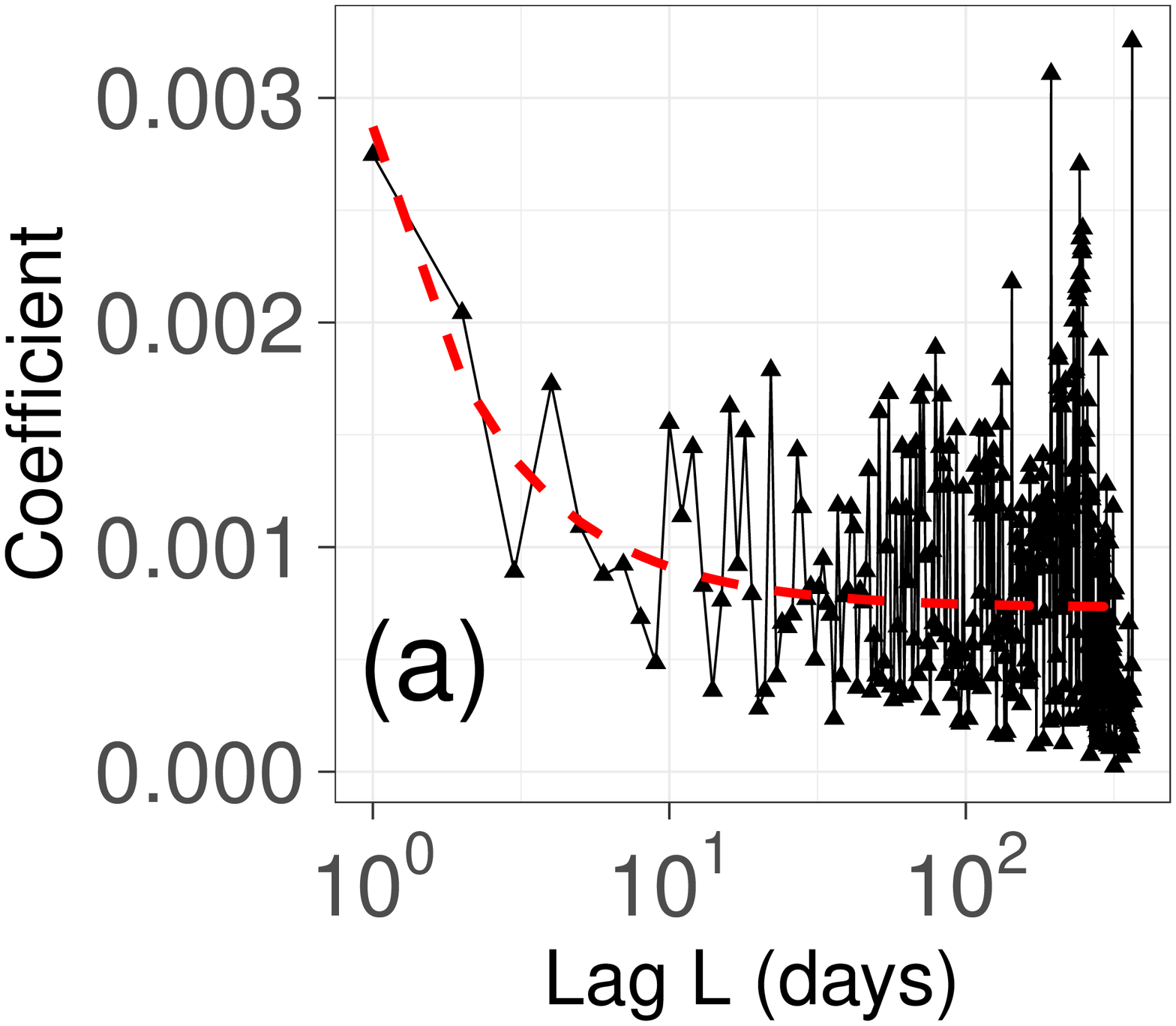}
\end{minipage}
\begin{minipage}{0.48\hsize}
\includegraphics[width=7cm]{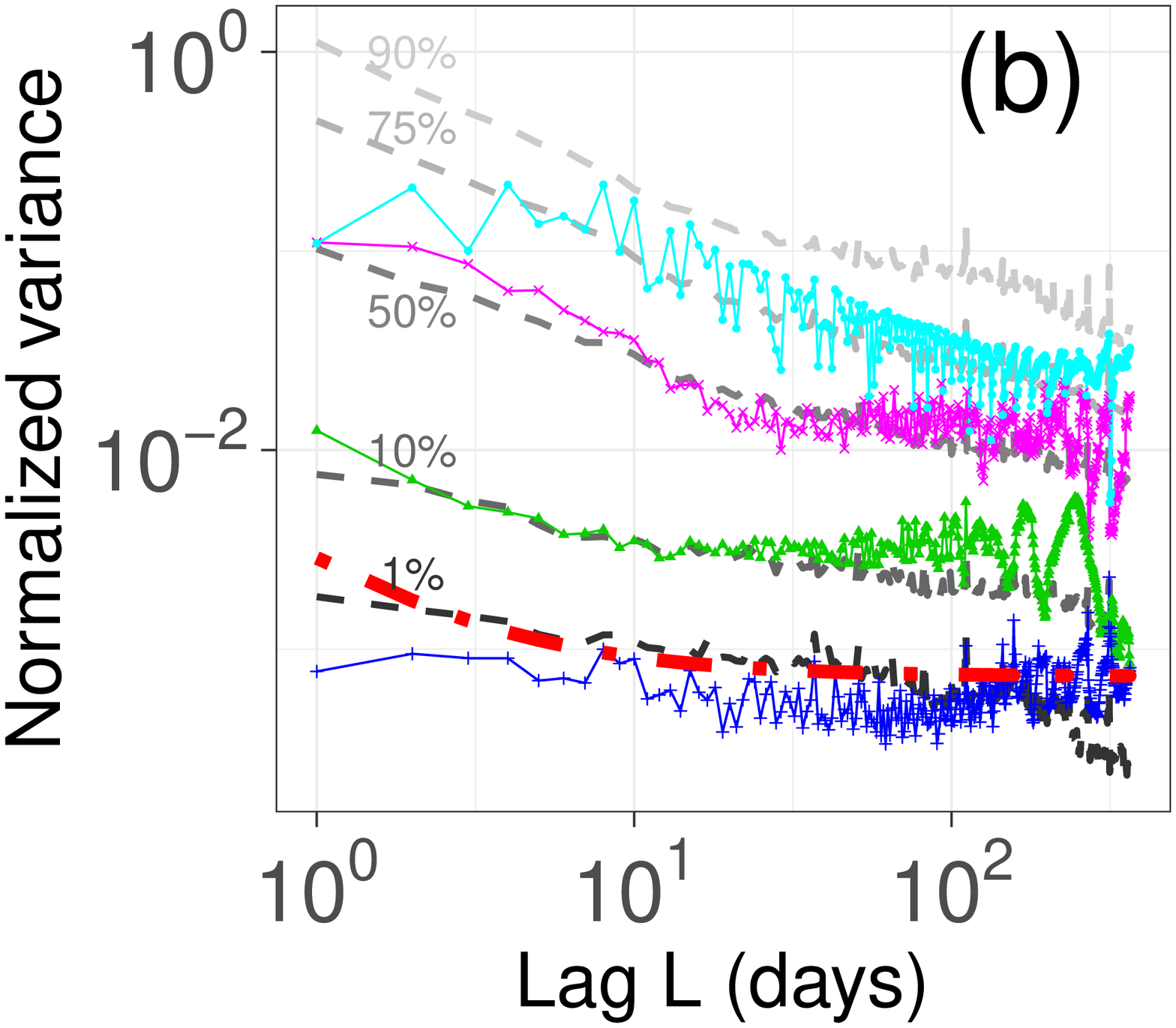}
\end{minipage}
\caption{(a) Coefficients of the FS, $b(L)$, given by Eq. \ref{V_df_emp}, which is directly estimated from the data using the method given in \ref{app_coef} (black triangles). In addition, the red dashed line shows the theoretical curve calculated by the random diffusion model based on the power-law forgetting given in Eqs. \ref{u_ans} and \ref{b_LL2} ($\check{\eta}_{j}=0.029$, $\check{\delta}_{j}=0.030$). 
From panel (a), we can see that $b(L)$ does not have a clear dependence on  lag $L$ for a large lag $L$, namely, $b(L) \approx {\rm constant}$ ($L>>1$). 
(b) Examples and the distribution of $\hat{b}_j(L)=V[\delta F_j^{(L)}]/E[F_j^{(L)}]^2$ nearly correspond to $b(L)$ for the particular words with a \textcolor{black}{large $E[F_j^{(L)}]$} (see Eq. \ref{V_df_emp}). The points show the empirical \textcolor{black}{data for ``ooi'' (``many'' in English; blue pluses), ``otonashii'' (``quiet personality'' in English; green triangles), ``tuyoi'' (``strong'' in English; pink crosses),  and ``kowai'' (``scary'' in English; cyan circles).}
 The gray dashed lines indicate the percentiles of the distribution $\hat{b}_j(L)$ as a function of $L$. \textcolor{black}{The red dash-dotted line indicates the same theoretical curve with panel (a).}
Panel (b) implies that the particular words also obey the similar FS to the lower bound shown in panel (a) 
 because the points in the plot indicate an approximately $\hat{b}_j(L) \approx O(1)$ for a large $L$. 
In addition, 
this result suggests common dynamics that is independent of the types of words.
} 
\label{fig_coef}
\end{figure*}

\begin{figure*}
\begin{minipage}{0.48\hsize}
\begin{overpic}[width=7cm,clip]{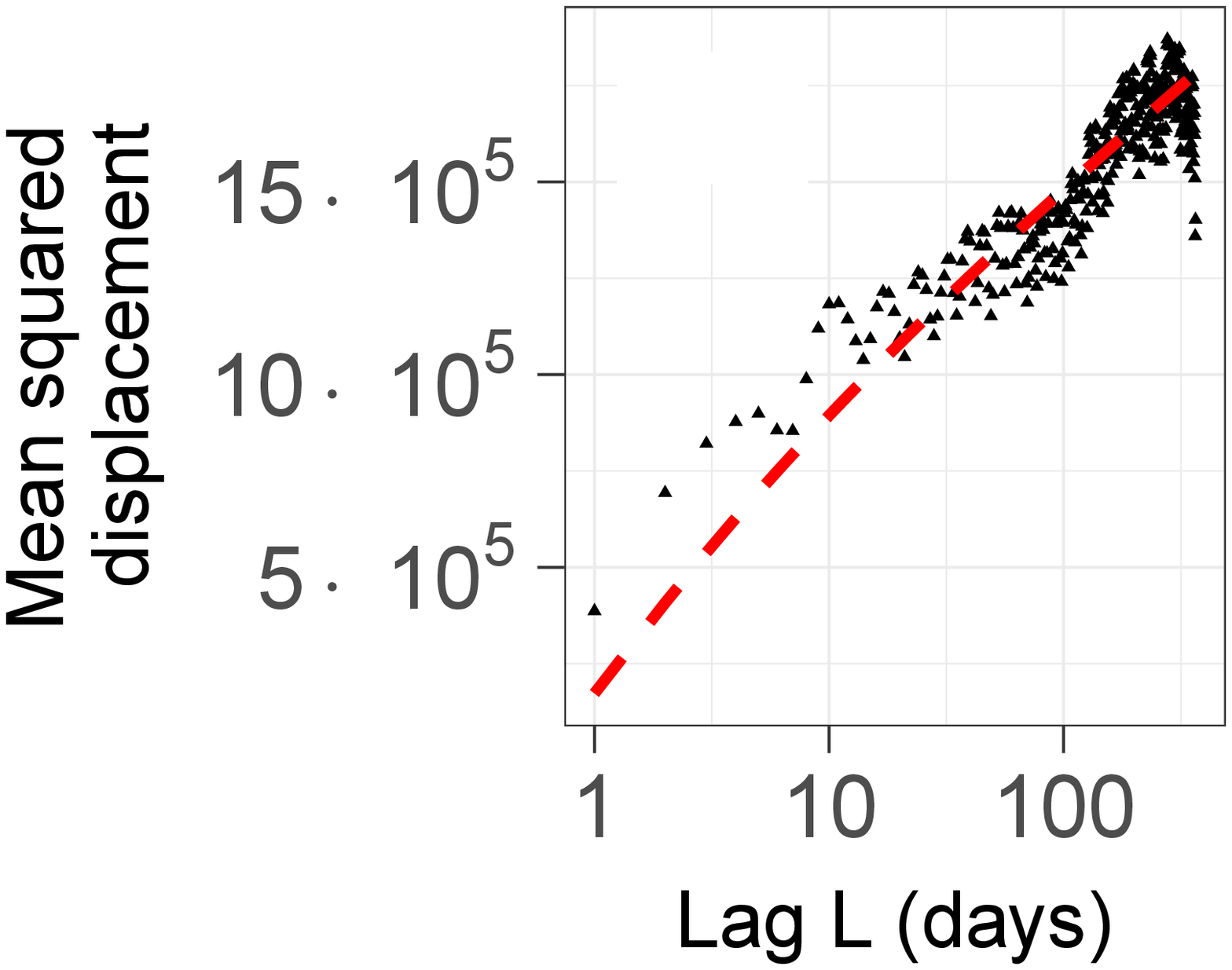}
\put(53,65){(a)}
\end{overpic}
\end{minipage}
\begin{minipage}{0.48\hsize}
\begin{overpic}[width=6cm,clip]{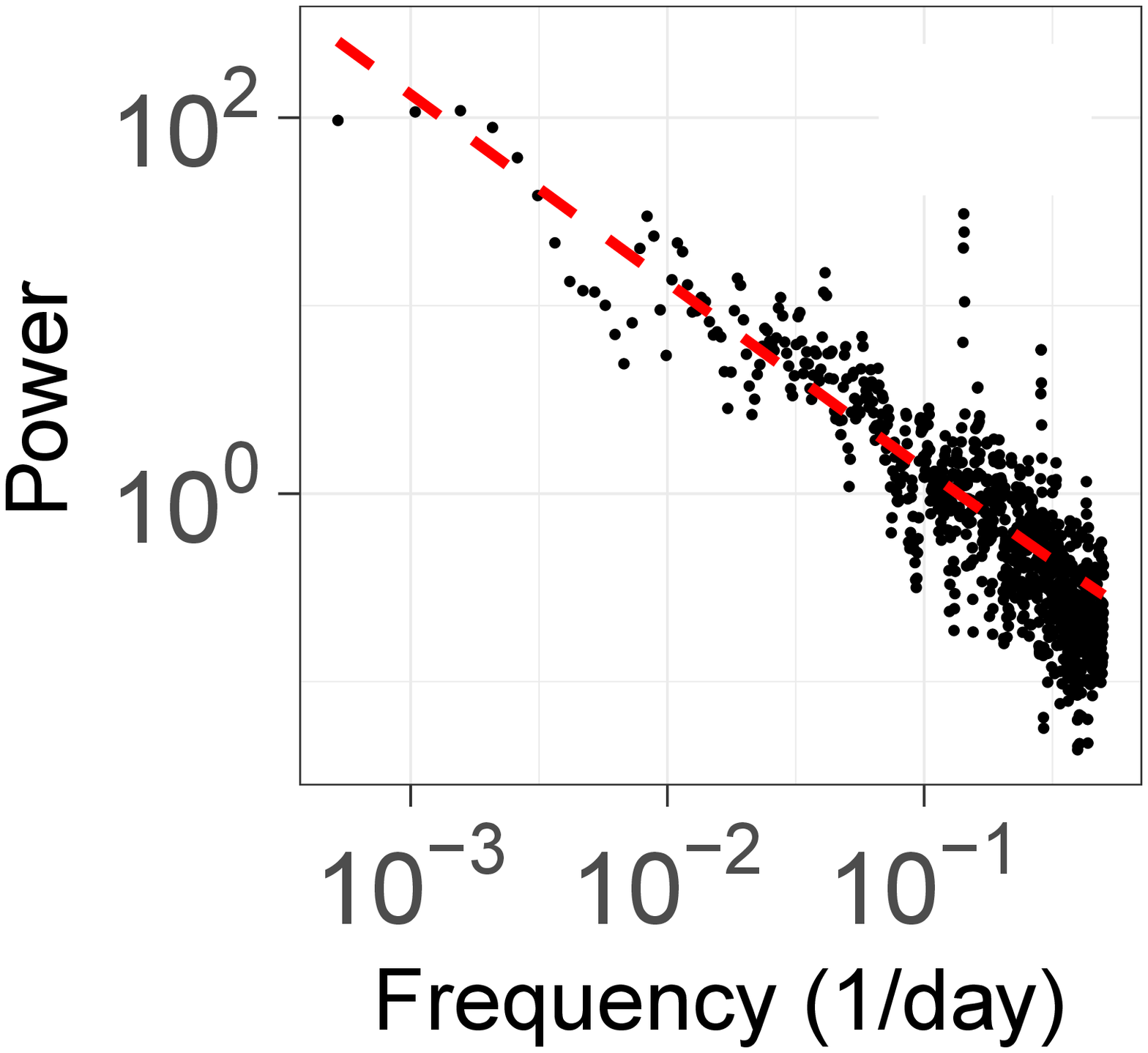}
\put(80,78){(b)}
\end{overpic}
\end{minipage}
\caption{
(a) Empirical MSD for the time series of word counts of  ``ooi'' in Japanese, $<f_j(L)^2>$ (``many'' in English; $\check{c}_j=22369.8$m $\check{\eta}_j=0.026$, $\check{\delta}_j= 0.012$). 
The dashed red line corresponds to the theoretical MSD of the model given by Eq. \ref{eq_rd}.  
Specifically, the theoretical curve is $<f_j(L)> =p_i <r_j(L)> + q_j$  given by Eq. 30 in Ref. \cite{watanabe2018empirical}, where  $p_{j}=2  \check{c}_j^2 \check{\eta}_j^2 /\Gamma(1/2)^2$, $q_j= \check{c}_j^2 \check{\eta_j}^2 (-2\log(4)-2\psi^{(0)}(\Gamma(1/2)^{-2}))/\Gamma(1/2)^2 +2 \check{c}_j^2 \check{\delta}_j+2 \check{c}_j$, and $\psi^{(0)}(x)$ is the digamma function.  
(b) Corresponding empirical PSD $P(\nu;f_j)$. The dashed red line corresponds to the theoretical PSD given of the model given by Eq. \ref{eq_rd}.
Specifically, the theoretical curve is $P(\nu,f_j) =v_j P(\nu,r_j(L)) + w_j$  given by Eq. 31 in Ref. \cite{watanabe2018empirical}, where $v_j=\check{c}_j \check{\eta}_j/(2 \pi)$ and $w_j= \check{c}_j^2 \check{\delta}_j^2+\check{c}_j$. 
 From panels (a) and (b), we can confirm that the theoretical curve is in good agreement with the empirical data. 
}
\label{fig_spect0}
\end{figure*}
\section{FS in empirical data}
\textcolor{black}{First, we discuss temporal FS (TFS), which is the main observable of this study (where, for simplicity, we call TFS just FS in this paper). }
 FS of the difference in the time series $\{A(t)\}$ ($t=1,2,\cdots, T$) 
is defined by the scaling between a temporal mean $E[A]$ and a temporal variance of the difference in time series $V[\delta A]$, 
\begin{equation}
V[\delta A] \propto E[A]^{\alpha}. \label{TFS_of_A}
\end{equation}
Here, the temporal mean $E[A]$ and temporal variance $V[A]$ are defined by 
\begin{equation}
E[A] \equiv \sum^{T}_{t=1}A(t)/T,
\end{equation}
\begin{equation}
V[\delta A] \equiv \sum^{T-1}_{t=1}(\delta A(t)-E[\delta A])^2/(T-1), 
\end{equation}
where $\delta A(t)$ means the difference at time $t$: $\delta A(t)=A(t)-A(t-1)$.
 Note that the above definition of FS in Eq. \ref{TFS_of_A} is expressed in terms of the variance, although the standard deviation is usually used in observations, where FS expressed by the standard deviation can be written as $V[\delta A]^{1/2} \propto E[A]^{\alpha/2}$. 
\textcolor{black}{ In addition, we assume in this section that $T>>1$ for simplicity.}
\subsection{FS for the daily time scale}
Herein, we investigate FS of the daily differential of the number of word appearances, $\delta f_j(t)=f_j(t)-f_j(t-1)$ (see Fig. \ref{tseries}(d)), 
which has been already studied intensively in Ref. \cite{RD_base}. 
From the black triangles in Fig. \ref{fig_TFS}, we can confirm the scaling with two exponents: 
\begin{equation}
V[\delta f_j] \propto
\begin{cases}
 E[ f_j]^{1}, & \text{$E[f_j] < \mu^{*}$}, \\
 E[ f_j]^{2}, & \text{$E[f_j] > \mu^{*}$}, 
\end{cases} 
\end{equation}
where $\mu^{*} \approx 100$. 
In addition,
we can also calculate the theoretical lower bound of this scaling by using the random diffusion model, which is mentioned in Section \ref{sec_model} \cite{RD_base},  
\textcolor{black}{
\begin{eqnarray}
V[\delta f_j]& \approx & 2 \cdot E[f_j] \cdot E[\frac{1}{m}]+E[f_j]^2 \cdot \{ 2 \cdot  \check{\delta}_0^2 \}, \nonumber \\ \label{v_delta_tilda_f2}
\end{eqnarray}} where $\check{\delta}_0$ is the parameter depending on the system and here \textcolor{black}{ $\check{\delta}_0^2=0.030^2$.}
This lower bound is shown by the gray dashed line in Fig. \ref{fig_TFS}(a). 
\par
\subsection{Analysis of the rescaling of the FS of word appearance data}
In this section, we investigate the time-scale dependence of FS (i.e., perform an analysis of the rescaling) to extract essential information of the dynamics of the time series.
In particular, we use the box means for the time-scale coarse-graining: 
\begin{eqnarray}
F_j^{(L)}(\tau^{(L)}) \equiv \sum^{L-1}_{i=0}f_j(L \cdot \tau^{(L)}+i)/L \quad (t=1,2,\cdots, T^{(L)}), \nonumber \\
\label{eq_F}
\end{eqnarray}
where $\tau^{(L)}=1,2,T^{(L)}$ is an index of time for the $L$-day scale, namely, $t=\tau^{(L)} \times L$. 
For example, $L \times F_j^{(L)}(\tau^{(L)})$ corresponds closely to a (normalized) weekly word-appearance time series for $L=7$, a monthly time series for $L=30$, and a yearly time series for $L=365$. 
\textcolor{black}{Fig. \ref{tseries}(e) shows an example of the box means time series of the time-scale coarse-graining for $L=30$ (i.e., monthly time series), \{$F_j^{(30)}(\tau^{(30)})\}$, and Fig. \ref{tseries}(f)  shows the corresponding differential time series, $\{\delta F_j^{(30)}(\tau^{(30)})\}$.  }
\par
The mean and variance of the difference of this value are defined in the same way as shown in Eq. \ref{TFS_of_A}: 
\begin{equation}
E[F_j^{(L)}]=\sum^{T^{(L)}}_{\tau^{(L)}=1}F_j(\tau^{(L)})/T
\end{equation}
and
\begin{equation}
V[\delta F_j^{(L)}] \equiv \sum^{T^{(L)}-1}_{\tau^{(L)}=1} (\delta F_j(\tau^{(L)})-E[\delta F_j(\tau^{(L)})])^2/(T^{(L)}-1).  
\end{equation}
The FS of the coarse-grained time series $\{F_j^{(L)}(\tau^{(L)})\}$ for $L=1$, $L=7$, $L=30$, and $L=365$ is plotted in Fig \ref{fig_TFS}(a).
From this figure, we can observe that the scaling with two exponents (i.e., kinked lower bounds) is similar to the time scale of a day ($L=1$).   
The lines in Fig \ref{fig_TFS} indicate the theoretical curve 
\begin{equation}
V[\delta F_j^{(L)}] = a(L) \cdot E[F_j^{(L)}]+ b(L) \cdot E[F_j^{(L)}]^2.   \label{V_df_emp}
\end{equation}
This equation for $L=1$ is consistent with Eq. \ref{v_delta_tilda_f2}. 
In Fig \ref{fig_TFS}(a), we set $a(L)=2 \times a_0/L$ and $b(L)=2 \times b_0/L$, which are obtained using the central limit theorem under the assumption that $\{f_j(t)\}$ is independent 
(and we use \textcolor{black}{$a_0=E[1/m] \approx 1$ and $b_0=\check{\delta}_0^2=0.030^2$}). 
These lines are in good agreement with the empirical lower bounds for a small mean $E[F^{(L)}]$ \textcolor{black}{(the right part of panel (a))}. However, they are in disagreement for a large mean $E[F^{(L)}]$ \textcolor{black}{(the left part of panel (a))}.  These results imply that the assumption that  $\{f_j(t)\}$ is independent does not hold. \par
Fig. \ref{fig_coef}(a) shows a result in which $b(L)$ is directly estimated from data using the method described in \ref{app_coef}.
From this figure, we can see that $b(L)$ does not have a clear dependence on $L$ for large $L$, namely, 
\begin{equation}
b(L) \approx {\rm constant} \quad (L>>1). \label{emp_b_new}
\end{equation}
\par
\textcolor{black}{
 Note that Fig. \ref{fig_coef}(b) shows the examples and the distribution of $b_j(L) \equiv V[\delta F_j^{(L)}]/E[F_j^{(L)}]^2$.  
 Because of Eq. \ref{V_df_emp}, $b_j(L)$ roughly corresponds to $b(L)$ for words with large $E[F_j^{(L)}]^2$. From this figure, we can see $b_j(L)=\rm constant$ holds not only for the whole set of words but also for individual words. 
 It can also be seen that the theoretical curve of $b(L)$ shown in Fig \ref{fig_coef}(a) and given in Eqs. \ref{u_ans} and \ref{b_LL2} roughly corresponds to the lower bound of the curve of $\{b_j(L)\}$ of individual words.
}
%
%
%

\begin{table*}
\centering
\begin{tabular}{ccc}
\hline
 Speed of forgetting & MSD &  FS   \\
 \hline
 $\beta=0$  & $L$ & $L$ \\
 $0< \beta <0.5$  & $L^{1-2\beta}$ & $L^{1- 2\beta}$  \\
 $\beta=0.5$ (blog data)  & $\log(L)$ & $O(1)$ \\
 $0.5 <\beta< 1$ & $O(1)$ &  $L^{1-2\beta}$  \\
 $\beta =1$  & $O(1)$ & $\log(L)^2 L^{-1}$ \\
 $\beta >1$  & $O(1)$ & $L^{-1}$ \\
\hline
\end{tabular}
\caption{Comparison between the dominant term of MSD and $b(L)$ of FS}
 \label{table1}
\end{table*}

\section{Model}
\textcolor{black}{
	In this section we explain \textcolor{black}{the aforementioned time-scale dependence of FS} by a \textcolor{black}{combination of two probabilistic models: (i) the} random walk model with power-law forgetting and (ii) the random diffusion model (i.e., a kind of Poisson point process) \cite{watanabe2018empirical}.
  The random walk model describes the latent concerns of the focused word $r_j(t)$ and it can essentially explain the ultraslow diffusion. The random diffusion model expresses the connection between the latent concern $r_j(t)$ described in the above-mentioned random walk model and the observable word counts $g_j(t)$ or $f_j(t)$.} \par
   \textcolor{black}{
 In a previous study \cite{watanabe2018empirical}, the model given by Eq. \ref{eq_rd} was shown to be able to reproduce well the characteristics of time series of word counts of well-established words. 
 In fact, Fig. \ref{fig_spect0} indicates that the model can reproduce the primary properties of the word count time series (i) logarithmic diffusion and (ii) power-law spectrum. 
 The contribution of this study is to show that the model, which is already known to reproduce these two properties, can also reproduce the time-scale dependence of FS. } \par
Here, first we introduce and discuss the random walk model with power-law forgetting. Next, we introduce the word count model, which is a combination of the random walk model and the random diffusion model.  
\subsection{Power-law forgetting process}
\label{sec_rw}
	\textcolor{black}{Here}, we present a random walk with power-law forgetting, which is one of the most representative standard explanations of anomalous diffusion in previous studies. This approach is also equivalent to the fractional dynamics approach (in our case, the fractional Langevin equation approach). \par
	The random walk model with power-law forgetting is given by 
	\begin{equation}
	r_j(t)=\frac{1}{Z(\beta)} \sum_{s=0}^{\infty} (s+d_\beta)^{-\beta} \cdot \eta_j(t-s), \label{eq_rw}
\end{equation}
where $Z(\beta)>0$ is an arbitrary coefficient, $d_\beta=Z(\beta)^{-1/\beta}$, $\beta$ is a constant used to characterize the forgetting speed, and $\eta_j(s)$ is independent and identically distributed (i.i.d.) noise where the mean takes zero and the standard deviation is $\check{\eta}_j$; that is, we can write ${\eta}_j(s)=\check{\eta}_j \times \eta^{(0)}_j(s)$. Here, $\eta^{(0)}_j(s)$ is i.i.d. noise where the mean takes zero and the standard deviation is 1. \par
\par
From Ref. \cite{watanabe2018empirical}, the MSD of this model for $L>>1$ is calculated by using  
\textcolor{black}{
\begin{eqnarray}
\left<r_j(L)^2 \right> \propto 
\begin{cases}
L^{1-2\beta}  & (0 \leq \beta <0.5), \\
\log(L) &(\beta=0.5),\\
O(1) &(\beta > 0.5), \\
\end{cases}
\label{th_msd}
\end{eqnarray}
and the power spectral density (PSD) in the case in which  $Z(\beta)=\Gamma(1-\beta)$ and $0<\beta<1$ is calculated as 
\begin{equation}
P(\nu;r_j) \approx \check{\eta}_j^2 (2 \sin(2 \pi \nu /2))^{-2(1-\beta)},
\end{equation}
where $\Gamma(x)$ is the gamma function.
Here, the MSD for the data is defined by
\begin{equation}
\left<r_j(L) \right> = \frac{\sum^{T-L}_{t=1} (r_j(t+L)-r_j(t))^2}{T-L}, \label{emp_MSD} 
\end{equation}
and the PSD is defined by
\begin{equation}
P(\nu;r_j) = \frac{1}{T} \left|\sum^{T}_{t=1}\exp(-i 2 \pi \nu t) r_j(t) \right|^2  ,
\end{equation}
where $\nu$ is the frequency [1/days].}
For $\beta = 0.5$, which corresponds to the blog data, the MSD is approximated by 
\begin{equation}
\left<r_j(L)^2 \right> \propto  \log(L) 
\end{equation}
and the PSD is approximated by
\begin{equation}
P(\nu;r_j) \propto 1/\nu. \label{psd_1_f}
\end{equation}
The PSD, which is inversely proportional to the frequency (i.e., $\propto 1/\nu$ ) indicates that this time series is a type of the well-known $1/f$ noise.
\textcolor{black}{
We can see the logarithmic-like MSD from the approximate straight line in the semi-log plot in Fig. \ref{fig_spect0}(a) and the $1/f$ noise-like PSD from the approximate straight line in the log-log plot in Fig. \ref{fig_spect0}(b).}
\textcolor{black}{The details of why logarithmic diffusion is derived from the model given by Eq. \ref{eq_rw} are discussed in Appendix B in Ref. \cite{watanabe2018empirical} and a rough derivation is given in the \ref{app_MSD} in this paper.} \par
\textcolor{black}{
Note that the continuous version of Eq. \ref{eq_rw} corresponds to the fractional Langevin equation, which is the expansion of the Langevin equation  \cite{eab2011fractional,magdziarz2007fractional,watanabe2018empirical}. 
In the case of the word counts, namely, $\beta=0.5$ and $Z(\beta)=\Gamma(1-\beta)$, the continuous version of Eq. \ref{eq_rw} is written by the half-order fractional Langevin equation
\begin{equation}
\sqrt{\frac{d}{dt}}r_j(t) \approx  \eta_j (t), \label{eq_sqrd_rw}
\end{equation}
where the operator is the special case of the Riemann--Liouville fractional derivative operator satisfying $\sqrt{\frac{d}{dt}} \cdot \sqrt{\frac{d}{dt}} x(t)= \frac{d}{dt} x(t)$ \cite{watanabe2018empirical}.
The details of the model are provided in Ref. \cite{watanabe2018empirical}. 
} 
\par
\textcolor{black}{Additionally, the empirical MSD given by Eq. \ref{emp_MSD} and the theory Eq. \ref{th_msd} are based on the time-averaged MSD of a single particle \cite{cherstvy2021scaled}.  
  Refs. \cite{cherstvy2017time} and \cite{cherstvy2021scaled} pointed out the time-averaged MSD is different the ensemble MSD in financial time series and the geometric Brownian motion and Ref. \cite{liang2019survey} in some kinds of ultraslow diffusion. 
  The difference between the time-averaged MSD and the ensemble MSD of the random walk model with the power law forgetting given by Eq. \ref{eq_rw} are briefly discussed in \ref{app_MSD} and we were able to confirm that two MSDs are different.
Although the empirical ensemble fluctuation related to the ensemble MSD for $L=1$ is discussed in Ref. \cite{RD_base}, the details of the ensemble MSD of word count time series are still uncertain and left for future research. }

\subsection{Model of word counts}
\label{sec_model}
\textcolor{black}{We use the RD model introduced in \cite{PhysRevLett.100.208701,PhysRevE.87.012805,sano2009,RD_base} to sample $g_j(t)$ or $f_j(t)$.  This model connects the essential dynamics of the concern of word $r_j(t)$ given by Eq. \ref{eq_rw} (i.e., the latent value) with the time series of word counts, $g_j(t)$ or $f_j(t)$ (i.e., the observed value).} 
The RD model is a kind of point process, which can be deduced from the simple model of the writing activity of independent bloggers \cite{RD_base}.
\textcolor{black}{In this model, values are sampled from the Poisson distribution in which the rate (or intensity) function is determined by a random variable or a stochastic process (i.e., the doubly stochastic Poisson process \cite{lowen2005fractal}).} In the case of blogs, the rate function is connected to the latent concern of word $r_j(t)$.
Particularly, the RD model is given by \cite{RD_base}
\begin{equation}
g_j(t) \sim Poi( \Lambda_j(t)), \label{eq_rd} 
\end{equation}
and its rate function of the Poisson distribution (denoted by $Poi(.)$), $\Lambda_j(t)$, is determined by the following definition of the product:  
\begin{equation}
\Lambda_j(t) \sim  m(t) \times \check{c}_j \times r_j(t)  \times (1+\check{\delta}_j \times \Delta^{(0)}_j(t)).\label{Lambda} 
\end{equation}
In this equation, the terms are as follows: 
\begin{itemize}
\item $m(t)$ is the (normalized) total number of blogs, dound by assuming that $\sum_{t=1}^{T}m(t)/T \approx 1$ for normalization (see Fig. \ref{tseries}(b)), where $m(t)$ is estimated by the ensemble median of the number of words at time $t$, as described in \ref{app_m}. 
\item $\check{c}_j$ is the scale of the $j$th word, namely, the temporal means of the $j$th word, where we estimate the mean of the raw word count of data $\check{c}_j \approx \sum_{t=1}^{T}g_j(t)/T$. 
\item $r_j(t)$ is the scaled time variation of concern of the $j$th word sampled from Eq. \ref{eq_rw} for our model, where we set $r_j(1)=1$.
\item $\check{\delta}_j$ is the magnitude (i.e., the standard deviation) of the ensemble fluctuation, which may be related to the magnitude of the heterogeneity of bloggers \cite{RD_base}. 
\item $\Delta^{(0)}_j(t)$ is the normalized ensemble fluctuation, which is sampled from the system-dependent random variable with a mean of $0$ and a standard deviation of $1$. 
\end{itemize}
\textcolor{black}{
More detailed explanations of the model are given in Ref. \cite{watanabe2018empirical}  (e.g., for the case $r_j(t)<0$, where the model is no longer valid, etc.). }
\par
\section{FS of the RD model}
\label{sec_fs_model}
Here, we calculate the FS of the RD model.
First, we introduce random variables $w_j(t)$, $\check{C}_j^{(L)}$, $R_j^{(L)}(t^{(L)})$, and $W_j^{(L)}(t^{(L)})$ for simplicity. \par
$w_j(t)$ is defined as
\begin{equation}
w_j(t) \equiv g_j(t)/m(t)-c_j(t)=f_j(t)-c_j(t). \label{base_u}
\end{equation}
Using this variable, we can write 
\begin{equation}
f_j(t)=\check{c}_j r_j(t)+w_j(t). \label{base_u2}
\end{equation}
From the definition, the mean of $w(t)$ is  
\begin{equation}
<w_j(t)>_w=0
\end{equation}
and, from Ref. \cite{RD_base}, the variance of $w_j(t)$ is 
\begin{eqnarray}
&&<(w_j(t)-<w_j(t)>)^2>_w=<w_j(t)^2>_w \nonumber \\
&& = 1/m(t) \cdot c_j(t) + {\check{\delta}_j}^2 \cdot c_j(t)^2,  \label{ww} 
\end{eqnarray}
where $<A>_w$ is the mean with respect to $w$, $<A>_w=\int A(w) p_w(w)dw$.
 \par
We also define the box means $\check{C}_j^{(L)}$, $R_j^{(L)}(t^{(L)})$s and $W_j^{(L)}(t^{(L)})$, corresponding to $\check{c}_j$, $r_j(t)$s and $w_j(t)$ as 
\begin{eqnarray}
&&\check{C}_j = \check{c}_j, \\
&&R_j^{(L)}(t^{(L)}) \equiv \frac{1}{L}\sum^{L-1}_{t=0}r_j(L  t^{(L)}+t), \label{sd_L_0}
\end{eqnarray}
\begin{equation}
W_j^{(L)}(t^{(L)}) \equiv \frac{1}{L}  \sum_{t=0}^{L-1}w_j(L  t^{(L)}+t).  \label{W0}
\end{equation}
Using these values, we can write the time-scale coarse-grained equation corresponding to Eq. \ref{base_u2} as 
\begin{equation}
F_j^{(L)}(t^{(L)})=\check{C}_j R_j^{(L)}(t^{(L)})+W_j^{(L)}(t^{(L)}).  \label{large_f}
\end{equation}
\par
Second, calculating the variance $V[\delta F_j^{(L)}]$, we can obtain 
\begin{eqnarray}
&&V[\delta F_j^{(L)}] \approx a(L) E[ F_j^{(L)}] +b(L) E[ F_j^{(L)}]^2,  \label{df_ans0}
\end{eqnarray}
where 
\begin{equation}
a(L)=\frac{2}{L} a^{(0)},
\end{equation}
\begin{eqnarray}
b(L)&=&V[\delta R_j^{(L)}]+\frac{2b^{(0)}_j(1+V[r_j])}{L} \label{b_LL} \\
     &\geq&V[\delta R_j^{(L)}]+\frac{2b^{(0)}_j}{L}, \label{b_LL2}
\end{eqnarray}
$a^{(0)}=E[1/m]$, and $b^{(0)}_j={\check{\delta}_{j}}^2$. 
The details of the derivation of the variance are provided in \ref{app_long_rd}. 
From Eq. \ref{df_ans0}, we can confirm that the RD model reproduces the empirical properties for a small mean $E[F^{(L)}]$, that is, $a(L) \propto 1/L$ (the left side in Fig. \ref{fig_TFS}(a)). In addition, we can also confirm that $b(L)$, which dominates the properties for a large mean $E[F^{(L)}]$ (i.e., the right side in Fig. \ref{fig_TFS}(a)), is determined by $V[\delta R_j^{(L)}]$, which is determined based on the  property of the dynamics of $\{R_j(t)\}$. 
\par
\subsection{Relation between TFS and the dynamics of $\{r_j(t)\}$}
\textcolor{black}{What are the dynamics or conditions that make $b(L) \approx \rm constant$ for $L>>1$, which is an empirical finding shown in Fig. \ref{fig_coef}(a)? 
To clarify this question, we studied the relation between $b(L)$ or $V[R^{(L)}_j]$ and the dynamics of $\{r_j(t)\}$ given by Eq. \ref{eq_rw}. }
\par
An approximation formula for $V[\delta R^{(L)}_j]$ for this model is given in Eq. \ref{u_ans} in \ref{app_forget}.
For $L>>1$, the main terms of Eq. \ref{u_ans_l} are given by
\begin{eqnarray}
&&V[\delta R^{(L)}_j] \approx \frac{\check{\eta}_j^2}{Z(\beta)^2} \nonumber \\
&&\begin{cases}
u_1(\beta)L^{1-2\beta}+u_2(\beta)L^{-\beta}+u_3(\beta)L^{-1} &\text{($0<\beta<1$)}, \\
u_a \log(L)^2 L^{-1} + u_b \log(L) L^{-1}+u_c L^{-1} &\text{($\beta=1$)}, \\
u_3(\beta)L^{-1} & \text{($\beta>1$)}, 
\end{cases}
\label{eq_b}
\end{eqnarray}
where $u_1(\beta)$, $u_2(\beta)$, $u_3(\beta)$, $u_a$, $u_b$, and $u_c$ are $L$-independent coefficients given by Eqs. \ref{u_ans_l_start}--\ref{u_ans_l_end} in \ref{app_l_forget}.
Thereby, the maximum term of a series can be written as 
\begin{eqnarray}
b(L) \approx V[\delta R^{(L)}_j] \propto 
\begin{cases}
L^{1-2\beta} &\text{($0<\beta<1$)}, \\
\log(L)^2 L^{-1} &\text{($\beta=1$)}, \\
L^{-1} & \text{($\beta>1$)} .
\end{cases}
\label{b_max}
\end{eqnarray}
\par
From this result, we can confirm that the empirical result given by Eq. \ref{emp_b_new},  
 $b(L) \approx \rm constant$, 
is reproduced  under the condition 
\begin{eqnarray}
\beta \approx 0.5.
\end{eqnarray}
\textcolor{black}{This condition, $\beta=0.5$, is consistent with the parameter that can explain both the empirical ultraslow diffusion and the $1/f$ noise discussed in Section \ref{sec_rw} or Ref. \cite{watanabe2018empirical}.}
\par
The red dashed line in Fig. \ref{fig_coef}(a) indicates the theoretical curve in which we insert Eq. \ref{u_ans} into Eq. \ref{b_LL2} for the parameter $\beta=0.5$ and $Z(\beta)=\Gamma(1-\beta)$.
From this figure, we can confirm that the theoretical curve is in accordance with the empirical lower bound. 
 In addition, the corresponding theoretical curve in Fig \ref{fig_TFS}(b) is in 
agreement with the empirical data (\textcolor{black}{for $\check{\eta}_{j}=0.029$, $\check{\delta}_{j}=0.030$}). \par
\textcolor{black}{Note that the dynamics given in Eq. \ref{eq_rd} is the common background dynamics of the word count time series of all words. When there are other additional effects such as breakthrough news or seasonalities, $V[\delta F_j^{(L)}]$ becomes larger than that from the theory based on Eq. \ref{eq_rd}. This is one of the reasons why the theoretical curve corresponds to the lower bound of the graph. Another reason is the word dependency of $\check{\eta}_j$.} \par
\subsection{Another dynamics model}
Finally, we confirm that an example of a typical time series model other than the proposed model does not satisfy the empirical properties of FS.
In the case of a random walk with dissipation $\kappa \geq 0$ and an external force $u_j(t) \geq 0$, 
\begin{equation}
r_j(t+1)= \kappa \cdot r_j(t)+u_j(t)+\eta_j(t), \label{eq_ar}
\end{equation}
where $\eta_j(t)$ is an i.i.d. random variable whose mean is 0, and we assume that the variance is $\check{\eta}_j^2<<1$ and that $r_j(t)>0$ takes nearly $1$. \par
From Eqs. \ref{app_rand_b0}, \ref{app_rand_b1}, and \ref{app_rand_b2} in \ref{app_random_walk}, we can obtain the variance for $T>>1$ and $L<<T$ as 
\begin{eqnarray}
V[\delta R_j^{(L)}] \approx 
\begin{cases}
 \check{\eta}_j^2 \cdot (\kappa-1)^{-2} \cdot L^{-1} & (0<\kappa<1), \\
\check{\eta}_j^2 \cdot 1/3 \cdot (2L+1/L) & (\kappa=1),\\
 O( \frac{\kappa^{2T}}{L(T-L)}) &  (\kappa>1).\\  
\end{cases}  
\end{eqnarray}
Thereby, the maximum term of a series for $L>>1$ is
\begin{eqnarray}
b(L) \approx V[\delta R_j^{(L)}] \propto 
\begin{cases}
  L^{-1} & (0< \kappa <1), \\
 L & (\kappa=1),\\
 O( \frac{\kappa^{2T}}{L(T-L)}) &  (\kappa>1).\\  
\end{cases} 
\end{eqnarray}
In all cases, these results disagree with the empirical result, $b(L) \approx \rm constant$, given by Eq. \ref{emp_b_new}. 
The details of the derivations and the results of the variance $V[\delta R_j^{(L)}]$ for the case of a random walk are provided in \ref{app_random_walk}.  \par
\textcolor{black}{Note that the random walk with dissipation given by \ref{eq_ar} for $u_j=0$ and $k \leq 1$ can be rewritten in the same form as Eq. \ref{eq_rw}, 
\begin{equation}
r_j(t)=\frac{1}{Z(\beta)} \sum_{s=0}^{\infty} \kappa^{s} \cdot \eta_j(t-s) \label{eq_ar2}.
\end{equation}
This equation corresponds to an exponential forgetting process.  The exponential decay is the most typical way of forgetting, however, it does not reproduce for all of the statistics of word counts time series we are focusing on: (i)time-scale-independent FS, (ii)logarithmic MSD, and (iii)$1/f$ noise-like PSD. }

\section{Discussion and conclusions }
In this study, we investigated the relationship between FS and anomalous diffusion driven by the random walk model with power-law forgetting given by Eq. \ref{eq_rd}. 
 By analyzing both the word count data and the theoretical  model, we showed that the ultraslow diffusion is linked to the fact that FS is fundamentally independent of the time scale $L$ (see Figs. \ref{fig_TFS} and \ref{fig_coef};  $b(L) =\rm constant$).  
 Furthermore, we theoretically derived the general relationship between FS and anomalous diffusion that is derived by using the model (Eq. \ref{b_max} and Table \ref{table1}). \textcolor{black}{In the context of modeling word count time series, it was demonstrated that the model given by Eq. \ref{eq_rd} for $\beta=0.5$ can explain the time-scale dependence of FS in addition to the previously known ultraslow diffusion and the PSD.  \textcolor{black}{Note that the results are limited to word count time series data of written language. Validation for languages in general is a future task. }
 } \par 
The fact that $b(L) \approx \rm constant$ implies the existence of ultraslow diffusion.
 In other words, by observing FS, we can indirectly verify the existence of ultraslow diffusion, which we cannot easily observe directly.
The advantages of indirect observations using FS are as follows:  
\begin{itemize}
 \item The implementation of the method is simple.  Basically, we only need to compute the means and the variances of time series by changing the time scale.  In the terminology of statistics, the method is a kind of analysis of the overdispersion for the Poisson process.
 \item The method allows for the integrated analysis of multiple time series. We use a graph that reflects all the data by mapping each time series to a single point, such as Fig. \ref{fig_TFS}. From the graph, we can visually and comprehensively read the information about the dynamics shared by the time series.
 \item The resolution of FS to distinguish the value of $\beta$ of the time series is higher than that of the MSD. From Table \ref{table1}, for $\beta > 0.5$, the MSD takes $O(1)$. However, in FS, the power-law exponent varies continuously for $\beta>0.5$. For example, it is difficult distinguish the time series with $\beta=0.51$ (i.e., nearly ultraslow diffusion) from that with $\beta=0.99$ (i.e., nearly i.i.d. noise) by using the MSD because both MSDs have $<f_j(L)^2> \propto O(1)$.  However, it is easy to distinguish by using FS, because $b(L) \propto L^{-0.02}$ for $\beta=0.51$ and  $b(L) \propto L^{-0.98}$ for $\beta=0.99$. 
\end{itemize}
Note that this condition, $b(L) \approx \rm constant$, alone does not determine the ultraslow diffusion---it only suggests the possibility of it.
Therefore, we need to add additional evidence, such as through careful direct observation, to identify it completely. \par
Ultraslow diffusion has rarely been observed in the real world.  We hope that our findings (i.e.,  the detection by using $b(L) \approx \rm constant$) will contribute to the discovery of ultraslow diffusion in the real world, which has previously been overlooked because of it being mixed with noise and other complicating factors. 
In reality, the empirical ultraslow diffusion reported in Ref. \cite{watanabe2018empirical} were indirectly found from the analysis of FS as described in this paper, although we did not write about it in the previous paper to focus the theme of the paper on anomalous diffusion.
In particular, time series data with $1/f$ noise, which has been observed in various systems, may offer a particularly high potential for such discovery,  
because time series driven by the random walk model with power law forgetting is characterized by $1/f$ noise (Eq. \ref{psd_1_f} and Fig. \ref{fig_spect0}(b)) as well as the FS discussed in this study.
\par
\section*{acknowledgments}
The authors would like to thank Hottolink, Inc., for providing the data. This work was supported by Leading Initiative for Excellent Young
Researcher (LEADER) of the Ministry of Education, Culture, Sports, Science and Technology in
Japan and JSPS KAKENHI Grant Numbers \textcolor{black}{17K13815} and 21K04529. We would like to thank Editage (www.editage.com) for English language editing.

\section*{Contributions}
H.W. conceived the presented idea. H.W. performed the data analys and theoretical calculation. H.W. discussed the results and contributed to the final manuscript.

\bibliography{adj6c_pnas}
\appendix
\renewcommand{\theequation}{A\arabic{equation}}
\renewcommand{\thefigure}{A-\arabic{figure}}
\renewcommand{\thetable}{S-A\arabic{table}}
\setcounter{figure}{0}
\setcounter{equation}{0}
\section{Estimation of $b(L)$}
\label{app_coef}
We use the following procedure to estimate $b(L)$ given by Eq. \ref{V_df_emp} from the actual data: 
\begin{enumerate}
\item We fix $L$.
\item We calculate $E[F_j^{(L)}]$ and $V[\delta F_j^{(L)}]$ for all words $(j=1,2,\cdots,W)$. 
\item We minimize $Q^{(L)}$ with respect to $b(L)$ under the condition $b(L)>0$. 
\end{enumerate}
Here, $Q^{(L)}$ is defined by
\begin{eqnarray}
&&Q^{(L)} \equiv s \cdot \{ \sum_{j \in \{j|Q_j^{(L)}>0,\; C_j \geq 100\}}\frac{{Q_j^{(L)}}^2}{N_p}\} \nonumber \\
&& +(1-s) \cdot \{\sum_{j \in \{j|Q_j^{(L)}<0,\; C_j \geq 100\}} 
 \frac{{Q_j^{(L)}}^2}{N_m} \},
\end{eqnarray}
where  
\begin{equation}
Q_j^{(L)}=\log{(V[\delta F_j^{(L)}(\tau^{L})]^{1/2})}-\log{((\frac{2}{L} \cdot C_j+b(L) \cdot C_j^2)^{1/2})} \label{QjL},
\end{equation}
$C_j=E[F_j^{(L)}]$, $N_p=\sum_{j \in \{j|Q_j^{(L)}>0, C_j \geq 100\}} 1$, and $N_m=\sum_{j \in \{j|Q_j^{(L)}<0,\; C_j \geq 100\}}1$. 
Note that the minimization of the first term of Eq. \ref{QjL} corresponds to a reduction of the data beneath the theoretical lower bound in Eq. \ref{V_df_emp}. 
However, when we use only the first term, the estimation of $b(L)$ is strongly affected by outliers.  
Therefore, we use the second term to accept the data beneath the theoretical curve, and $s$ is the parameter used to control the ratio of acceptance.  
Here, we use $s=0.9$ in our analysis. In addition, the reason why we only use $\check{c}_j \geq 100$ is that we neglect words with a small $c_j$, as these do 
not affect the estimation $b(L)$ (see Eq. \ref{V_df_emp} and Fig. \ref{fig_TFS}). 
\par
\renewcommand{\theequation}{B\arabic{equation}}
\renewcommand{\thefigure}{B-\arabic{figure}}
\renewcommand{\thetable}{S-B\arabic{table}}
\setcounter{figure}{0}
\setcounter{equation}{0}
\section{$V[\delta F^{(L)}_j]$ for given $\{r_j(t)\}$} 
\label{app_long_rd}
We calculate $V[ \delta F_j^{(L)}]$ for given $\{r_j(t)\}$. 
Using Eq. \ref{large_f}, we can decompose $V[ \delta F_j^{(L)}]$ as 
\begin{eqnarray}
&&V[ \delta F_j^{(L)} (\tau^{(L)})]=\check{C_j}^2 V[\delta R_j^{(L)}]+ V[\delta W_j^{(L)}] \nonumber \\
&&+ 2 \cdot \check{C}_j E[ \delta R_j^{(L)}  \delta W_j^{(L)}]. \label{deltaF_d}
\end{eqnarray}
\par
First, we calculate the second term in Eq. \ref{deltaF_d}.  
The second term is written as $V[\delta W_j^{(L)}]=E[\delta {W_j^{(L)}}^2]-E[\delta {W_j^{(L)}}]^2$,
where $E[\delta {W_j^{(L)}}^2]$ is given by 
\begin{eqnarray}
&&E[\delta {W_j^{(L)}}^2]= \frac{1}{T^{(L)}-1} \sum^{T^{(L)}-1}_{t^{(L)}=1} \{W_j^{(L)}(t^{(L)}+1) \nonumber \\
&&-W^{(L)}(t^{(L)})\}^2  \\
&\approx& \frac{1}{L} \{ \sum^{T}_{t=L+1}\frac{w_j(t)^2}{T-L} + \sum^{T-L}_{t=1} \frac{w_j(t)^2}{T-L} \},  \label{E_W3}
\end{eqnarray}
where we use the assumption that $T>>1$, $<w_j(t)>=0$, and $\{w_j(t)\}$ are independently distributed random variables.
Approximating the sums in Eq. \ref{E_W3} by Eq. \ref{ww}  (using the assumption $T>>1$ and $L \leq T/2$), we write 
\begin{equation}
E[\delta {W_j^{(L)}}^2] \approx \frac{2}{L}\{ E[1/m] \cdot \check{c}_j + {\check{\delta}_j}^2  (1+V[r_j]) \check{c}_j^2 \}.   \label{EWW}
\end{equation}
In the calculation, we also use the approximations  
\begin{eqnarray}
&& \sum^{T}_{t=L+1} \frac{r_j(t)}{T-L} \approx 1, \\
&& \sum^{T-L}_{t=1} \frac{r_j(t)}{T-L} \approx 1, 
\end{eqnarray}
and
\begin{eqnarray}
&&\frac{1}{2}\{\sum^{T}_{t=L+!}\frac{\{r_j(t)-\sum^{T}_{t=L+1}\frac{r_j(t)}{T-L}\}^2}{T-L} \nonumber \\
&&+ \sum^{T-L}_{t=1}\frac{\{r_j(t)-\sum^{T-L}_{t=1}\frac{r_j(t)}{T-L}\}^2}{T-L}\} \nonumber \\
&\approx & V[r_j].
\end{eqnarray}
These approximations are based on the assumption that $T>>L$ and $\{r_j(t)\}$ do not have a particular trend.
 \par
Next, we calculate $E[\delta {W_j^{(L)}}]^2$. Using Eq. \ref{W0}, we can estimate $E[\delta {W_j^{(L)}}]^2$ as follows:
\begin{eqnarray}
E[\delta {W_j^{(L)}}]^2&=&\{\frac{W_j^{(L)}(T^{(L)})-W_j^{(L)}(1)}{T^{(L)}-1} \}^2 \nonumber  \\
&& \approx \{\frac{O(1/\sqrt{L})}{T^{(L)}-1} \}^2 \approx O(\frac{L}{(T-L)^2}) \\
&&\approx
\begin{cases}
 O(1/T^2) &(L<<T), \\
 O(1/T)  & (L \approx T).
\end{cases} 
\label{EWEW}
\end{eqnarray}
Therefore, we can neglect this term for $T>>1$. \par
Consequently, inserting Eq. \ref{EWW} and Eq. \ref{EWEW} into Eq. \ref{deltaF_d}, and using $E[\delta C_j \delta W_j] \approx O(\sqrt{1/T}) \approx 0$ $(T>>1)$, 
we can obtain 
\begin{eqnarray}
&&V[\delta F_j^{(L)}] \approx a(L) \check{C}_j +b(L) \check{C}_j^2  \label{df_ans0}, 
\end{eqnarray}
where 
\begin{equation}
a(L)=\frac{2}{L} a_0
\end{equation}
and 
\begin{equation}
b(L)= V[\delta R_j^{(L)}]+\frac{2b_0(1+V[r_j])}{L}.
\end{equation}
Here, \textcolor{black}{$a_0=E[1/m]$ and $b_0=\check{\delta}_j^2$}.
%
\renewcommand{\theequation}{C\arabic{equation}}
\renewcommand{\thefigure}{C-\arabic{figure}}
\renewcommand{\thetable}{S-C\arabic{table}}
\setcounter{figure}{0}
\setcounter{equation}{0}
\section{$V[\delta R^{(L)}_j]$ for a random walk}
\label{app_random_walk}
We calculate $V[\delta R^{(L)}_j]$ for the following random walk with dissipation $\kappa \geq 0$ and external force $u(t)$: 
\begin{equation}
r(t+1)=\kappa  \cdot r(t)+u(t)+\eta(t),  \label{s_ran_a}
\end{equation}
where $<\eta(t)>_{\eta}=0$, $<(\eta(t)-<\eta(t)>_{\eta})^2>_{\eta}=\check{\eta}^2<<1$, $u(t)>0$, and $\kappa \geq 0$ and we omit the subscript $j$. 
By using Eq. \ref{s_ran_a}, $R(I)$, defined by \ref{sd_L_0}, is written as 
\begin{eqnarray}
&& R(I) = \frac{1}{L} \sum^{(LI-1)}_{t'=(I-1)L}[\sum^{t'-1}_{k=0} \kappa^k u(1+t'-k)  \nonumber \\
&& + \kappa^{t'} r(1) + \sum^{t'-1}_{k=0}\kappa^k \eta(1+t'-k)] \nonumber \\
&& = P_1(I)+P_2(I)+P_3(I).
\end{eqnarray}
Here, we define $P_1(I)$, $P_2(I)$, and $P_3(I)$ as follows:
\begin{eqnarray}
P_1(I) \equiv \frac{1}{L}  \sum^{(LI-1)}_{t'=(I-1)L} \sum^{t-1}_{k=0} \kappa^k  \cdot u(1+t'-k), \label{p1i}
\end{eqnarray}
\begin{eqnarray}
P_2(I) \equiv \frac{1}{L} \sum^{(LI-1)}_{t'=(I-1)L} \kappa^{t'} \cdot r(1), \label{p2i}
\end{eqnarray}
\begin{eqnarray}
P_3(I) \equiv \frac{1}{L}  \sum^{(LI-1)}_{t'=(I-1)L} \sum^{t-1}_{k=0} \kappa^k \cdot \eta(1+t'-k).  
\end{eqnarray}
Because $V[\delta R^{(L)}]$ can be decomposed as
\begin{equation}
V[\delta R^{(L)}_j] =V[\delta R]=E[\delta R^2]-E[\delta R]^2, \label{VR_rand}
\end{equation}
we calculate $E[\delta R^2]$ and $E[\delta R]^2$, respectively. \par
{\bf Calculation of $E[\delta R^2]$.}
Here, we calculate the first term of Eq. \ref{VR_rand}, $E[\delta R^2]$.
$\delta R(I)^2$ is denoted by 
\begin{eqnarray}
&& \delta R_j(I)^2  \nonumber \\
&=& (\delta P_1(I)+ \delta P_2(I)+ \delta P_3(I))^2  \nonumber \\
&\approx& \delta P_1(I)^2+ \delta P_2(I)^2+\delta P_3(I)^2+ 2 \cdot \delta P_1(I) \cdot \delta P_2(I) \nonumber. \\ \label{RJ2}
\end{eqnarray}
\par
We estimate the effects of the first term in Eq. \ref{RJ2}, $\delta P_1(I)^2$.
$\delta P_1(I)$ can be written as
\begin{equation}
\delta P_1(I)=\frac{1}{L^2} \cdot \sum^{L(I+1)}_{t=2}g_0(t,I) u(t), 
\end{equation}
where 
\begin{eqnarray}
&&g_0(t,T) = \nonumber \\ 
&&\begin{cases}
\frac{ (\kappa^L-1)^2 \cdot \kappa^{-t+1} \cdot \kappa^{(I-1)L} }{\kappa-1} , & 2 \leq t \leq (I-1)L+1 ,\\ 
\frac{(\kappa^L-2) \cdot \kappa^{(-t+1)} \cdot \kappa^{I \cdot L}+1}{\kappa-1}, & (I-1)L+2 \leq t \leq LI+1 , \\
\frac{\kappa^{(-t+1)} \cdot \kappa^{(I+1)L}-1}{\kappa-1}, & IL+2 \leq t \leq L(I+1). \\
\end{cases} 
\nonumber \\
\end{eqnarray}  
In addition, using the variables
\begin{equation}
u_0=\sum^{T}_{t=1}u(t)/T
\end{equation}
and
\begin{equation}
\delta u'(t)=u(t)-u_0, 
\end{equation}
we can write
\begin{eqnarray}
&&\delta P_1(I)^2  \nonumber \\
&=& \frac{1}{L^2} \{ \sum^{L(I+1)}_{t=2}g_0(t,I) u(t) \}^2 \\
&=& \frac{1}{L^2} \{u_0^2 \cdot [\sum^{L(I+1)}_{t=2}g_0(t,I)]^2+ [\sum^{L(I+1)}_{t=2}g_0(t,I) u'(t)]^2 \}^2 \nonumber \\
&\approx& \frac{1}{L^2} u_0^2 \cdot \{\sum^{L(I+1)}_{t=2}g_0(t,I) \}^2 \\
&=& \frac{1}{L^2} u_0^2 \cdot [\frac{\kappa^{(I-1)L} \cdot (\kappa^L-1)^2}{(\kappa-1)^2}]^2.
\end{eqnarray}
Hence, the temporal average of $\delta P_1(I)^2$ is obtained by 
\begin{eqnarray}
&&\bar{P}_1 \equiv \sum^{T^{(L)}-1}_{I=1} \frac{\delta P_1(I)^2}{T^{(L)}-1} \\
 &\approx& \frac{1}{T^{(L)}-1} \cdot \sum^{T^{(L)}-1}_{I=1}  \frac{1}{L^2} u_0^2 \cdot [\frac{\kappa^{(I-1)L} \cdot (\kappa^L-1)^2}{(\kappa-1)^2}]^2 \nonumber \\
&=& \frac{u_0^2}{L^2 \cdot(T^{(L)}-1)} \frac{(\kappa^L-1)^3  (\kappa^{2L(T^{(L)}-1)}-1)}{(\kappa-1)^4 \cdot (\kappa^L+1)}.  \label{P1_m} 
\end{eqnarray}
\par
Similarly, we estimate the effects of $\delta P_2(I)^2$ as  
\begin{eqnarray}
\delta P_2(I)^2 &=&  \frac{r(1)}{L^2} \cdot \frac{\kappa^{2IL}(\kappa^L+\kappa^{-L}-2)^2}{(\kappa-1)^2}.
\end{eqnarray}
Therefore, the temporal average of $\delta P_2(I)^2$ is obtained by 
\begin{eqnarray}
&&\bar{P}_2 \equiv \sum^{T^{(L)}-1}_{I=1}\frac{\delta P_2(I)^2}{T^{(L)}-1}  \\
&=& \frac{r(1)^2}{L^2 \cdot (T^{(L)}-1)} \frac{(\kappa^L-1)^3 (\kappa^{2L(T^{(L)}-1)}-1)}{(\kappa-1)^2 \cdot (\kappa^L+1)}.  \nonumber \\
 \label{P2_m}
\end{eqnarray}
 \par 
Lastly, we investigate the effects of $\delta P_3(I)^2$, i.e., 
\begin{eqnarray}
\delta P_3(I)^2 &=&  \frac{1}{L^2} \{ \sum^{L(I+1)}_{t=2}g_0(t,I) \eta(t) \}^2 \\
&\approx& \frac{1}{L^2} \eta_0^2 \cdot \sum^{L(I+1)}_{t=2}g_0(t,I)^2,   
\end{eqnarray}
where, from the definition, 
\begin{equation}
\eta_0^2 \approx \sum^{T}_{t=1}\eta(t)^2/T.
\end{equation}
%
We can calculate the sum of $g(t,I)$ with respect to $t$ as 
\begin{equation}
 \sum^{L(I+1)}_{t=2}g_0(t,I)^2=Q_1(I)+Q_2+Q_3, 
\end{equation}
where 
\begin{eqnarray}
Q_1(I)\equiv \sum^{(I-1)L+1}_{t=2}g_0(t,I)^2  =\frac{(\kappa^L-1)^4 \cdot(\kappa^{2(I-1)L}-1)}{(\kappa-1)^3 \cdot (\kappa+1)}, \nonumber \\
\end{eqnarray}
\begin{eqnarray}
&&Q_2 \equiv \sum^{LI+1}_{(I-1)L+2}g_0(t,I)^2  \nonumber \\
&&=\frac{L(\kappa^2-1)+(\kappa^{2L}-3\kappa^L+2)(\kappa^{2L}-\kappa^{L}+2 \kappa)}{(\kappa-1)^3 \cdot (\kappa+1)} ,\nonumber \\
\end{eqnarray}
\begin{eqnarray}
&&Q_3 \equiv \sum^{(I+1)L}_{LI+2}g_0(t,I)^2   \nonumber \\
&&=\frac{(\kappa^L-1)(\kappa^L-2\kappa-1)+L(\kappa^2-1)}{(\kappa-1)^3 \cdot (\kappa+1)}. 
\end{eqnarray}
From these results, we can obtain the temporal average of $\delta P_3(I)^2$ as 
\begin{eqnarray}
&&\bar{P}_3 \equiv \sum^{T^{(L)}-1}_{I=1}\frac{\delta P_3(I)^2}{T^{(L)}-1}  \\
&=&\frac{\check{\eta}^2}{L^2(T^{(L)}-1)}\cdot(\bar{P}_{3a}+\bar{P}_{3b}+\bar{P}_{3c}).  \label{P3_m}
\end{eqnarray}
Here, 
\begin{eqnarray}
&&\bar{P}_{3a}= \nonumber \\
&&\frac{(\kappa^L-1)^3 \cdot (\kappa^{2L(T^{(L)}-1)}-(T^{(L)}-1)\kappa^{2L}+T^{(L)}-2)}{(\kappa-1)^3 (\kappa+1) (\kappa^L+1)}, \nonumber \\
\end{eqnarray}
\begin{eqnarray}
&&\bar{P}_{3b}=(T^{(L)}-1) \nonumber \\
&&\times \frac{L \cdot (\kappa^2-1) +(\kappa^{2L}-3\kappa^L+2)(\kappa^{2L}-\kappa^L+2\kappa )}{(\kappa-1)^3 (\kappa+1)}, \nonumber \\
\end{eqnarray}
\begin{eqnarray}
\bar{P}_{3c}=(T^{(L)}-1) \cdot \frac{L \cdot (\kappa^2-1) + (\kappa^L-1)(\kappa^L-2\kappa-1)}{(\kappa-1)^3 (\kappa+1)}. \nonumber \\
\end{eqnarray}
\par
{\bf Calculation of $E[\delta R]^2$.}
Next, we calculate $E[\delta R]^2$.
 $E[\delta R]^2$ can be decomposed as follows:
\begin{eqnarray}
&&E[\delta R]^2=(E[\delta P_1]+E[\delta P_2]+E[\delta P_3])^2 \\
&=& E[\delta P_1]^2+E[\delta P_2]^2+2 \cdot E[\delta P_1] \cdot E[\delta P_2] \nonumber \\
&+&2E[\delta P_3](E[\delta P_1]+E[\delta P_2]+E[\delta P_3]) \nonumber \\
&\approx& E[\delta P_1]^2+E[\delta P_2]^2+2 \cdot E[\delta P_1] \cdot E[\delta P_2],  \label{EQ2}
\end{eqnarray} 
where we use $E[\delta P_3] \approx 0$. \par
$E[\delta P_1]$ and $E[\delta P_2]$ are obtained as 
\begin{eqnarray}
&&E[\delta P_{1}]=P_{1}[T^{(L)}]-P_{1}[1] \\
&=&\frac{u_0}{T^{(L)}-1} \cdot \frac{1}{L}[\frac{-\kappa^{(T^{(L)}-1)L}+\kappa^{T^{(L)}L}-\kappa^L+1}{(\kappa-1)^2}],  \label{EQ1} \nonumber \\
\end{eqnarray}
\begin{eqnarray}
&&E[\delta P_{2}]=P_{2}[T^{(L)}]-P_{2}[1] \\
&=&\frac{r(1)}{T^{(L)}-1} \cdot \frac{1}{L}[\frac{-\kappa^{(T^{(L)}-1)L}+\kappa^{T^{(L)}L}-\kappa^L+1}{\kappa-1}].  \label{EQ2} \nonumber \\
\end{eqnarray}
{\bf Calculation of $V[R^{(L)}]$.} 
Lastly, we calculate $V[R^{(L)}]$.
Substituting Eqs. \ref{RJ2} and \ref{EQ2} for Eq. \ref{VR_rand}, we can obtain
\begin{eqnarray}
&&V[R^{(L)}] \approx \nonumber \\
&&R^{(1)} \cdot u_0^2 +R^{(2)} \cdot \check{\eta}^2 +R^{(3)} \cdot r(1)^2+R^{(4)} \cdot u_0 \cdot r(1), \nonumber \\
\end{eqnarray}
Where, from Eq. \ref{P1_m} and Eq. \ref{EQ1},
\begin{eqnarray}
R^{(1)} =R_1^{(1)}+R_2^{(1)}, 
\end{eqnarray}
\begin{eqnarray}
R_1^{(1)}=  \frac{1}{L^2 \dot (T^{(L)}-1)}[ \frac{(\kappa^L-1)^3 \cdot (\kappa^{2L(T^{(L)}-1)}-1)}{(\kappa-1)^4 \cdot (\kappa^L+1)}], \nonumber \\  
\end{eqnarray}
\begin{eqnarray}
R_2^{(1)}=-\frac{1}{L^2 \cdot (T^{(L)}-1)^2} \cdot W_1^2, 
\end{eqnarray}
\begin{eqnarray}
W_1=\frac{\kappa^{T^{(L)}L}-\kappa^{(T^{(L)}-1)L}+1-\kappa^L}{(\kappa-1)^2};
\end{eqnarray}
from Eq. \ref{P3_m},  
\begin{eqnarray}
R^{(2)} =R_1^{(2)}+R_2^{(2)}+R_3^{(2)}, 
\end{eqnarray}
\begin{eqnarray}
&&R_{1}^{(2)}= \frac{1}{L^2 \cdot (T^{(L)}-1)}  \nonumber  \\
&& \times \frac{(\kappa^L-1)^3 \cdot (\kappa^{2L(T^{(L)}-1)}-(T^{(L)}-1)\kappa^{2L}+T^{(L)}-2)}{(\kappa-1)^3 (\kappa+1) (\kappa^L+1)}, \nonumber \\
\end{eqnarray}
\begin{eqnarray}
&&R_{2}^{(2)}=\nonumber  \\
&&\frac{1}{L^2} \cdot \frac{L \cdot (\kappa^2-1) +(\kappa^{2L}-3\kappa^L+2)(\kappa^{2L}-\kappa^L+2\kappa)}{(\kappa-1)^3 (\kappa+1)}, \nonumber \\
\end{eqnarray}
\begin{eqnarray}
{R}_{3}^{(2)}=\frac{1}{L^2} \cdot \frac{L \cdot (\kappa^2-1) + (\kappa^L-1)(\kappa^L-2\kappa-1)}{(\kappa-1)^3 (\kappa+1)};
\nonumber \\
\end{eqnarray}
from Eqs. \ref{P2_m} and \ref{EQ2}, 
\begin{eqnarray}
R^{(3)} =R_1^{(3)}+R_2^{(3)}, 
\end{eqnarray}
\begin{eqnarray}
R_{1}^{(3)} = \frac{1}{L^2 \cdot (T^{(L)}-1)} \frac{(\kappa^L-1)^3 \cdot (\kappa^{2L(T^{(L)}-1)}-1)}{(\kappa-1)^2 \cdot (\kappa^L+1)}, \nonumber \\
\end{eqnarray}
\begin{eqnarray}
R_{2}^{(3)}=-\frac{1}{L^2 \cdot (T^{(L)}-1)^2} \cdot W_2^2, 
\end{eqnarray}
\begin{eqnarray}
W_2=\frac{\kappa^{T^{(L)}L}-\kappa^{(T^{(L)}-1)L}+1-\kappa^L}{\kappa-1}; 
\end{eqnarray}
and, from Eqs. \ref{EQ1} and \ref{EQ2}, 
\begin{eqnarray}
R^{(4)}=-\frac{2}{L^2 \cdot (T^{(L)}-1)^2} \cdot W_1 \cdot W_2.
\end{eqnarray}
\par


\subsection{Calculation of $V[{R^{(L)}}^2]$ for large $L$}
We calculate $V[{R^{(L)}}^2]$ for $L>>1$. Here, we assume $L/2 \geq T$. \\
{\bf Case of $\kappa<1$:} \\
Using $\kappa^L \approx 0$ and $\kappa^T \approx 0$, we can obtain 
\begin{equation}
R_1^{(1)} \approx \frac{1}{L(T-L)} \frac{1}{(\kappa-1)^4},
\end{equation}
\begin{equation}
R_2^{(1)} \approx \frac{1}{(T-L)^2} \frac{1}{(\kappa-1)^4},
\end{equation}
\begin{equation}
R_1^{(2)} \approx \frac{2L-T}{L^2 (T-L)} \cdot  \frac{1}{(\kappa-1)^3 \cdot (\kappa+1)} ,
\end{equation}
\begin{equation}
R_2^{(2)} \approx \frac{1}{L} \cdot \frac{1}{(\kappa-1)^2 }+\frac{1}{L^2} \cdot \frac{4k}{(\kappa-1)^3 \cdot (\kappa+1)},
\end{equation}
\begin{equation}
R_3^{(2)} \approx \frac{1}{L} \cdot \frac{1}{(\kappa-1)^2 }+\frac{1}{L^2} \cdot \frac{2\kappa+1}{(\kappa-1)^3 \cdot (\kappa+1)},
\end{equation}
\begin{equation}
R_1^{(3)} \approx \frac{1}{L(T-L)} \frac{1}{(\kappa-1)^2},
\end{equation}
\begin{equation}
R_2^{(3)} \approx \frac{1}{(T-L)^2} \frac{1}{(\kappa-1)^4},
\end{equation}
\begin{equation}
R^{(4)} \approx \frac{1}{(T-L)^2} \frac{1}{(\kappa-1)^3}. 
\end{equation}
Considering only dominant terms, we can obtain
\begin{equation}
V[\delta R^{(L)}]\approx    \frac{2}{L}  \frac{ \check{\eta}^2 }{(\kappa-1)^2 } + \frac{1}{L(T-L)}( \frac{u_0^2}{(\kappa-1)^4}+ \frac{r(1)^2}{(\kappa-1)^2}). \nonumber \\
\end{equation}
In the case of $T-L>>1$, we can obtain the simpler form 
\begin{equation}
V[\delta R^{(L)}] \approx \check{\eta}^2  \frac{2}{L} \cdot \frac{1}{(\kappa-1)^2 } \propto \frac{1}{L}. \label{app_rand_b0}
\end{equation}
\par
{\bf Case of $\kappa=1$:} 
Next, we calculate the case of $\kappa=1$. 
Taking the limit of $\kappa \to1$, we get 
\begin{equation}
R_1^{(1)} \approx  L^2,
\end{equation}
\begin{equation}
R_2^{(1)} \approx -L^2,
\end{equation}
\begin{equation}
R_1^{(2)} \approx 0,
\end{equation}
\begin{equation}
R_2^{(2)} \approx \frac{2L^3+3L^2+L}{6} \cdot \frac{1}{L^2},
\end{equation}
\begin{equation}
R_3^{(2)} \approx \frac{2L^3-3L^2+L}{6} \cdot \frac{1}{L^2},
\end{equation}
\begin{equation}
R_1^{(3)} \approx 0,
\end{equation}
\begin{equation}
R_2^{(3)} \approx 0,
\end{equation}
\begin{equation}
R^{(4)} \approx 0. 
\end{equation}
From these results, we can obtain
\begin{equation}
V[R^{(L)}] \approx \frac{1}{3} \cdot \check{\eta}^2 \cdot (2L+1/L) \propto  L.  \label{app_rand_b1}
\end{equation}
\par 
{\bf Case of $ \kappa >1$:} 
Lastly, we calculate the case of $ \kappa >1$ for $L>>1$. \par
Calculation of $R^{(1)}$. 
For $L>>1$, in the case of $ \kappa >1$, a dominant term of $R_1^{(1)}$ is given by 
\begin{eqnarray}
R_1^{(1)} \approx \frac{1}{L(T-L)} \frac{k^{2T}}{(k-1)^4}.
\end{eqnarray}
In a similar way, 
\begin{equation}
R_2^{(1)} \approx  \frac{-1}{(T-L)^2}\frac{k^{2T}}{(k-1)^4}. 
\end{equation}
\par
Calculation of $R^{(2)}$. \\
For $L>>1$,  
\begin{eqnarray}
R_1^{(2)} &\approx&  \frac{\kappa^{2T}}{L(T-L)} \frac{1}{(\kappa-1)^3 \cdot (\kappa+1)}. \\
\end{eqnarray}
Similarly,
\begin{equation}
R_2^{(2)} \approx  \frac{\kappa^{4L}}{L^2} \frac{1}{(\kappa-1)^3 \cdot (\kappa+1)},  
\end{equation}
\begin{equation}
R_3^{(2)} \approx  \frac{\kappa^{2L}}{L^2} \frac{1}{(\kappa-1)^3 \cdot (\kappa+1)}.
\end{equation}
Therefore,
\begin{equation}
R^{(2)} \approx \frac{\kappa^{2T}}{L(T-L)} \frac{1}{(\kappa-1)^3 \cdot (\kappa+1)}.
\end{equation}
\par
Calculation of  $R^{(3)}$ and  $R^{(4)}$. 
For a large $L$, 
\begin{equation}
R_1^{(3)} \approx \frac{\kappa^{2T}}{L(T-L)} \frac{1}{(\kappa-1)^2},
\end{equation}
\begin{equation}
R_2^{(3)} \approx-\frac{\kappa^{2T}}{(T-L)^2} \frac{1}{(\kappa-1)^2}. 
\end{equation}
Therefore, 
\begin{equation}
R^{(3)} \approx \frac{\kappa^{2T}}{(T-L) \cdot (\kappa-1)^2} [\frac{1}{L}-\frac{1}{T-L}]. 
\end{equation}
$R^{(4)}$ is also approximated as 
\begin{equation}
R^{(4)} \approx -\frac{2 \cdot \kappa^{2T}}{(T-L)^2} [\frac{1}{(\kappa-1)^3}]. 
\end{equation}
Consequently, for $T>>L>>1$, we can obtain
\begin{equation}
V[\delta R^{(L)}] \propto \frac{\kappa^{2T}}{L(T-L)}. \label{app_rand_b2}
\end{equation}
%

\renewcommand{\theequation}{D\arabic{equation}}
\renewcommand{\thefigure}{D-\arabic{figure}}
\renewcommand{\thetable}{S-D\arabic{table}}
\setcounter{figure}{0}
\setcounter{equation}{0}
\section{$V[\delta R_j^{(L)}]$ for the power-law forgetting}
\label{app_forget}
We calculate $V[\delta R_j^{(L)}]$ for the power-law forgetting process given by  Eq. \ref{eq_rw}.  
Here, we consider $L \geq 2$ and omit the suffix $j$ for simplification. 
$R(I)$ is defined by 
\begin{equation}
R(I)=\sum_{t=LI+1}^{L(I+1)}\frac{r(t)}{L}. 
\end{equation}
From the definition, $r(t)$ is written as
\begin{equation}
r(t)=\sum_{s=0}^{\infty} \theta(s) \cdot \eta(t-s), \label{app_model_theta}
\end{equation}
where 
\begin{equation}
\theta(s) \equiv \frac{1}{Z(\beta)}(s+ d_\beta)^{-\beta}.  \label{app_theta}
\end{equation}
Then, we can calculate 
\begin{equation}
R(I)=\frac{1}{L}\sum^{LI+1}_{t=-\infty}\theta_1'(t) \eta(t) + \sum^{L(I+1)}_{t=LI+2}\theta_2'(t)\eta(t), 
\end{equation}
where 
\begin{eqnarray}
\theta_1'(t)&=&\sum^{L-1}_{k=0} \theta(k+LI+1-t), \label{theta_1d} \\
\theta_2'(t)&=&\sum^{L-(t-LI)}_{k=0} \theta(k). \label{theta_2d} 
\end{eqnarray}
\par
In a similar way, we can also calculate $R(I+1)$ as follows: 
\begin{equation}
R(I+1)=\frac{1}{L}\sum^{L(I+1)+1}_{t=-\infty}\theta_3'(t) \eta(t) + \sum^{L(I+2)}_{t=L(I+1)+2}\theta_4'(t)\eta(t), 
\end{equation}
where
\begin{eqnarray}
\theta_3'(t)&=&\sum^{L-1}_{k=0} \theta(k+L(I+1)+1-t), \label{theta_3d} \\
\theta_4'(t)&=&\sum^{L-(t-L(I+1))}_{k=0} \theta(k). \label{theta_4d} \\
\end{eqnarray}
From these results, $\delta R(I)=R(I+1)-R(I)$ can be calculated as 
\begin{eqnarray}
&&\delta R(I)= R(I+1)-R(I) \\
&=&\frac{1}{L} \sum^{LI+1}_{t=-\infty}(\theta_3'(t)-\theta_1'(t)) \eta(t) \nonumber  
+\sum^{L(I+1)}_{t=LI+2}(\theta_3'(t)-\theta_2'(t)) \eta(t) \nonumber  \\
&+&\sum^{L(I+2)}_{t=L(I+1)+1}\theta_4'(t) \eta (t).  
\end{eqnarray}
\par
Taking the average of $\delta R(I)$ with respect to $\eta$ gives   
\begin{eqnarray}
&&V_{\eta}[\delta R] \approx E_{\eta}[(R(I+1)-R(I))^2] \nonumber \\
&&= \frac{\check{\eta}^2}{L^2} \{\sum^{LI+1}_{t=-\infty}(\theta_3'(t)-\theta_1'(t))^2 \nonumber  \\
&&+\sum^{L(I+1)}_{t=LI+2}(\theta_3'(t)-\theta_2'(t))^2  
+\sum^{L(I+2)}_{t=L(I+1)+1}\theta_4'(t)^2 \}  \nonumber \\
&&=\frac{\check{\eta}^2}{L^2 Z(\beta)^2} (U_1(\beta,L)+U_2(\beta,L)+U_3(\beta,L)),  \label{u_ans} \nonumber \\
\end{eqnarray}
where we use $E[\delta R] \approx 0$ because $<\eta(t)>=0$.  \par
Here, $U_1(\beta,L)$, $U_2(\beta,L)$, and $U_3(\beta,L)$ are defined as
\begin{equation}
U_1(\beta,L)=\sum^{LI+1}_{t=-\infty}\frac{ Z(\beta)^2}{\check{\eta}^2} (\theta_3'(t)-\theta_1'(t))^2, \label{U_1}
\end{equation}
\begin{equation}
U_2(\beta,L)=\sum^{L(I+1)}_{t=LI+2} \frac{ Z(\beta)^2}{\check{\eta}^2}(\theta_3'(t)-\theta_2'(t))^2, \label{U_2}
\end{equation}
and
\begin{equation}
U_3(\beta,L)=\sum^{L(I+2)}_{t=L(I+1)+1} \frac{ Z(\beta)^2}{\check{\eta}^2} (\theta_4'(t))^2. \label{U_3}
\end{equation}
We factor out $Z(\beta)^2$ for later calculations. 
$U_1(\beta,L)$ is given by Eq. \ref{U_1_ans}, $U_2(\beta,L)$ is given by Eq. \ref{u2_ans}, and $U_3(\beta,L)$ is given by Eq. \ref{u3_ans}.
The details of the derivations of these equations are given in the following section and beyond.
 \par
\subsection{Calculation of $U_1(\beta,L)$}
Here, we calculate $U_1(\beta,L)$ defined by Eq. \ref{U_1}: 
\begin{eqnarray}
&&U_1(\beta,L)= \nonumber \\
&&\frac{ Z(\beta)^2}{\check{\eta}^2} \sum^{LI+1}_{t=-\infty}(\theta_3'(t)-\theta_1'(t))^2 \\
&&=\frac{ Z(\beta)^2}{\check{\eta}^2} \sum^{LI+1}_{t=-\infty}(\sum^{L}_{i=1}\theta(k+L(I+1)+1-t) \nonumber \\
&&-\sum^{L}_{i=1}\theta(k+LI+1-t))^2. \nonumber \\
\end{eqnarray}
By replacing the index $t$ with a new index $t'=t+LI+1$, $U_1(\beta,L)$ can be written as
\begin{eqnarray}
U_1(\beta,L)= \sum^{\infty}_{t'=0}\frac{ Z(\beta)^2}{\check{\eta}^2}(\sum^{L-1}_{k=0}\theta(t'+k+L)-\sum^{L-1}_{k=0}\theta(t'+k))^2. \nonumber \\
\end{eqnarray}
\par
Using the Euler--Maclaurin formula \cite{abramowitz1964handbook}, we can obtain 
\begin{eqnarray}
&&\sum^{b}_{k=a}g(k) \approx  \nonumber \\
&&\int^{b}_{a}g(x)dx+\frac{1}{2}(g(b)+g(a))+\frac{1}{12}(\frac{d}{dx}g(x)|_{b}-\frac{d}{dx}g(x)|_{a}), \nonumber \\ 
\label{euler}
\end{eqnarray}
and then we make the approximation 
\begin{eqnarray}
&&U_1 \approx  \nonumber \\
&&\frac{ Z(\beta)^2}{\check{\eta}^2} \sum^{\infty}_{t'=0} \left\{ \theta(t'+L)+\frac{\theta(t'+2L-1)+\theta(t'+L+1)}{2} \right. \nonumber  \\
&+&\frac{\theta'(t'+2L-1)-\theta'(t'+L+1)}{12} \nonumber \\
&+&\int^{L-1}_{1}\theta(t'+k+L)dk \nonumber \\
&-&  \theta(t')-\frac{\theta(t'+L-1)+\theta(t'+1)}{2} \nonumber \\ 
&-&\left. \frac{\theta'(t'+L-1)-\theta'(t'+1)}{12}-\int^{L-1}_{1} \theta(t'+k)  dk \right\}^2. \nonumber \\
\end{eqnarray}
Substituting Eq. \ref{app_theta} into $\theta(t')$ and performing some calculations, 
for $\beta>0$ and $\beta \neq 1$, we obtain
\begin{eqnarray}
&&U_1 \approx \nonumber \\
&&\sum^{\infty}_{t'=0}\left\{  \sum^{3}_{k=1}J^{(-1)}_{A,k} (t+A^{(-1)}_k)^{-(\beta-1)} \right. \nonumber \\
&&+\left. \sum^{6}_{k=1}J^{(0)}_{A,k} ((t+A^{(0)}_k)^{-\beta}+ \sum^{3}_{k=1}J^{(+1)}_{A,k}(t+A^{(+1)}_k)^{-(\beta+1)}\right\}^2. \nonumber \\
\end{eqnarray}
For $\beta=1$, 
\begin{eqnarray}
&&U_1 \approx \nonumber \\
&&\sum^{\infty}_{t'=0}\left\{  \sum^{3}_{k=1}J^{(-1)}_{A,k} \log(t+A^{(-1)}_k) \right. \nonumber \\
&&\left. + \sum^{6}_{k=1}J^{(0)}_{A,k} ((t+A^{(0)}_k)^{-\beta}+ \sum^{3}_{k=1}J^{+1}_{A,k}(t+A^{(+1)}_k)^{-(\beta+1)}\right\}^2. \nonumber \\
\end{eqnarray}
We combine the two equations into  
\begin{eqnarray}
&&U_1 \approx \nonumber \\
&&\sum^{\infty}_{t'=0}\left\{\sum^{+1}_{m=-1} \sum_{k=1}^{p_m}J^{(m)}_{A,k} f_A(t,\beta+m,A^{(m)}_k) \right\}^2,  \nonumber \\
\end{eqnarray}
where 
\begin{eqnarray}
f_A(x,\alpha,U)= 
\begin{cases}
(x+U)^{-\alpha} & (\alpha \neq 0), \\
\log(x+U) &  (\alpha=0),
\end{cases}
\end{eqnarray}
\begin{eqnarray}
A_1^{(-1)}&=&=A_1^{(+1)}=2L-1-d_{\beta}, \\
A_2^{(-1)}&=&A_2^{(+1)}=L+1+d_{\beta} ,\\
A_3^{(-1)}&=&A_3^{(+1)}=L-1+d_{\beta}, \\
A_4^{(-1)}&=&A_4^{(+1)}=1+d_{\beta} ,\\
A_1^{(0)}&=&2L-1+d_{\beta}, \\
A_2^{(0)}&=&1+L+d_{\beta}, \\
A_3^{(0)}&=&L+d_{\beta}, \\
A_4^{(0)}&=&L-1+d_{\beta}, \\
A_5^{(0)}&=&1+d_{\beta}, \\
A_6^{(0)}&=&d_{\beta}, 
\end{eqnarray}
$p_{-1}=p_{1}=4$, $p_{0}=6$, 
\begin{eqnarray}
J^{(-1)}_{A,1}=J^{(-1)}_{A,4}=
\begin{cases}
1/(1-\beta) & \text{($\beta \neq 1$)}, \\
1 & \text{($\beta = 1$)}, \\
\end{cases}
\end{eqnarray}
\begin{eqnarray}
J^{(-1)}_{A,2}=J^{(-1)}_{A,3}=
\begin{cases}
-1/(1-\beta) & \text{($\beta \neq 1$)} ,\\
-1 & \text{($\beta = 1$)} ,\\
\end{cases} 
\end{eqnarray}
\par
\begin{eqnarray}
J^{(0)}_{A,1}&=&J^{(0)}_{A,2}=1/2 ,\\
J^{(0)}_{A,3}&=&1,  \\
J^{(0)}_{A,4}&=&J^{(0)}_{A,5}=-1/2, \\
J^{(0)}_{A,6}&=&-1, 
\end{eqnarray}
and 
\begin{eqnarray}
J^{(+1)}_{A,1}&=&J^{(+1)}_{A,3}=1/12 \cdot \beta ,\\
J^{(+1)}_{A,2}&=&J^{(+1)}_{A,4}=-1/12 \cdot \beta.  
\end{eqnarray}
We expand $U_1$ as 
\begin{eqnarray}
&&U_1 \approx \nonumber \\
&&\sum^{\infty}_{t'=0}\left\{\sum^{+1}_{m=-1} \sum_{k=1}^{p_m}J^{(m)}_{A,k} f_A(t,\alpha+m,A^{(m)}_k) \right\}^2 \nonumber \\
&&=\sum^{+1}_{m_1=-1}U_1^{(2)}(t;\beta,m_1,m_1)+2 \sum_{m_1>m_2}U_1^{(2)}(t;\beta,m_1,m_2),  \nonumber \\
\end{eqnarray}
where 
\begin{eqnarray}
&&U^{(2)}_1(t;\beta,m_1,m_2) \nonumber \\
&& =\sum_{i=1}^{p_{m_1}} \sum_{j=1}^{p_{m_2}} J_{Ai}^{(m_1)} J_{Aj}^{(m_2)} \nonumber \\
&& \lim_{Q \to \infty} \sum^{Q}_{t=0}f_A(t,\beta+m_i,A^{(m)}_i)f_A(t,\beta+m_j,A^{(m)}_j). \nonumber \\
\end{eqnarray}
%
We calculate $U^{(2)}_1(t;\beta,m_1,m_2)$. 
Using the Euler--Maclaurin formula in Eq. \ref{euler}, we can obtain 
\begin{eqnarray}
&&\sum^{Q}_{t=0}f_A(t,\alpha_1,V_1)f_A(t,\alpha_2,V_2) \nonumber \\
&&\approx G_2^{(R)}(1,Q;\alpha_1,\alpha_2,V_1,V_2) \nonumber \\
&&+\int^Q_1 (t+V_1)^{-\alpha_1}(t+V_2)^{-\alpha_2} dt \nonumber \\
&&\approx G_2^{(R)}(1,Q;\alpha_1,\alpha_2,V_1,V_2)+G_2^{(I)}(Q;\alpha_1,\alpha_2,V_1,V_2) \nonumber \\
&&-G_2^{(I)}(1;\alpha_1,\alpha_2,V_1,V_2). \nonumber \\ \label{fafa}
\end{eqnarray}
(i) For $\alpha_1 \neq 1$, $\alpha_2 \neq 1$ $x_1=0$, and $x_2=0$, $G_2^{(R)}(x_1,x_2;\alpha_1,\alpha_2,V_1,V_2)$ is given by 
\begin{eqnarray}
&&G_2^{(R)}(x_1,x_2;\alpha_1,\alpha_2,V_1,V_2)=(V_1)^{-\alpha_1}(V_2)^{-\alpha_2} \nonumber \\
&&+\frac{1}{2}\{(V_1+x_1)^{-\alpha_1}(V_2+x_1)^{-\alpha_2} \nonumber \\
&&+(V_1+x_2)^{-\alpha_1}(V_2+x_2)^{-\alpha_2}\}  \nonumber \\
&&-\frac{1}{12}\{(\alpha_2(V_2+x_2)+\alpha_1(V_1+x_2)) \nonumber \\
&&\times(V_1+x_2)^{-\alpha_1-1}(V_2+x_2)^{-\alpha_2-1} \nonumber \\
&&-(\alpha_2(V_1+x_1)+\alpha_1(V_2+x_1)) \nonumber \\
&&\times(V_1+x_1)^{-\alpha_1-1}(V_2+x_1)^{-\alpha_2-1}\}. \nonumber \\ 
\end{eqnarray}
Here, because $Q$ approaches zero for $Q>>1$, $G_2^{(R)}(x,Q;0,\alpha_2,V_1,V_2)$ is not dependent on $Q$.
Accordingly, we can denote $G_2^{(R)}(x,Q;0,\alpha_2,V_1,V_2)$ as $G_2^{(R \to Q)}(x;\alpha_1,\alpha_2,V_1,V_2)$. \\
$G_2^{(R \to Q)}(x;\alpha_1,\alpha_2,V_1,V_2)$ is written as 
\begin{eqnarray}
&&G_2^{(R \to Q)}(x;\alpha_1,\alpha_2,V_1,V_2)=(V_1)^{-\alpha_1}(V_2)^{-\alpha_2} \nonumber \\
&&+ \frac{1}{2}\{(V_1+x_1)^{-\alpha_1}(V_2+x_1)^{-\alpha_2}\}  \nonumber \\
&&-\frac{1}{12}\{ 
-(\alpha_2(V_1+x_1)+\alpha_1(V_2+x_1)) \nonumber \\
&&\times(V_1+x_1)^{-\alpha_1-1}(V_2+x_1)^{-\alpha_2-1}\}. \nonumber \\ 
\end{eqnarray}
As a special point, for $x_1=0$ and $x_2=0$, we determine
\begin{eqnarray}
&&G_2^{(R)}(0,0;\alpha_1,\alpha_2,V_1,V_2)=G_2^{(R \to Q)}(x;\alpha_1,\alpha_2,V_1,V_2) \nonumber \\
&&=(V_1)^{-\alpha_1}(V_2)^{-\alpha_2}.
\end{eqnarray}
\\
(ii) In the case of $\alpha_1 = 1$ and $\alpha_2 = 1$, we can also calculate 
\begin{eqnarray}
&&G_2^{(R)}(x_1,x_2;0,0,V_1,V_2)=\log(V_1) \log(V_2)+ \nonumber \\
&&\frac{1}{2}\{\log(V_1+x_1)\log(V_2+x_1)  \nonumber \\
&&+\log(V_1+x_2)\log(V_2+x_2)\}  \nonumber \\
&+&\frac{1}{12}\{\frac{\log(V_1+x_2)}{V_2+x_2}+\frac{\log(V_2+x_2)}{V_1+x_2} \nonumber \\
&-&(\frac{\log(V_1+x_1)}{V_2+x_1}+\frac{\log(V_2+x_1)}{V_1+x_1})\},  \nonumber \\ 
\end{eqnarray}
and, for $Q>>1$, we can obtain
\begin{eqnarray}
&&G_2^{(R)}(x_1,x_2;0,\alpha_2,V_1,V_2)=\log(V_1) V_2^{-\alpha_2}+ \nonumber \\
&&\frac{1}{2}\{\log(V_1+x_1)V_2^{-\alpha_2}+\log(V_1+x_2)(V_2+x_2)^{-\alpha_2}\}  \nonumber \\
&&+\frac{1}{12} \times \{ (V_2+x_2)^{-\alpha_2-1} \times  \nonumber \\
&&\frac{(-\beta(V_1+x_2)\log(V_1+x_2)+V_2+x_2)}{V_1+x_2} \nonumber \\
&&- (V_2+x_1)^{-\alpha_2-1} \times  \nonumber \\
&&\frac{(-\beta(V_1+x_1)\log(V_1+x_1)+V_1+x_1)}{V_1+x_1}\}. \nonumber \\ 
\end{eqnarray}
As a special point, for $x_1=0$ and $x_2=0$, we determine
\begin{eqnarray}
&&G_2^{(R)}(0,0;0,0,V_1,V_2)=G_2^{(R \to Q)}(x;\alpha_1,\alpha_2,V_1,V_2) \nonumber \\
&&=\log(V_1) \log(V_2).
\end{eqnarray}
 \\
(iii) In the case of $\alpha_1 = 1$ and $\alpha_2 \neq 0$, we can also obtain
\begin{eqnarray}
&&G_2^{(R)}(x_1,x_2;0,\alpha_2,V_1,V_2)=\log(V_1) V_2^{-\alpha_2} \nonumber \\
&&+\frac{1}{2}\{\log(V_1+x_1)V_2^{-\alpha_2}+\log(V_1+x_2)(V_2+x_2)^{-\alpha_2}\}  \nonumber \\
&&+\frac{1}{12} \times \nonumber \\
&& \{\frac{(V_2+x_2)^{-\alpha_2-1}(-\beta(V_1+x_2)\log(V_1+x_2)+V_2+x_2)}{V_1+x_2}  \nonumber \\
&&-(V_2+x_1)^{-\alpha_2-1} \times  \nonumber \\
&&\frac{(-\beta(V_1+x_1)\log(V_1+x_1)+V_1+x_1)}{V_1+x_1}\}. \nonumber \\ 
\end{eqnarray}
As a special point, for $x_1=0$ and $x_2=0$, we determine
\begin{equation}
G_2^{(R)}(0,0;0,\alpha_2,V_1,V_2)= \log(V_1) V_2^{-\alpha_2}.
\end{equation}
\par
Next, we calculate the integration term of Eq. \ref{fafa}, $G_2^{(I)}(x;\alpha_1,\alpha_2,V_1,V_2)$. 
$G_2^{(I)}(x;\alpha_1,\alpha_2,V_1,V_2)$ is defined by 
\begin{eqnarray}
&&G_2^{(I)}(x;\alpha_1,\alpha_2,V_1,V_2) \nonumber \\
&&=
\begin{cases}
\int (x+V_1)^{-\alpha_1}(x+V_2)^{-\alpha_2} dx & (\alpha_1,\alpha_2>0), \\
\int \log(x+V_1) \log(x+V_2) dx & (\alpha_1,\alpha_2=0), \\
\int \log(x+V_1) (x+V_2)^{-\alpha_2} dx & (\alpha_1=0, \alpha_2>0), \\
\end{cases}, \nonumber \\ 
\end{eqnarray}
where we neglect an integral constant. 
\par
We calculate $G_2^{(I)}(x;\alpha_1,\alpha_2,V_1,V_2)$. \\
(i) In the case of $V_1=V_2$, $\alpha_1 \neq 0$, and $\alpha_2 \neq 0$, we can write 
\begin{eqnarray}
&&G_2^{(I)}(x;\alpha_1,\alpha_2,V_1,V_2) =\frac{1}{1-\alpha_1-\alpha_2}(x+V_1)^{-\alpha_1-\alpha_2+1}. \nonumber \\
\label{GI_AA}
\end{eqnarray}
(ii) When $\alpha_1$ and $\alpha_2$ are nonintegers, and $V_1 \neq V_1$, we obtain
\begin{eqnarray}
&&G_2^{(I)}(x;\alpha_1,\alpha_2,V_1,V_2)= (V_j+x)^{1-\alpha_j}(V_i-V_j)^{-\alpha_i} \times \nonumber  \\
&&\frac{{}_2F_1(\alpha_i,1-\alpha_j,2-\alpha_j,\frac{-(V_j+x)}{(V_i-V_j)})}{(1-\alpha_j)},  \nonumber \\
\label{GI_real}
\end{eqnarray}
where $i={\arg \max}_{k \in \{1,2\}}\{V_k\}$ and $j={\arg \min}_{k \in \{1,2\}}\{V_k\}$ (under this condition, $G_2^{(I)}$ takes a real number). \\
(iii) For $\alpha_1=1,2,3,\cdots$, $\alpha_2=1,2,3,\cdots$ and $V_1 \neq V_2$,  
%
%
%
\par
using a partial fraction decomposition, 
\begin{eqnarray}
\frac{1}{(z+q_1)^{n_1}(z+q_2)^{n_2}}=\sum_{k=1}^{n_1}\frac{h^{(1)}_k}{(z+q_1)^k}+\sum_{k=1}^{n_2}\frac{h^{(2)}_k}{(z+q_2)^k} ,\nonumber \\
\end{eqnarray}
with
\begin{eqnarray}
&&h^{(1)}_k=\frac{1}{(n_1-k)!}(-1)^{n_2}\frac{(n_1+n_2-k-1)!}{(n_2-1)!} \nonumber \\
&&\times (-q_2+q_1)^{-n_1-n_2+k}, \nonumber \\
&&h^{(2)}_k=\frac{1}{(n_2-k)!}(-1)^{n_1}\frac{(n_1+n_2-k-1)!}{(n_1-1)!}\nonumber \\
&&\times (-q_1+q_2)^{-n_1-n_2+k}, \nonumber \\
\end{eqnarray}
we can write
\begin{eqnarray}
&&G_2^{(I)}(x;\alpha_1,\alpha_2,V_1,V_2) \nonumber \\
&&=h_1^{(1)}\log(x+V_1)+h_1^{(2)}\log(x+V_2) \nonumber \\ 
&&+\sum_{k=2}^{\alpha_1}\frac{1}{1-k}h^{(1)}_k (x+V_1)^{1-k}+\sum_{k=2}^{\alpha_2} \frac{1}{1-k} h^{(2)}_k (x+V_2)^{1-k}. \nonumber \\
\label{GI_int}
\end{eqnarray}
(iv) For $\alpha_1=0$, $\alpha_2=0$, and $V_1 \neq V_2$, we can write
\begin{eqnarray}
&&G_2^{(I)}(x;0,0,V_1,V_2) \nonumber \\
&&=Li_{2}(\frac{V_1+x}{V_1-V_2}) (V_2-V_1) \nonumber \\
&&+\log(V_1+x)\{(V_1+x)\log(V_2+x) \nonumber \\
&&+(V_2-V_1)\log(-\frac{V_2+x}{V_1-V_2})-V_1-x\} \nonumber \\
&&+V_1-V_2\log(V_2+x)-x\log(V_2+x)+2x, \nonumber \\  \label{GI_00AB} 
\end{eqnarray}
where $Li_{2}(x)$ is the polylogarithm. \\
(v) For $\alpha_1=0$, $\alpha_2=0$, and $V_1=V_2$, we can write
\begin{eqnarray}
&&G_2^{(I)}(x;0,0,V_1,V_2)  \nonumber \\
&&=\log(V_1+x)^2 (V_1+x)+2\log(V_1+x)(-V_1-x) \nonumber \\
&&+2x+V_1.  \nonumber \\
\label{GI_00AA} 
\end{eqnarray}
(vi) For $\alpha_1=0$, $\alpha_2=1$, and $V_1 \neq V_2$, we can write
\begin{eqnarray}
&&G_2^{(I)}(x;0,1,V_1,V_2) \nonumber \\
&&=Li_{2}(\frac{V_1+x}{V_1-V_2}) + \log(V_1+x)\{\log(-\frac{V_2+x}{V_1-V_2})\}. \nonumber \\
\label{GI_01AB} 
 \end{eqnarray}
(vii) For $\alpha_1=0$, $\alpha_2=1$, and $V_1=V_2$, we can write
\begin{eqnarray}
&&G_2^{(I)}(x;0,1,V_1,V_2)=\frac{1}{2}\log^2(V_1+x).
\label{GI_01AA} 
\end{eqnarray}
(viii) For $\alpha_1=0$, $\alpha_2=2$, and $V_1 \neq V_2$, we can write
\begin{eqnarray}
&&G_2^{(I)}(x;0,2,V_1,V_2) \nonumber \\ 
&&=\frac{(V_2+x)\log(V_2+x)-(V_1+x)\log(V_1+x)}{(V_1-V_2)(V_2+x)}. \nonumber \\
\label{GI_02AB}
 \end{eqnarray}
(ix) For $\alpha_1=0$, $\alpha_2=2$, and $V_1= V_2$, we can write
\begin{eqnarray}
&&G_2^{(I)}(x;0,2,V_1,V_2)=-\frac{\log(V_1+x)+1}{V_1+x}.
\label{GI_02AA}
\end{eqnarray}
\par
Consequently, the summary of $G_2^{(I)}$ is given by  
\begin{eqnarray}
&&G_2^{(I)}(x;\alpha_1,\alpha_2,V_1,V_2) \nonumber \\
&&=\begin{cases}
\text{($\alpha_1 \neq 0$, $\alpha_2 \neq 0$, $V_1=V_2$): }\\
\frac{1}{1-\alpha_1-\alpha_2}(x+V_1)^{-\alpha_1-\alpha_2+1} \\
\text{($\alpha_2$ and $\alpha_2$ are noninteger, $V_1 \neq V_2$): }\\
\frac{(V_j+x)^{(1-\alpha_j)}(V_i-V_j)^{(-\alpha_i)}{}_2F_1(\alpha_i,1-\alpha_j,2-\alpha_j,\frac{-(V_j+x)}{(V_i-V_j)})}{(1-\alpha_j)} \\
\text{($\alpha_1, \alpha_2 \geq 2$, $\alpha_1$ and $\alpha_2$ are integers, $V_1 \neq V_2$): }\\
h_1^{(1)}\log(x+V_1)+h_1^{(2)}\log(x+V_2) \\
+\sum_{k=2}^{\alpha_1}\frac{1}{1-k}h^{(1)}_k (x+V_1)^{1-k} \\
+\sum_{k=2}^{\alpha_2} \frac{1}{1-k} h^{(2)}_k (x+V_2)^{1-k} \\
\text{($\alpha_1=0$, $\alpha_2=0$, $V_1= V_2$): }\\
\log(V_1+x)^2 (V_1+x)+2\log(V_1+x)(-V_1-x) \\
+2x+V_1 \nonumber  \\
\text{($\alpha_1=0$, $\alpha_2=0$, $V_1 \neq V_2$): }\\
Li_{2}(\frac{V_1+x}{V_1-V_2}) (V_2-V_1)  \\
+\log(V_1+x)\{(V_1+x)\log(V_2+x)  \\
+(V_2-V_1)\log(-\frac{V_2+x}{V_1-V_2})-V_1-x\}  \\
+V_1-V_2\log(V_2+x)-x\log(V_2+x)+2x \\
\text{($\alpha_1=0$, $\alpha_2=1$, $V_1 = V_2$): }\\
\frac{1}{2}\log^2(V_1+x) \\
\text{($\alpha_1=0$, $\alpha_2=1$, $V_1 \neq V_2$): }\\
Li_{2}(\frac{V_1+x}{V_1-V_2}) + \log(V_1+x)\{\log(-\frac{V_2+x}{V_1-V_2})\} \\
\text{($\alpha_1=0$, $\alpha_2=2$, $V_1=V_2$): }\\
-\frac{\log(V_1+x)+1}{V_1+x} \\
\text{($\alpha_1=0$, $\alpha_2=2$, $V_1 \neq V_2$): }\\
\frac{(V_2+x)\log(V_2+x)-(V_1+x)\log(V_1+x)}{(V_1-V_2)(V_2+x)}.  \\
\end{cases} \nonumber \\
\end{eqnarray}
\subsection{Calculation of $G_2^{(Q)}(\beta,m_1,m_2)$} 
$U_1(t;\beta,m_1,m_2)$ is decomposed into 
\begin{eqnarray}
&&U_1(t;\beta,m_1,m_2)= \nonumber \\
&&\sum_{i=1}^{p_{m_1}} \sum_{j=1}^{p_{m_2}} J_{Ai}^{(m_1)} J_{Aj}^{(m_2)}  G_2^{(R)}(1,Q;\beta+m_1,\beta+m_2,A^{(m_1)}_i,A^{(m_2)}_j) \nonumber \\&&- \sum_{i=1}^{p_{m_1}} \sum_{j=1}^{p_{m_2}} J_{Ai}^{(m_1)} J_{Aj}^{(m_2)}  G_2^{(I)}(0;\beta+m_1,\beta+m_2,A^{(m_1)}_i,A^{(m_2)}_j) \nonumber \\
&&+ \sum_{i=1}^{p_{m_1}} \sum_{j=1}^{p_{m_2}} J_{Ai}^{(m_1)} J_{Aj}^{(m_2)}  G_2^{(I)}(Q;\beta+m_1,\beta+m_2,A^{(m_1)}_i,A^{(m_2)}_j). \nonumber \\
\end{eqnarray}
We have already calculated $G_2^{(R)}(1,Q;\alpha_1,\alpha_2,A^{(m_1)}_i,A^{(m_2)}_j)$ as $G_2^{R \to Q}(x,\alpha_1,\alpha_2,V_1,V_2)$ in the previous sections.
Here, we calculate 
\begin{eqnarray}
&&G_2^{(Q)}(\beta,m_1,m_2) \equiv \nonumber \\
&&\sum_{i=1}^{p_{m_1}} \sum_{j=1}^{p_{m_2}} J_{Ai}^{(m_1)} J_{Aj}^{(m_2)}  G_2^{(I)}(Q;\beta+m_1,\beta+m_2,A^{(m_1)}_i,A^{(m_2)}_j) \nonumber \\
\label{G_Q}
\end{eqnarray}
for $Q>>1$.
\par
\paragraph{(i) When $\beta$ is a noninteger} \par
We study the asymptotic behavior of $G^{(I)}(Q;\alpha_1,\alpha_2,V_1,V_2)$ in Eq. \ref{G_Q} for $Q>>1$ in the case of $V_1 \neq V_2$. 
Here, we use the following formulas of the asymptotic behavior of the hypergeometric function \cite{hypergeom}. 
When $b-a$, $c-a$, $a$, $b$, and $c$ are nonintegers for a large $x$, 
\begin{eqnarray}
&&{}_2F_1(a,b,c,x) \nonumber \\
&&\approx \frac{\Gamma(b-a) \Gamma(c)}{ \Gamma(b) \Gamma(c-a)}\frac{1}{(-x)^{a}} \nonumber \\ 
&&+\frac{(-1)^a a(1+a-c) \Gamma(b-a) \Gamma(c)}{(1+a-b)\Gamma(b) \Gamma(c-a)}\frac{1}{x^{a+1}} \nonumber \\ 
&&+\frac{(-1)^a a(1+a)(1+a-c)(2+a-c) \Gamma(b-a) \Gamma(c)}{2(1-a-b)(2-a+b)\Gamma(b) \Gamma(c-a)}\frac{1}{x^{a+2}} \nonumber \\ 
&&+\frac{\Gamma(a-b)\Gamma(c)}{\Gamma(a)\Gamma(c-b)}\frac{1}{(-x)^b}. \label{f_ap1}
\end{eqnarray}
When both $b-a$ and $c-a$ are integers and $c-b>0$ for a large $x$,
\begin{eqnarray}
&&{}_2F_1(a,b,c,x)  \nonumber \\
&&\approx \frac{\Gamma(b-a) \Gamma(c)}{ \Gamma(b) \Gamma(c-a)}\frac{1}{(-x)^{a}} \nonumber \\ 
&&+\frac{(-1)^a a(1+a-c) \Gamma(b-a) \Gamma(c)}{(1+a-b)\Gamma(b) \Gamma(c-a)}\frac{1}{x^{a+1}} \nonumber \\ 
&&+\frac{(-1)^a a(1+a)(1+a-c)(2+a-c) \Gamma(b-a) \Gamma(c)}{2(1+a-b)(2+a-b)\Gamma(b) \Gamma(c-a)}\frac{1}{x^{a+2}} \nonumber \\
&&+(-1)^{b-a}\Gamma(c) \nonumber \\
&&\frac{(\log(-x)+\psi(b-a+1)-\psi(c-b)-\psi(b)-\gamma)}{\Gamma(a)\Gamma(b-a+1)\Gamma(c-b)}\frac{1}{(-x)^b}. \nonumber \\
\label{f_ap2}
\end{eqnarray}
When $b=a$ and $c-a$ are integers for a large $x$, 
\begin{eqnarray}
&&{}_2F_1(a,b,c,x) \nonumber \\
&&\approx \frac{ \Gamma(c)}{ \Gamma(a) \Gamma(c-a)}\frac{(\log(-z)-\psi(c-a)-\psi(a)-2\gamma)}{(-x)^{a}} \nonumber \\ 
&&+\frac{\Gamma(c)^2(-x)^{-c}}{\Gamma(a)^2(\Gamma(c-a+1))^2}. \nonumber \\
\label{f_ap3}
\end{eqnarray}
Here, $\gamma=0.5772$ is the Euler constant. \par
By using these formulas, the hypergeometric function in Eq. \ref{GI_real} is written as \par
\begin{eqnarray}
&& {}_2F_1(\alpha_i,1-\alpha_j,2-\alpha_j,\frac{-(V_j+x)}{(V_i-V_j)}))  \nonumber \\
&& \approx  P_1 (\frac{(V_i-V_j)}{Q+V_j})^{\alpha_i}+ P_2 (\frac{(V_i-V_j)}{Q+V_j})^{\alpha_i-1} \nonumber \\
&&+ P_3 (\frac{(V_i-V_j)}{Q+V_j})^{\alpha_i-2} \nonumber \\
&& + P_4 (\frac{(V_i-V_j) } {Q+V_j})^{1-\alpha_j}\log(Q) + P_5(V_i,V_j) (\frac{(V_i-V_j)}{Q+V_j})^{1-\alpha_j},  \nonumber \\
\end{eqnarray}
where
\begin{eqnarray}
P_1=
\begin{cases}
\frac{(1-\alpha_j)}{(1-\alpha_i-\alpha_j)} &  \text{($\alpha_i+\alpha_j \neq 1$)}, \\
0 & \text{($\alpha_i+\alpha_j =1$)},
\end{cases}
\end{eqnarray}
\begin{eqnarray}
P_2=
\begin{cases}
\frac{-\alpha_i(\alpha_i+\alpha_j-1)}{(\alpha_i+\alpha_j)}P_1 & \text{($\alpha_i+\alpha_j \neq 0$)}, \\
0 & \text{($\alpha_i+\alpha_j=0$)}, 
\end{cases}
\end{eqnarray}
\begin{eqnarray}
P_3=
\begin{cases}
\frac{(1+\alpha_i)(\alpha_i+\alpha_j)}{2(1+\alpha_i+\alpha_j)}P_2 &  \text{($\alpha_i+\alpha_j \neq -1$)} ,\\
0& \text{($\alpha_i+\alpha_j=-1$)} ,\\
\end{cases}
\end{eqnarray}
\begin{eqnarray}
&&P_4= 
\begin{cases}
0 & \text{($\alpha_i+\alpha_j$ is a noninteger)}, \\
\frac{\Gamma(2-\alpha_j)}{\Gamma(\alpha_i)}  \frac{(-1)^{1-\alpha_i-\alpha_j}}{\Gamma(2-\alpha_i-\alpha_j)} & \text{($\alpha_i+\alpha_j=0,-1,1$)},
\end{cases} \nonumber \\
\end{eqnarray}
\begin{eqnarray}
&&P_5(V_i,V_j)=\frac{\Gamma(2-\alpha_j)}{\Gamma(\alpha_i)} \times \nonumber \\  
&&\begin{cases}
\text{($\alpha_i+\alpha_j$ is a noninteger)}\\
\Gamma(\alpha_i+\alpha_j-1) \\
\text{($\alpha_i+\alpha_j=-1,0,1$)} \\
\frac{(-1)^{1-\alpha_i-\alpha_j}(-\log(V_i-V_j)+\psi(2-\alpha_i-\alpha_j)-\psi(1-\alpha_j))}{\Gamma(2-\alpha_i-\alpha_j)}. \\ 
\end{cases}
\end{eqnarray}
Substituting these results, we can obtain the following approximation: 
\begin{eqnarray}
&&G_2^{(I)}(Q;\alpha_1,\alpha_2,V_1,V_2) \approx  \frac{1}{1-\alpha_j}    \nonumber  \\
&& \times \left \{P_1 (Q+V_j)^{1-\alpha_i-\alpha_j} \right. \nonumber \\
&&+ P_2 (V_i-V_j)(Q+V_j)^{-\alpha_i-\alpha_j}  \nonumber \\
&&+ P_3 (V_i-V_j)^2 (Q+V_j)^{-\alpha_i-\alpha_j-1} \nonumber \\ 
&& + P_4 (V_i-V_j)^{1-\alpha_i-\alpha_j}\log(Q) \nonumber \\ 
&& \left. + P_5(V_i,V_j)\times (V_i-V_j)^{1-\alpha_i-\alpha_j} \right\}, \nonumber \\
\end{eqnarray}
where $i={\arg \max}_{k \in \{1,2\}}\{V_k\}$ and $j={\arg \min}_{k \in \{1,2\}}\{V_k\}$.  (Under this condition, $G_2^{(I)}$ takes a real number.) \par
In addition, using the asymptotic behavior of the power-law function,  
\begin{equation}
(Q+V)^{\alpha} \approx Q^{\alpha}+(\alpha)VQ^{\alpha-1}+\frac{(\alpha)(\alpha-1)}{2}V^2 Q^{\alpha-2}, \label{pow_a}
\end{equation}
we can obtain $G_2^{(I)}$ as a sum of the power-law function of $Q$, 
\begin{eqnarray}
&&G_2^{(I)}(Q;\alpha_1,\alpha_2,V_1,V_2) \approx  \frac{1}{1-\alpha_j}    \nonumber  \\
&&\times \left\{ P_1 Q^{1-\alpha_i-\alpha_j}+ (P_1(-\alpha_i-\alpha_j+1)V_j \right. \nonumber \\
&&+P_2(V_i-V_j))Q^{-\alpha_i-\alpha_j}  \nonumber \\
&&+ (\frac{P_1}{2}(-\alpha_i-\alpha_j+1)(-\alpha_i-\alpha_j)V_j^2 \nonumber \\
&&+P_2(-\alpha_i-\alpha_j)(V_i-V_j)V_j \nonumber \\
&&+P_3 (V_i-V_j)^2) Q^{-\alpha_i-\alpha_j-1} \nonumber \\ 
&& + P_4 (V_i-V_j)^{1-\alpha_i-\alpha_j}\log(Q) \nonumber \\ 
&& \left. + P_5(V_i,V_j)\times (V_i-V_j)^{1-\alpha_i-\alpha_j} \right\}. \nonumber \\
\end{eqnarray}
For $V_1=V_2$, by using Eq. \ref{pow_a}, Eq. \ref{GI_AA} is approximated as
\begin{eqnarray}
&&G_2^{(I)}(x;\alpha_1,\alpha_2,V_1,V_2) \nonumber \\
&&\approx \frac{1}{1-\alpha_1-\alpha_2}(Q+V_2)^{-\alpha_1-\alpha_2+1} \nonumber \\
&&\approx \frac{1}{1-\alpha_1-\alpha_2} (Q^{-\alpha_1-\alpha_2+1}+(-\alpha_1-\alpha_2+1)V_2Q^{-\alpha_1-\alpha_2} \nonumber \\
&&+\frac{(-\alpha_1-\alpha_2)(-\alpha_1-\alpha_2+1)}{2}V_2^2 Q^{-\alpha_1-\alpha_2-1}).
\end{eqnarray}
Hence, because $Q^{a} \to 0$ ($a<0$, $Q>>1$), 
\begin{eqnarray}
&&G_2^{(Q)}(\beta,m_1,m_2)  \approx \nonumber \\
&&\sum_{i=1}^{p_{m_1}} \sum_{j=1}^{p_{m_2}} J_{Ai}^{(m_1)} J_{Aj}^{(m_2)}  G^{(I)}(Q;\beta+m_1,\beta+m_2,A^{(m_1)}_i,A^{(m_2)}_j) \nonumber \\
&&\approx \sum_{i=1}^{p_{m_1}} \sum_{j=1}^{p_{m_2}} J_{Ai}^{(m_1)} J_{Aj}^{(m_2)} \left[ \right. \nonumber \\
&&
\begin{cases}
\frac{ P_5(V_q,V_r) \times (V_q-V_r)^{1-2\beta-m_1-m_2}}{1-\beta-m_s} & \text{($V_q \neq V_r$)}, \\
0 & \text{($V_q = V_r$)}, \\
\end{cases} 
\nonumber \\
&&+
\begin{cases}
0 & \text{($2\beta+m_1+m_2 \neq 0$)}, \\ 
V_r & \text{($2\beta+m_1+m_2 = 0$)}, \\ 
\end{cases}
\left. \right] \nonumber \\
\end{eqnarray}
where $V_q=\max\{A^{(m_1)}_i,A^{(m_2)}_j\}$, $V_r=\min\{A^{(m_1)}_i,A^{(m_2)}_j\}$, $s=\arg \min_{\{u=1,2\}}\{A_{q_u}^{(m_u)}\}$, and $(q_1,q_2)=(i,j)$. \par

\paragraph{Case of $\beta=1$} \par
We use the approximation formulas for $x>>1$, 
\begin{eqnarray}
Li_2(x) \approx -\frac{1}{2}\log(-x)^2-\frac{\pi^2}{6}, \label{li2_app}
\end{eqnarray}
\begin{eqnarray}
\log(x+A) \approx \log(x)+\frac{A}{x} \label{log_ap}. 
\end{eqnarray}
From Eqs. \ref{GI_00AB} and \ref{GI_00AA},  we can obtain 
\begin{eqnarray}
&&G_2^{(Q)}(1,-1,-1)  \nonumber \\
&& \approx \sum_{i=1}^{p_{-1}} \sum_{j=1}^{p_{-1}} J_{Ai}^{(-1)} J_{Aj}^{(-1)}  G^{(I)}(Q;0,0,A^{(-1)}_i,A^{(-1)}_j) \nonumber \\
&&\approx \sum_{i=1}^{p_{-1}} \sum_{j=1}^{p_{-1}} J_{Ai}^{(-1)} J_{Aj}^{(-1)} G_{(-1,-1)}^{(Q)}(A^{(-1)}_i,A^{(-1)}_j), 
\end{eqnarray}
where
\begin{eqnarray}
&&G_{(-1,-1)}^{(Q)}(V_1,V_2)  \nonumber \\
&&=\begin{cases}
(V_2-V_1)(-\frac{1}{2}\log(V_2-V_1)^2-\frac{\pi}{6})+V_1    & (V_1 \neq V_2), \\
V_1 & (V_1=V_2).\\
\end{cases} \nonumber \\
\end{eqnarray}
\par
Similarly, from Eqs. \ref{GI_01AB} and \ref{GI_01AA}, 
\begin{eqnarray}
&&G_2^{(Q)}(1,-1,0)  \nonumber \\
&& \approx \sum_{i=1}^{p_{-1}} \sum_{j=1}^{p_{0}} J_{Ai}^{(-1)} J_{Aj}^{(0)}  G^{(I)}(Q;0,0,A^{(-1)}_i,A^{(0)}_j) \nonumber \\
&&\approx \sum_{i=1}^{p_{-1}} \sum_{j=1}^{p_{0}} J_{Ai}^{(-1)} J_{Aj}^{(0)} G_{-1,0}^{(Q)}(A_i^{(-1)},A^{(0)}_j), \nonumber \\
\end{eqnarray}
where 
\begin{eqnarray}
G_{(-1,0)}^{(Q)}(V_1,V_2) =
\begin{cases}
-\frac{1}{2}\log(V_2-V_1)^2  & (V_1 \neq V_2) ,\\
0 & (V_1=V_2).\\
\end{cases}.
\end{eqnarray}
Because the order of Eqs. \ref{GI_02AB} and \ref{GI_02AA} is $\log(Q)/Q$, 
\begin{eqnarray}
G_2^{(Q)}(1,-1,2) \approx 0.
\end{eqnarray}
In addition, from Eq. \ref{GI_int}, 
\begin{eqnarray}
G_2^{(Q)}(1,m_1,m_2) \approx 0 \quad (m_1=0,1,m_2=0,1) .
\end{eqnarray}
Note that in  Eq. \ref{GI_int}, because $h_1^{(1)}=-h_1^{(2)}$, the terms of $\log(Q)$ approach zero for $Q>>1$ by cancelling each other out. 
\par
\paragraph{When $\beta$ is an integer and $\beta \neq 1$ } \par 
The terms of $\log(Q)$ in Eq. \ref{GI_int} approach zero for $Q>>1$ by cancelling each other out because $h_1^{(1)}=-h_1^{(2)}$.
We use the following result: 
\begin{equation}
G_2^{(I)}(Q;\alpha_1,\alpha_2,V_1,V_2)  \approx 0. 
\end{equation}
Therefore, 
\begin{equation}
G_2^{(Q)}(\beta,m_1,m_2) \approx 0. 
\end{equation}
We summarize $G_2^{(Q)}$ as follows:
\begin{eqnarray}
&&G_2^{(Q)}(\beta,m_1,m_2)  \nonumber \\
&&\approx \sum_{i=1}^{p_{m_1}} \sum_{j=1}^{p_{m_2}} J_{Ai}^{(m_1)} J_{Aj}^{(m_2)}G_2^{(I;Q)}(\beta,m_1,m_2,A^{(m_1)}_i,A^{(m_2)}_j),  \nonumber \\
\end{eqnarray}
where
\begin{eqnarray}
&&G_2^{(I;Q)}(\beta,m_1,m_2,A^{(m_1)}_i,A^{(m_2)}_j)=  \nonumber \\
&&\begin{cases}
\text{($2\beta+m_1+m_2 \neq 1, 0,-1$, $A^{(m_1)}_i \neq A^{(m_2)}_j$)} \\
\frac{\Gamma(2\beta+m_1+m_2-1)\Gamma(2-\beta-m_{s})}{(1-\beta-m_{s}) \Gamma(\beta+m_{k})} (A_{q}-A_{r})^{1-2\beta-m_1-m_2}   \\
\text{($\beta$ is noninteger $2\beta+m_1+m_2=-1,1$)} \\
 \frac{(-1)^{1-2\beta-m_1-m_2} \Gamma(2-\beta-m_s)}{(1-\beta-m_s)\Gamma(\beta+m_k)} (A_{q}-A_{r})^{1-2\beta-m_1-m_2} \times \\
 \frac{(-\log(A_{q}-A_{r})+\psi(2-2\beta-m_1-m_2)-\psi(1-\beta-m_s))}{\Gamma(2-2\beta-m_1-m_2)}   \\
\text{($\beta$ is noninteger, $2\beta+m_1+m_2=0$)} \\
 \frac{(-1)^{1-2\beta-m_1-m_2} \Gamma(2-\beta-m_s)}{(1-\beta-m_s)\Gamma(\beta+m_k)} (A_{q}-A_{r})^{1-2\beta-m_1-m_2} \times \\
 \frac{(-\log(A_{q}-A_{r})+\psi(2-2\beta-m_1-m_2)-\psi(1-\beta-m_s))}{\Gamma(2-2\beta-m_1-m_2)}   \\
+A_{r} \\
\text{($\beta=1$, $m_1=-1,m_2=-1$, $A^{(m_1)}_i \neq A^{(m_2)}_j$)} \\
(A^{(m_2)}_j-A^{(m_1)}_i)(-\frac{1}{2}\log(A^{(m_2)}_j-A^{(m_1)}_i)^2-\frac{\pi}{6}) \\
+A^{(m_1)}_i  \\
\text{($\beta=1$, $m_1=-1,m_2=-1$, $A^{(m_1)}_i = A^{(m_2)}_j$)} \\
A^{(m_1)}_i \\
\text{($\beta=1$, $(m_1,m_2) \in \{(-1,0),(0,-1)\}$, $A^{(m_1)}_i \neq A^{(m_2)}_j$)} \\ 
-\frac{1}{2}\log(A^{(m_2)}_j-A^{(m_1)}_i)^2 \\
\text{($\beta=1$ and [$m_1=1$ or $m_2=1$])} \\
0 \\
\text{($\beta=2,3,4, \cdots$)} \\
0\\
\text{($A^{(m_1)}_i = A^{(m_2)}_j$, $2\beta+m_1+m_2 \neq 0$)}\\
0 \\ 
\text{($A^{(m_1)}_i = A^{(m_2)}_j$, $2\beta+m_1+m_2=0$, $\beta \neq 1$)}\\
A_{r}. \\ 
\end{cases} \nonumber \\
\end{eqnarray}
Here, $A_q=\max\{A^{(m_1)}_i,A^{(m_2)}_j\}$, $A_r=\min\{A^{(m_1)}_i,A^{(m_2)}_j\}$, $t=\arg \max_{\{u=1,2\}}\{A_{q_u}^{(m_u)}\}$, $s=\arg \min_{\{u=1,2\}}\{A_{q_u}^{(m_u)}\}$, and $(q_1,q_2)=(i,j)$.  The reason why we change the suffix is to avoid a constant of the integrations from becoming a complex number.  \par
Consequently, $U_1$ is obtained by 
\begin{eqnarray}
&&U_1(\beta,L) \approx 
\sum^{+1}_{m_1=-1}\sum^{+1}_{m_2=-1} \sum_{i=1}^{p^{(A)}_{m_1}} \sum_{j=1}^{p^{(A)}_{m_2}}  J_{A_i}^{(m_1)}J_{A_j}^{(m_2)} \nonumber \\
&&\times \{ G_2^{(R;Q)}(1,\beta+m_1,\beta+m_2; A_i^{(m_1)},A_j^{(m_2)}) \nonumber \\
&&+G_2^{(I;Q)}(\beta,m_1,m_2,A_i^{(m_1)},A_j^{(m_2)}) \nonumber \\
&&-G_2^{(I)}(1,\beta+m_1,\beta+m_2; A_i^{(m_1)},A_j^{(m_2)}) \}. 
 \nonumber \label{U_1_ans} \\
\end{eqnarray}

\subsection{Calculation of $U_2(\beta,L)$}
$U_2$ is defined by Eq. \ref{U_2} as 
\begin{equation}
U_2(\beta,L)=\sum^{L(I+1)}_{t=LI+2}\frac{ Z(\beta)^2}{\check{\eta}^2} (\theta_3'(t)-\theta_2'(t))^2. \label{U2_app}
\end{equation}
Substituting Eqs. \ref{theta_2d} and \ref{theta_3d} into Eq. \ref{U2_app}, 
we obtain \begin{eqnarray}
&&U_2(\beta,L) \nonumber \\
&&=\sum^{L(I+1)}_{t=LI+2}\frac{ Z(\beta)^2}{\check{\eta}^2} (\sum^{L-1}_{k=0}\theta(k+L(I+1)+1-t) \nonumber \\
&&-\sum^{L(I+1)-t}_{k=0}\theta(k))^2. \nonumber \\
\end{eqnarray}
By using a shifted index $t'=t+LI+2$, $U_2$ is written as  
\begin{eqnarray}
&&U_2=\sum^{L-2}_{t'=0}\frac{ Z(\beta)^2}{\check{\eta}^2} (\sum^{L-1}_{k=0}\theta(k+L-1-t') \nonumber \\
&&-\sum^{L-2-t'}_{k=0}\theta(k))^2.  \label{U2_app2}
\end{eqnarray}
Using the Euler--Maclaurin formula in Eq. \ref{euler}, we can obtain 
\begin{eqnarray}
&&U_2 \approx  \nonumber \\
&&\frac{ Z(\beta)^2}{\check{\eta}^2} \sum^{L-3}_{t'=0} \left\{\theta(L-1-t')+\frac{\theta(L-t')+\theta(2L-2-t')}{2} \right. \nonumber  \\
&+&\frac{\theta(2L-2-t')-\theta'(L-1-t')}{12}  \nonumber \\
&+&\int^{L-1}_{1}\theta(k+L-1-t')dk \nonumber \\
&-&  \theta(0)-\frac{\theta(1)+\theta(L-2-t)}{2} \nonumber \\ 
&-&\left. \frac{\theta'(L-2-t)-\theta'(1)}{12}-\int^{L-2-t}_{1} \theta(k)  dk \right\}^2 \nonumber \\
&+&u_2^{(0)}(\beta)^2, 
\end{eqnarray}
where $u_2^{(0)}(\beta)^2$ is given by 
\begin{eqnarray}
&&u_2^{(0)}(\beta)  \nonumber \\
&&= (1+d_{\beta})^{-\beta}+\frac{1}{2}((2+d_{\beta})^{-\beta}+(L+d_{\beta})^{-\beta}) \nonumber \\
&&+(-\beta)\frac{1}{12}((L+d_{\beta})^{-\beta-1}-(2+d_{\beta})^{-\beta-1}) \nonumber \\
&&+\begin{cases}
\frac{1}{-\beta+1}((L+d_{\beta})^{-\beta+1}-(2+d_{\beta})^{-\beta+1} & \text{($\beta \neq 1$)}, \\
\log(L+d_{\beta})-\log(2+d_{\beta})  & \text{($\beta=1$)}.\\
\end{cases} \nonumber \\
\end{eqnarray}
Here, $u_2^{(0)}(\beta)$ corresponds to the term of $t'=L-2$ in Eq. \ref{U2_app2}. We separated and directly calculated this term to improve the accuracy. 
\par
Substituting Eq. \ref{app_theta} into $\theta(t')$ and performing some calculations, 
for $\beta>0$ and $\beta \neq 1$, we obtain
\begin{eqnarray}
&&U_2 \approx \nonumber \\
&&\sum^{L-3}_{t=0}\left\{B_0+  \sum^{p_B^{(-1)}}_{k=1}J^{(-1)}_{B,k} (-t+B^{(-1)}_k)^{-(\beta-1)} \right. \nonumber \\
&&+\sum^{p_B^{(0)}}_{k=1}J^{(0)}_{B,k} ((-t+B^{(0)}_k)^{-\beta} \nonumber  \\
&&+\left.  \sum^{p_B^{(+1)}}_{k=1}J^{(+1)}_{B,k}(-t+B^{(+1)}_k)^{-(\beta+1)}\right\}^2 \nonumber \\
&&+u_2^{(0)}(\beta)^2, 
\end{eqnarray}
and, for $\beta=1$, 
\begin{eqnarray}
&&U_2 \approx \nonumber \\
&&\sum^{L-3}_{t=0}\left\{B_0+ \sum^{p_B^{(-1)}}_{k=1}J^{(-1)}_{B,k} \log(-t+B^{(-1)}_k) \right. \nonumber \\
&& + \sum^{p_B^{(0)}}_{k=1}J^{(0)}_{B,k} ((-t+B^{(0)}_k)^{-\beta} \nonumber \\
&&\left. + \sum^{p_B^{(+1)}}_{k=1}J^{+1}_{B,k}(-t+B^{(+1)}_k)^{-(\beta+1)}\right\}^2 \nonumber \\
&&+u_2^{(0)}(\beta)^2.
\end{eqnarray}
By combining these results, $U_2$ can be written basy  
\begin{eqnarray}
&&U_2 \approx \nonumber \\
&&\sum^{L-3}_{t=0}\left\{B_0+\sum^{+1}_{m=-1} \sum_{k=1}^{p^{(m)}_B}J^{(m)}_{B,k} f_A(-t,\beta+m,B^{(m)}_k) \right\}^2 \nonumber \\
&&+u_2^{(0)}(\beta)^2, 
\end{eqnarray}
\par
%
%
%
%
%
where 
\begin{eqnarray}
&&B_1^{(-1)}=B_1^{(0)}=B_1^{(+1)}=2L-2-d_{\beta}, \\
&&B_2^{(-1)}=B_2^{(0)}=B_2^{(+1)}=L+d_{\beta}, \\
&&B_3^{(-1)}=B_4^{(0)}=B_3^{(+1)}=L-2+d_{\beta}, \\
&&B_3^{(0)}=L-1+d_{\beta}, 
\end{eqnarray}
$p^{(-1)}_B=3$, $p^{(0)}_B=4$, $p^{(+1)}_B=3$, 
\begin{eqnarray}
&&B_0= \nonumber \\
&&\begin{cases}
-a^{-\beta}-\frac{1}{2}(1+d_{\beta})^{-\beta}-\frac{\beta}{12}(1+d_{\beta})^{-\beta-1}+\frac{(d_{\beta}+1)^{-\beta+1}}{-\beta+1},\\
-a^{-\beta}-\frac{1}{2}(1+d_{\beta})^{-\beta}-\frac{\beta}{12}(1+d_{\beta})^{-\beta-1}+\log(d_{\beta}+1), \\
\end{cases} \nonumber \\
\end{eqnarray}
\begin{eqnarray}
&&J_{B,1}^{(-1)}=
\begin{cases}
1/(-\beta+1)& (\beta \neq 1), \\
1 & (\beta = 1) ,
\end{cases} \\
&&J_{B,2}^{(-1)}=J_{B,3}^{(-1)}=
\begin{cases}
-1/(-\beta+1)& (\beta \neq 1), \\
-1 & (\beta = 1) ,
\end{cases}  \nonumber  \\ 
\end{eqnarray}
\begin{eqnarray}
&&J_{B,1}^{(0)}=J_{B,3}^{(0)}=1/2, \\
&&J_{B,2}^{(0)}=1, \\
&&J_{B,4}^{(0)}=-1/2,  
\end{eqnarray}
and 
\begin{eqnarray}
&&J_{B,1}^{(+1)}=-\beta/12,\\
&&J_{B,2}^{(+1)}=J_{B,2}^{(+1)}=\beta/12.
\end{eqnarray}
 \par
Expanding the squared term gives 
\begin{eqnarray}
&&U_2 \nonumber \\
&&\approx \sum^{L-3}_{t'=0}\left\{B_0+\sum^{+1}_{m=-1} \sum_{k=1}^{p_m}J^{(m)}_{B,k} f_A(t,\alpha+m,B^{(m)}_k) \right\}^2 \nonumber \\
&&=B_0^2(L-2)+2B_0\sum^{+1}_{m=-1}U^{(1)}_2(-t;\beta,m) \nonumber \\
&&+\sum^{+1}_{m_1=-1}\sum^{+1}_{m_2=-1}U^{(2)}_2(t;\beta,m_1,m_2) \nonumber \\
&&+u_2^{(0)}(\beta)^2, \label{u2_expand}
\end{eqnarray}
\par
where 
\begin{eqnarray}
&&U^{(1)}_2(t;\beta,m_1)=  \sum_{i=1}^{p^{(B)}_{m_1}}  J_{Bi}^{(m_1)}  
  \sum^{L-2}_{t=0}f_A(-t,\beta+m_i,B^{(m)}_i) 
 \nonumber \\
\end{eqnarray}
and 
\begin{eqnarray}
&&U^{(2)}_2(t;\beta,m_1,m_2) \nonumber \\
&& =\sum_{i=1}^{p^{(B)}_{m_1}} \sum_{j=1}^{p^{(B)}_{m_2}} J_{Bi}^{(m_1)} J_{Bj}^{(m_2)} \nonumber \\
&& \sum^{L-2}_{t=0}f_A(-t,\beta+m_i,B^{(m)}_i)f_A(-t,\beta+m_j,B^{(m)}_j).
 \nonumber \\
\end{eqnarray}
%
Using the Euler--Maclaurin formula in Eq. \ref{euler}, we can approximate the sum in $U^{(1)}_2(t;\beta,m_1,m_2))$  
for $L \geq 3$ as 
\begin{eqnarray}
&&\sum^{L-3}_{t=0}f_A(t,\alpha_1,V_1) \\
&&\approx G_1^{(R)}(3-L,\max\{-1,3-L\};\alpha_1,V_1) \nonumber \\
&&+\int^{\max\{L-3,1\}}_1 (-t+V_1)^{-\alpha_1} dt \nonumber \\
&&\approx G_1^{(R)}(3-L,\max\{-1,3-L\},;\alpha_1,V_1) \nonumber \\
&&-G_1^{(I)}(\min\{-L+3,-1\};\alpha_1,V_1) \nonumber \\
&&+G_1^{(I)}(-1;\alpha_1,V_1), \nonumber \\
\end{eqnarray}
where $G_1^{(R)}(x_1,x_2;\alpha_1,V_1)$, 
for $\alpha_1 \neq 1$ and $\alpha_2 \neq 1$, is given by
\begin{eqnarray}
&&G_1^{(R)}(x_1,x_2;\alpha_1,V_1)=(V_1)^{-\alpha_1} \nonumber \\
&&+\frac{1}{2}\{(V_1+x_1)^{-\alpha_1}+(V_1+x_2)^{-\alpha_1}\}  \nonumber \\
&&-\frac{1}{12}\{(-\alpha_1)(V_1+x_2)^{-\alpha_1-1} \nonumber \\
&&-(-\alpha_1)(V_1+x_1)^{-\alpha_1-1}\},  \nonumber \\  \label{u21_ans1}
\end{eqnarray}
And, for $x_1=0$ and $x_2=0$, we define
\begin{eqnarray}
&&G_1^{(R)}(x_1,x_2;\alpha_1,V_1)=V_1^{-\alpha_1}.
\end{eqnarray}
For $\alpha_1 = 0$, the corresponding term is also written as 
\begin{eqnarray}
&&G_1^{(R)}(x_1,x_2;\alpha_1,V_1) \nonumber \\
&&=\log(V_1)+ \frac{1}{2}\{\log(V_1+x_1)+\log(V_1+x_2)\}  \nonumber \\
&&-\frac{1}{12}\{(V_1+x_2)^{-1} \nonumber \\
&&-(V_1+x_1)^{-1}\},  \nonumber \\ 
\end{eqnarray}
And, for $x_1=0$ and $x_2=0$, we define
\begin{eqnarray}
&&G_1^{(R)}(x_1,x_2;\alpha_1,V_1)=\log(V_1).
\end{eqnarray}
In addition, $G_1^{(I)}(x;\alpha,V)$ is calculated as 
\begin{eqnarray}
&&G_1^{(I)}(x;\alpha,V)\nonumber \\
&&=
\begin{cases}
\text{($\alpha \neq 0$, $\alpha \neq 1$)} \\
\int (x+V)^{-\alpha} dx = \frac{1}{-\alpha+1} (x+V)^{-\alpha+1} \\
\text{($\alpha=1$)} \\
\int (x+V)^{-1} dx= \log(x+V)  \\
 \text{($\alpha=0$)} \\
\int \log(x+V) dx= (x+V)\log(x+V)-x.  \\ 
\end{cases} \label{u21_ans2}
\end{eqnarray}
Here, we omit a constant of integration.  \par
Next, we calculate $U^{(2)}_2(t;\beta,m_1,m_2))$ in Eq. \ref{u2_expand}. 
Using the Euler--Maclaurin formula in Eq. \ref{euler}, we can approximate the sum in $U^{(2)}_2(t;\beta,m_1,m_2))$ as 
\begin{eqnarray}
&&\sum^{L-3}_{t=0}f_A(-t,\alpha_1,V_1)f_A(-t,\alpha_2,V_2) \\
&&\approx G_2^{(R)}(3-L,\max\{-1,3-L\};\alpha_1,\alpha_2,V_1,V_2) \nonumber \\
&&+\int^{\max\{L-3,1\}}_1 (-t+V_1)^{-\alpha_1}(-t+V_2)^{-\alpha_2} dt \nonumber \\
&&\approx G_2^{(R)}(3-L,\max\{-1,3-L\},;\alpha_1,\alpha_2,V_1,V_2) \nonumber \\
&&-G_2^{(I)}(\min\{3-L,1\};\alpha_1,\alpha_2,V_1,V_2) \nonumber \\
&&+G_2^{(I)}(-1;\alpha_1,\alpha_2,V_1,V_2) . \label{u22_ans}
\end{eqnarray}
Therefore, substituting Eqs. \ref{u21_ans1}, \ref{u21_ans2}, and \ref{u22_ans} into Eq. \ref{u2_expand}, for $L>3$, we have 
\begin{eqnarray}
&&U_2 \approx 
(L-2)B_0^2 \nonumber \\
&&+2B_0\sum^{+1}_{m=-1}\sum_{i=1}^{p^{(B)}_{m}}  J_{Bi}^{(m)} \{ 
G_1^{(R)}(-1,3-L,\beta+m,B_i^{(m)}) \nonumber \\
&&-G_1^{(I)}(-L+3,\beta+m,B_i^{(m)})  \nonumber \\
&&+G_1^{(I)}(-1,\beta+m,B_i^{(m)}) \} \nonumber \\
&&+\sum^{+1}_{m_1=-1}\sum^{+1}_{m_2=-1} \sum_{i=1}^{p^{(B)}_{m_1}} \sum_{j=1}^{p^{(B)}_{m_2}}  J_{Bi}^{(m_1)}J_{Bj}^{(m_2)} \{ \nonumber \\
&&G_2^{(R)}(3-L,-1,\beta+m_1,\beta+m_2; B_i^{(m_1)},B_j^{(m_2)}) \nonumber \\
&&-G_2^{(I)}(3-L,\beta+m_1,\beta+m_2; B_i^{(m_1)},B_j^{(m_2)}) \nonumber \\
&&+G_2^{(I)}(-1,\beta+m_1,\beta+m_2; B_i^{(m_1)},B_j^{(m_2)}) \} \nonumber \\
&+& u^{(0)}_2(\beta)^2 \label{u2_ans}
\end{eqnarray}
and, for $L=3$, we have 
\begin{eqnarray}
&&U_2 \approx 
(L-2)B_0^2 \nonumber \\
&&+2B_0\sum^{+1}_{m=-1}\sum_{i=1}^{p^{(B)}_{m}}  J_{Bi}^{(m)} 
G_1^{(R)}(0,0,\beta+m,B_i^{(m)}) \nonumber \\
&&+\sum^{+1}_{m_1=-1}\sum^{+1}_{m_2=-1} \sum_{i=1}^{p^{(B)}_{m_1}} \sum_{j=1}^{p^{(B)}_{m_2}}  J_{Bi}^{(m_1)}J_{Bj}^{(m_2)} \{ \nonumber \\
&&
G_2^{(R)}(0,0,\beta+m_1,\beta+m_2; B_i^{(m_1)},B_j^{(m_2)}) \} \nonumber \\
&+& u^{(0)}_2(\beta)^2. 
\end{eqnarray}

For $L=2$, from Eq. \ref{u21_ans2}, we have 
\begin{eqnarray}
&&U_2 \approx  u^{(0)}_2(\beta)^2. 
\end{eqnarray}
Note that, because we cannot use the integral approximation method, we calculated directly from the sums for $L=2$.
\par
\subsection{Calculation of $U_3(\beta,L)$}
$U_3$ is defined by Eq. \ref{U_3} as
\begin{equation}
U_3= \frac{ Z(\beta)^2}{\check{\eta}^2} \sum^{L(I+2)}_{t=L(I+1)+1}(\theta_4'(t))^2. \label{U3_app} 
\end{equation}
Substituting Eqs. \ref{theta_2d} and \ref{theta_3d} into Eq. \ref{U3_app} gives 
\begin{eqnarray}
U_3&=&  \sum^{L(I+2)}_{t=L(I+1)+1}\sum^{L-(t-L(I+1))}_{k=0}(k+d_{\beta})^{-\beta}.
\end{eqnarray}
Using a shifted index $t'=t-L(I+1)-1$, we can write
\begin{eqnarray}
U_3&=& \sum^{L-1}_{t'=0}(\sum^{L-(1+t')}_{k=0}(k+d_{\beta})^{-\beta})^2. 
\end{eqnarray}
Using the Euler--Maclaurin formula in Eq. \ref{euler}, we can obtain 
\begin{eqnarray}
&&U_3 \approx  \nonumber \\
&&\frac{ Z(\beta)^2}{\check{\eta}^2} \sum^{L-2}_{t'=0} \left\{\theta(0)+\frac{\theta(1)+\theta(L-1-t))}{2} \right. \nonumber  \\
&+&\left. \frac{\theta'(L-1-t)-\theta'(1)}{12} +\int^{L-1-t}_{1}\theta(k)dk \right\}^2 \nonumber \\
&+&\theta(0)^2.
\end{eqnarray}
Here, because we cannot use the integral approximation method, we calculated directly from the sums for $t'=L-1$.
\par 
Substituting Eq. \ref{app_theta} into $\theta(t')$, 
for $\beta>0$ and $\beta \neq 1$, we get 
\begin{eqnarray}
&&U_3 \approx \nonumber \\
&&\sum^{L-2}_{t=0}\left\{C_0+  \sum^{p_C^{(-1)}}_{k=1}J^{(-1)}_{C,k} (-t+C^{(-1)}_k)^{-(\beta-1)} \right. \nonumber \\
&&+ \sum^{p_C^{(0)}}_{k=1}J^{(0)}_{C,k} ((-t+C^{(0)}_k)^{-\beta} \nonumber \\
&&\left. + \sum^{p_C^{(+1)}}_{k=1}J^{(+1)}_{C,k}(-t+C^{(+1)}_k)^{-(\beta+1)}\right\}^2 \nonumber \\
&&+d_{\beta}^{-2\beta},  
\end{eqnarray}
and, for $\beta=1$, we get
\begin{eqnarray}
&&U_3 \approx \nonumber \\
&&\sum^{L-2}_{t=0}\left\{C_0+ \sum^{p_C^{(-1)}}_{k=1}J^{(-1)}_{C,k} \log(-t+C^{(-1)}_k) \right. \nonumber \\
&& + \sum^{p_C^{(0)}}_{k=1}J^{(0)}_{C,k} ((-t+C^{(0)}_k)^{-\beta} \nonumber \\
&&\left.+ \sum^{p_C^{(+1)}}_{k=1}J^{+1}_{C,k}(-t+C^{(+1)}_k)^{-(\beta+1)}\right\}^2 \nonumber \\
&&+d_{\beta}^{-2\beta}. 
\end{eqnarray}
Combining these two equations gives  
\begin{eqnarray}
&&U_3 \approx \nonumber \\
&&\sum^{L-2}_{t=0}\left\{C_0+\sum^{+1}_{m=-1} \sum_{k=1}^{p^{(m)}_C}J^{(m)}_{C,k} f_A(-t,\beta+m,C^{(m)}_k) \right\}^2 \nonumber \\
&&+d_{\beta}^{-2\beta}, 
\end{eqnarray}
\par
%
%
%
%
%
where
\begin{eqnarray}
&&C_1^{(-1)}=C_1^{(0)}=C_1^{(-1)}=L-1+a, \\
\end{eqnarray}
$p^{(-1)}_C=1$, $p^{(0)}_C=1$, $p^{(+1)}_C=1$, 
\begin{eqnarray}
&&C_0= \nonumber \\
&&\begin{cases}
d_{\beta}^{-\beta}+\frac{1}{2}(1+d_{\beta})^{-\beta}+\frac{\beta}{12}(1+d_{\beta})^{-\beta-1}-\frac{(d_{\beta}+1)^{-\beta+1}}{-\beta+1},\\
d_{\beta}^{-\beta}+\frac{1}{2}(1+d_{\beta})^{-\beta}+\frac{\beta}{12}(1+d_{\beta})^{-\beta-1}-\log(d_{\beta}+1), \\
\end{cases} \nonumber \\
\end{eqnarray}
\begin{eqnarray}
&&J_{C,1}^{(-1)}=
\begin{cases}
1/(-\beta+1)& (\beta \neq 1), \\
1 & (\beta = 1) ,
\end{cases} 
\end{eqnarray}
\begin{eqnarray}
&&J_{C,1}^{(0)}=1/2,  
\end{eqnarray}
and
\begin{eqnarray}
&&J_{C,1}^{(+1)}=-\beta/12.
\end{eqnarray}
\par
Expanding the squared term gives 
\begin{eqnarray}
&&U_3 \approx \nonumber \\
&&\sum^{L-2}_{t'=0}\left\{C_0+\sum^{+1}_{m=-1} \sum_{k=1}^{p_m}J^{(m)}_{B,k} f_A(t,\alpha+m,B^{(m)}_k) \right\}^2 \nonumber \\
&&=C_0^2+2C_0\sum^{+1}_{m=-1}U^{(1)}_3(-t;\beta,m) \nonumber \\
&&+\sum^{+1}_{m_1=-1}\sum^{+1}_{m_2=-1}U^{(2)}_3(t;\beta,m_1,m_2) \nonumber \\
&&+d_{\beta}^{-2\beta}, \label{u3_expand}
\end{eqnarray}
\par
where 
\begin{eqnarray}
&&U^{(1)}_3(t;\beta,m)  \nonumber \\ 
&&=  \sum_{i=1}^{p^{(C)}_{m}}  J_{Ci}^{(m)}  \sum^{L-2}_{t=0}f_A(-t,\beta+m,C^{(m)}_i)
 \nonumber \\
\end{eqnarray}
and
\begin{eqnarray}
&&U^{(2)}_3(t;\beta,m_1,m_2) \nonumber \\
&& = \sum_{i=1}^{p^{(C)}_{m_1}} \sum_{j=1}^{p^{(C)}_{m_2}} J_{Ci}^{(m_1)} J_{Cj}^{(m_2)} \nonumber \\
&& \sum^{L-2}_{t=0}f_A(-t,\beta+m_i,C^{(m)}_i)f_A(-t,\beta+m_j,C^{(m)}_j). \nonumber \\
\end{eqnarray}
%
Consequently, as with $U_2$, for $L>2$, $U_3$ is also calculated by 
\begin{eqnarray}
&&U_3 \approx 
(L-1)C_0^2 \nonumber \\
&&+2C_0\sum^{+1}_{m=-1}\sum_{i=1}^{p^{(C)}_{m}}  J_{C_i}^{(m)} \{ 
G_1^{(R)}(-1,2-L,\beta+m,C_i^{(m)}) \nonumber \\
&&-G_1^{(I)}(-L+2,\beta+m,C_i^{(m)})  \nonumber \\
&&+G_1^{(I)}(-1,\beta+m,C_i^{(m)}) \} \nonumber \\
&&+\sum^{+1}_{m_1=-1}\sum^{+1}_{m_2=-1} \sum_{i=1}^{p^{(C)}_{m_1}} \sum_{j=1}^{p^{(C)}_{m_2}}  J_{C_i}^{(m_1)}J_{C_j}^{(m_2)} \{ \nonumber \\
&&G_2^{(R)}(2-L,-1,\beta+m_1,\beta+m_2; B_i^{(m_1)},B_j^{(m_2)}) \nonumber \\
&&-G_2^{(I)}(2-L,\beta+m_1,\beta+m_2; C_i^{(m_1)},C_jj^{(m_2)}) \nonumber \\
&&+G_2^{(I)}(-1,\beta+m_1,\beta+m_2; C_i^{(m_1)},C_j^{(m_2)}) \} \label{u3_ans}
 \nonumber \\
&&+d_{\beta}^{-2\beta} , 
\end{eqnarray}
and, for $L=2$, 
\begin{eqnarray}
&&U_3 \approx 
(L-1)C_0^2 \nonumber \\
&&+2C_0\sum^{+1}_{m=-1}\sum_{i=1}^{p^{(C)}_{m}}  J_{C_i}^{(m)} 
G_1^{(R)}(0,0,\beta+m,C_i^{(m)}) \nonumber \\
&&+\sum^{+1}_{m_1=-1}\sum^{+1}_{m_2=-1} \sum_{i=1}^{p^{(C)}_{m_1}} \sum_{j=1}^{p^{(C)}_{m_2}}  J_{C_i}^{(m_1)}J_{C_j}^{(m_2)} \{ \nonumber \\
&&
G_2^{(R)}(0,0,\beta+m_1,\beta+m_2; C_i^{(m_1)},C_j^{(m_2)}) \} \nonumber \\
 \nonumber \\
&&+d_{\beta}^{-2\beta}. 
\end{eqnarray}
\par
\subsection{Calculation of $V[R_j^{(L)}]$}
\label{app_sec_ans_u}
We can obtain $V[R_j^{(L)}]$ by substituting $U_1(\beta,L)$ (Eq. \ref{U_1_ans}), $U_2(\beta,L)$ (Eq. \ref{u2_ans}), and $U_3(\beta,L)$ (Eq. \ref{u3_ans}) into Eq. \ref{u_ans} approximately.
\renewcommand{\theequation}{E\arabic{equation}}
\renewcommand{\thefigure}{E-\arabic{figure}}
\renewcommand{\thetable}{S-E\arabic{table}}
\setcounter{figure}{0}
\setcounter{equation}{0}
\section{Asymptotic behavior of  $V[R_j^{(L)}]$ in the case of the power-law forgetting process for $L>>1$}
\label{app_l_forget}
In this section, we calculate the asymptotic behavior of  $V[R_j^{(L)}]$  for $L>>1$.
Because  $V[R_j^{(L)}]$ is decomposed into $U_1(\beta,L)$, $U_2(\beta,L)$, and $U_3(\beta,L)$ (Eq. \ref{u_ans}), we calculate the asymptotic behaviors of $U_1(\beta,L)$ (Eq. \ref{U_1_ans}), $U_2(\beta,L)$ (Eq. \ref{u2_ans}), and $U_3(\beta,L)$ (Eq. \ref{u3_ans}), respectively.
 \par
\subsection{Asymptotic behavior of $U_1(\beta,L)$ for $L>>1$}
The dominant terms of Eq. \ref{U_1_ans} are the cases of $m_1=-1$ and $m_2=-1$, namely, 
\begin{equation}
\sum^{p^{(A)}_{(-1)}}_{i=1} \sum^{p^{(A)}_{(-1)}}_{j=1} J^{(-1)}_{A_i}J^{(-1)}_{A_j} G_2^{(I;Q)}(\beta,-1,-1,A_i^{(-1)},A_j^{(-1)})
\end{equation}
and
\begin{equation}
\sum^{p^{(A)}_{(-1)}}_{i=1} \sum^{p^{(A)}_{(-1)}}_{j=1} J^{(-1)}_{A_i}J^{(-1)}_{A_j} G_2^{(I)}(1,\beta-1,\beta-1,A_i^{(-1)},A_j^{(-1}).
\end{equation}
Calculating these terms for $L>>1$, we can write 
\begin{eqnarray}
&&U_1(\beta,L) \approx 
u_1^{(1)}(\beta)L^{3-2\beta},
\label{u1_l}
\end{eqnarray}
where, for $\beta \neq 1$ and $\beta>0$, 
\begin{eqnarray}
&&u_1^{(1)}(\beta)=\frac{1}{(2-\beta)(1-\beta)^2} \nonumber \\
&&\times \biggl\{ 4 {}_2F_1(\beta-1,2-\beta,3-\beta,-1) -\frac{(2-\beta)(4+2^{3-2\beta})}{3-2\beta} \nonumber \\
&&+2 \frac{\Gamma(3-\beta)}{\Gamma(\beta-1)}(2^{3-2\beta}q_1(\beta,L)-4q_2(\beta,L)) \biggr\} \nonumber \\
\end{eqnarray}
and, for $\beta=1$, 
\begin{eqnarray}
&&u_1^{(1)}(\beta) \nonumber \\
&&=-\frac{4\pi^2}{6}-4\psi(2)-4\log(2)^2-2\log(-1)^2 
\end{eqnarray}
%
In addition, 
\begin{eqnarray}
&&q_1(\beta,L) \nonumber \\
&&=\begin{cases}
\frac{-\log(2)+\psi(3-2\beta)+\psi(2-\beta)}{\Gamma(4-2\beta)} & \text{($\beta=0.5$)}, \\
\Gamma(2\beta-3) & \text{($\beta \neq 0.5$)},\\
\end{cases}
\end{eqnarray}
and 
\begin{eqnarray}
&&q_2(\beta,L) \nonumber \\
&&=\begin{cases}
\frac{(\psi(3-2 \beta)+\psi(2-\beta))}{\Gamma(4-2\beta)} & \text{($\beta=0.5$)}, \\
\Gamma(2 \beta-3) & \text{($\beta \neq 0.5$)}. \\
\end{cases}
\end{eqnarray}
\par
Here, for the calculations, we use the following approximation formulas of the hypergeometric function \cite{hypergeom}: 
\begin{equation}
{}_2F_1(a,b,c;x) \to 1  \quad (x \to 0),  \label{f_ap0}
\end{equation}
in Eqs. \ref{f_ap1}, \ref{f_ap2}, and \ref{f_ap3} and for the logarithmic function of Eq. \ref{log_ap}. In addition, we replace $L+\rm constant$ with $L$.
\par 
\subsection{Asymptotic behavior of $U_2(\beta,L)$ for $L>>1$} \par
As with $U_1(\beta,L)$, we can calculate the asymptotic behavior of $U_2(\beta,L)$.
We focus on terms of higher order than $O(L)$ in Eq. \ref{u2_ans}.
The highest order terms are 
\begin{equation}
\sum^{p^{(B)}_{(-1)}}_{i=1} \sum^{p^{(B)}_{(-1)}}_{j=1} J^{(-1)}_{B_i}J^{(-1)}_{B_j} G_2^{(I;Q)}(3-L,\beta-1,\beta-1,B_i^{(-1)},B_j^{(-1)})
\end{equation}
and
\begin{equation}
\sum^{p^{(B)}_{(-1)}}_{i=1} \sum^{p^{(B)}_{(-1)}}_{j=1} J^{(-1)}_{B_i}J^{(-1)}_{B_j} G_2^{(I)}(-1,\beta-1,\beta-1,B_i^{(-1)},B_j^{(-1)}). 
\end{equation}
The second highest order terms are  
\begin{equation}
2B_0 \sum^{p^{(B)}_{(-1)}}_{i=1} J^{(-1)}_{B_i} G_1^{(I;Q)}(3-L,\beta-1,B_i^{(-1)})
\end{equation}
and
\begin{equation}
2B_0 \sum^{p^{(B)}_{(-1)}}_{i=1} J^{(-1)}_{B_i} G_1^{(I;Q)}(-1,\beta-1,B_i^{(-1)}).
\end{equation}
The term of $O(L)$ is 
\begin{equation}
(L-1)B_0^2.
\end{equation}
Calculating these terms, we can obtain 
\begin{eqnarray}
&&U_2(\beta,L) \nonumber \\
&&\approx \begin{cases}
u^{(2)}_1(\beta)L^{3-2\beta} u^{(2)}_2(\beta)L^{2-\beta}+u^{(2)}_3(\beta)L &\text{($0<\beta<1$)}, \\
u^{(2)}_a \log(L)^2 L + u^{(2)}_b \log(L)L+u^{(2)}_cL &\text{($\beta=1$)}, \\
u^{(2)}_3(\beta)L & \text{($\beta>1$)}, \label{u2_l} 
\end{cases} \nonumber \\
\end{eqnarray}
where 
\begin{eqnarray}
u^{(2)}_1(\beta)&=&\frac{-4 {}_2F_1(\beta-1,2-\beta,3-\beta,-1)}{(2-\beta)(1-\beta)^2} \nonumber \\
&+&\frac{(2^{3-2\beta}+3)}{(3-2\beta)(1-\beta)^2},  \nonumber \\
\end{eqnarray}
\begin{eqnarray}
&&u^{(2)}_2(\beta)=2 B_0 \frac{2^{2-\beta}-3}{(2-\beta)(1-\beta)}, 
\end{eqnarray}
\begin{eqnarray}
&&u^{(2)}_3(\beta)= B_0^2 , 
\end{eqnarray}
\begin{eqnarray}
&&u^{(2)}_a=1  ,
\end{eqnarray}
\begin{eqnarray}
&&u^{(2)}_b=-4\log(2)-2-2B_0,
\end{eqnarray}
\begin{eqnarray}
&&u^{(2)}_c= 2+4\psi(2)-4\psi(1) \nonumber  \\
&&+\log(2)(4+2\log(2)-4\log(-1)+4B_0) .
\end{eqnarray}
Here, for the calculations, we use these approximation formulas of the hypergeometric functions in Eqs. \ref{f_ap0}, \ref{f_ap1}, \ref{f_ap2}, and  \ref{f_ap3} 
and in the logarithmic function of Eq. \ref{log_ap}. In addition, we replace $L+ \rm constant$ with $L$. 
\subsection{Asymptotic behavior of $U_3(\beta,L)$ for $L>>1$} \par
As with $U_2(\beta,L)$, we can calculate the asymptotic behavior of $U_2(\beta,L)$.
We focus on terms of higher order than $O(L)$ in Eq. \ref{u3_ans}.
The highest order terms are 
\begin{equation}
\sum^{p^{(C)}_{(-1)}}_{i=1} \sum^{p^{(C)}_{(-1)}}_{j=1} J^{(-1)}_{C_i}J^{(-1)}_{C_j} G_2^{(I;Q)}(2-L,\beta-1,\beta-1,C_i^{(-1)},C_j^{(-1)}) 
\end{equation}
and 
\begin{equation}
\sum^{p^{(C)}_{(-1)}}_{i=1} \sum^{p^{(C)}_{(-1)}}_{j=1} J^{(-1)}_{C_i}J^{(-1)}_{C_j} G_2^{(I)}(-1,\beta-1,\beta-1,C_i^{(-1)},C_j^{(-1)}).
\end{equation}
The second highest order terms are  
\begin{equation}
2C_0 \sum^{p^{(C)}_{(-1)}}_{i=1} J^{(-1)}_{C_i} G_1^{(I;Q)}(2-L,\beta-1,C_i^{(-1)}) 
\end{equation}
and
\begin{equation}
2C_0 \sum^{p^{(C)}_{(-1)}}_{i=1} J^{(-1)}_{C_i} G_1^{(I;Q)}(-1,\beta-1,C_i^{(-1)}). 
\end{equation}
The term of $O(L)$ is 
\begin{equation}
(L-1)C_0^2.
\end{equation}
Calculating these terms, we can obtain 
\begin{eqnarray}
&&U_3(\beta,L) \approx \nonumber \\
&&\begin{cases}
u^{(3)}_1(\beta)L^{3-2\beta} u^{(3)}_2(\beta)L^{2-\beta}+u^{(3)}_3(\beta)L &\text{($0<\beta<1$)}, \\
u^{(3)}_a \log(L)^2 L + u^{(3)}_b \log(L)L+u^{(3)}_cL &\text{($\beta=1$)} ,\\
u^{(3)}_3(\beta)L & \text{($\beta>1$)} ,\label{u3_l} 
\end{cases}  \nonumber \\
\end{eqnarray}
where 
\begin{eqnarray}
&&u^{(3)}_1(\beta)=\frac{1}{(1-\beta)^2 (3-2\beta)} , \nonumber \\
\end{eqnarray}
\begin{eqnarray}
&&u^{(3)}_2(\beta)=2 C_0 \frac{2-\beta}{1-\beta},
\end{eqnarray}
\begin{eqnarray}
&&u^{(3)}_3(\beta)= C_0^2  ,
\end{eqnarray}
\begin{eqnarray}
&&u^{(3)}_a=1  ,
\end{eqnarray}
\begin{eqnarray}
&&u^{(3)}_b=(-2+2C_0),
\end{eqnarray}
\begin{eqnarray}
&&u^{(3)}_c= -1+C_0^2+2C_0. 
\end{eqnarray}
Here, for the calculations, we use the approximation formulas of the hypergeometric functions in Eqs. \ref{f_ap0}, \ref{f_ap1}, \ref{f_ap2}, and \ref{f_ap3} 
and for the logarithmic function in Eq. \ref{log_ap}. In addition, we replace $L+ \rm constant$ with $L$. \par
\subsection{Asymptotic behavior of $V[R_j^{(L)}]$ for $L>>1$}
Substituting Eqs. \ref{u1_l}, \ref{u2_l}, and \ref{u3_l} into Eq. \ref{u_ans}, we can obtain
\begin{eqnarray}
&&V[\delta R_j^{(L)}] \approx \frac{\check{\eta}^2}{Z(\beta)^2} \times  \nonumber \\
&&\begin{cases}
u_1(\beta)L^{1-2\beta}+u_2(\beta)L^{-\beta}+u_3(\beta)L^{-1} &\text{($0<\beta<1$)}, \\
u_a \log(L)^2 L^{-1} + u_b \log(L) L^{-1}+u_c L^{-1} &\text{($\beta=1$)}, \\
u_3(\beta)L^{-1} & \text{($\beta>1$)},  
\end{cases} \nonumber  \\ 
 \label{u_ans_l}
\end{eqnarray}
\begin{eqnarray}
&&u_1(\beta)=u_1^{(1)}(\beta)+u_1^{(2)}(\beta)+u_1^{(3)}(\beta),  \nonumber \\
\label{u_ans_l_start}
\end{eqnarray}
\begin{eqnarray}
&&u_2(\beta)=u_2^{(2)}(\beta)+u_2^{(3)}(\beta),
\end{eqnarray}
\begin{eqnarray}
&&u_3(\beta)= u_3^{(2)}(\beta)+u_3^{(3)}(\beta) ,
\end{eqnarray}
\begin{eqnarray}
&&u_a=1  ,
\end{eqnarray}
\begin{eqnarray}
&&u_b=(-2+2C_0),
\end{eqnarray}
\begin{eqnarray}
&&u_c= -1+C_0^2+2C_0. \label{u_ans_l_end}
\end{eqnarray}
\par
Consequently, the highest order term is obtained by  
\begin{eqnarray}
V[\delta R_j^{(L)}] \propto 
\begin{cases}
L^{1-2\beta} &\text{($0<\beta<1$)}, \\
\log(L)^2 L^{-1} &\text{($\beta=1$)}, \\
L^{-1} & \text{($\beta>1$)}. 
\end{cases}
\end{eqnarray}

%
%
%
%
\renewcommand{\theequation}{S.H\arabic{equation}}
\renewcommand{\thefigure}{H-\arabic{figure}}
\renewcommand{\thetable}{S-H\arabic{table}}
\setcounter{figure}{0}
\setcounter{equation}{0}
\section{Estimation of scaled total number of blogs, $m(t)$, from the data}
\label{app_m}
  We estimate the scaled total number of blogs, $m(t)$, by using the moving median as follows: 
\textcolor{black}{
\begin{enumerate}
\item[1.] We create a set $S$ consisting of indexes of words such that $\check{c}_j$ takes a value larger than the threshold $\check{c}_j \geq 100$, where $\check{c}_j=\sum^{T}_{t=1}g_j(t)/T$.
\item[2.] We estimate $m(t)$ as the median of $\{g_j(t)/\check{c}_j:j \in S \}$ with respect to $j$.
\item[3.] For $t=1,2,\cdots,T$, we calculate $m(t)$ using step 2. 
\end{enumerate}
} \par
Here, we use only words with $\check{c}_j \geq 100$ in step 1 because we neglect the discreteness. In step 2, we apply the median because of its robustness to outliers.

\renewcommand{\theequation}{G\arabic{equation}}
\renewcommand{\thefigure}{G-\arabic{figure}}
\renewcommand{\thetable}{G-A\arabic{table}}
\setcounter{figure}{0}
\setcounter{equation}{0}
\section{MSD of the power-law forgetting process for $\beta=0.5$ and $L>>1$}
\label{app_MSD}
By rough approximate calculations, we show that the logarithmic diffusion can be derived by the power-law forgetting process given by Eq. \ref{eq_rw} or Eq. \ref{app_model_theta} for $\beta=0.5$ and $L>>1$. More detailed and accurate derivation can be found in Ref.\cite{watanabe2018empirical}. \par
The MSD of the model given by Eq. \ref{app_model_theta} can be calculated as 
\begin{eqnarray}
&&\left<(r(t+L)-r(t))^2 \right> \nonumber \\
&=&\left<(\sum^{\infty}_{s=0} \theta(s)  \eta(t+L-s) - \sum^{\infty}_{s'=0}\theta(s')  \eta(t-s'))^2 \right> \nonumber \\
&=&\left< \left \{\sum^{-1}_{s=-L} \theta(s+L) \eta(t-s) +\sum^{\infty}_{s=0} (\theta(s+L) \right.  \right. \nonumber \\
&-& \left. \left. \theta(s) ) \eta(t-s) \right \}^2 \right> \\ 
\end{eqnarray}
By using $\sum^{B}_{t=A} \eta(t)^2/(B-A) \approx \hat{\eta}^2$ for $B-A>>1$ and $L>>1$, 
\begin{eqnarray}
&&\left<(r(t+L)-r(t))^2 \right> \nonumber \\
&\approx& \left<\hat{\eta}^2  (\sum^{-1}_{s=-L} \theta(s+L)^2 +\sum^{\infty}_{s=0} (\theta(s+L) - \theta(s) )^2) \right> \nonumber \\ 
&=&\hat{\eta}^2  (S_1+S_2), \label{app_tmsd}
\label{ss}
\end{eqnarray}
where for the case of $\beta=0.5$ and $L>>1$, 
\begin{eqnarray}
S_1 &\equiv& \sum^{-1}_{s=-L} \theta(s+L)^2 \approx \int^{-1}_{-L} \theta(s+L)^2 ds \nonumber \\
&=& \int^{-1}_{-L} \frac{1}{Z(\beta)^2}(s+L+d_\beta)^{-1}ds \nonumber \\
&=&\frac{1}{Z(\beta)^2}\left[\log(s+L+d_\beta) \right]_{-L}^{-1} \nonumber \\
&\propto& \log(L) \quad(L>>1)
\end{eqnarray}
and 
\begin{eqnarray}
&&S_2 \equiv \sum^{\infty}_{s=0} (\theta(s+L) - \theta(s) )^2) \approx \int^{\infty}_{0} (\theta(s+L) - \theta(s) )^2 ds \nonumber \\
&&= \int^{\infty}_{0} \frac{1}{Z(\beta)^2}((s+L+d_\beta)^{-0.5}-(s+d_\beta)^{-0.5})^2ds \nonumber \\ 
&&= \frac{1}{Z(\beta)^2} \left[\log(\frac{(s+d_\beta+L)(s+d_\beta)}{(\sqrt{\frac{s+d_\beta}{L}}+\sqrt{\frac{s+d_\beta}{L}+1})^4})\right]_{0}^{\infty} \nonumber \\
&&\propto 2\log(L)-\log(L) = \log(L) \quad (L>>1).
\end{eqnarray}
Therefore for $\beta=0.5$ we can obtaion 
\begin{equation}
\left<(r(t+L)-r(t))^2 \right> \propto  \log(L) \quad (L>>1).
\end{equation}
This approximation is simple but very rough. Hence, it is not a good approximation for the time scale of blog data ($L<1000$).
More accurate approximations that hold for small $L$ are discussed in Ref. \cite{watanabe2018empirical}.
\par
Note that the ensemble MSD of the power-law forgetting process on a finite time scale can be roughly calculated as follows:
\begin{eqnarray}
&&\left<r'(L)_j^2 \right>_{j} \nonumber \\
&&=\left<\sum_{s=0}^{(L-1)} \theta(s) \cdot \eta_j(L-s) \right>_{j} \\
&&\approx \check{\eta}^2 \sum_{s=0}^{L-1} \theta(s)^2=\check{\eta}^2 S_1,  
\end{eqnarray}
where $\left<A_j(t) \right>_j= \sum^{W}_{j}A_j(t)/W$ is the ensemble average for $W>>1$ and we defined the power-law forgetting process for a finite time scale as
\begin{eqnarray}
r_j'(t)=\sum_{s=0}^{t-1} \theta(s) \cdot \eta_j(t-s). \label{app_emsd}
\end{eqnarray}
From these rough calculations, we may say that the difference between the time-averaged MSD given by Eq. \ref{app_tmsd} and the emsamble MSD given by Eq. \ref{app_emsd} is $S2$ term in Eq. \ref{app_tmsd}.  \par
\begin{figure}
\begin{minipage}{0.98\hsize}
\centering
\begin{overpic}[width=6cm,clip]{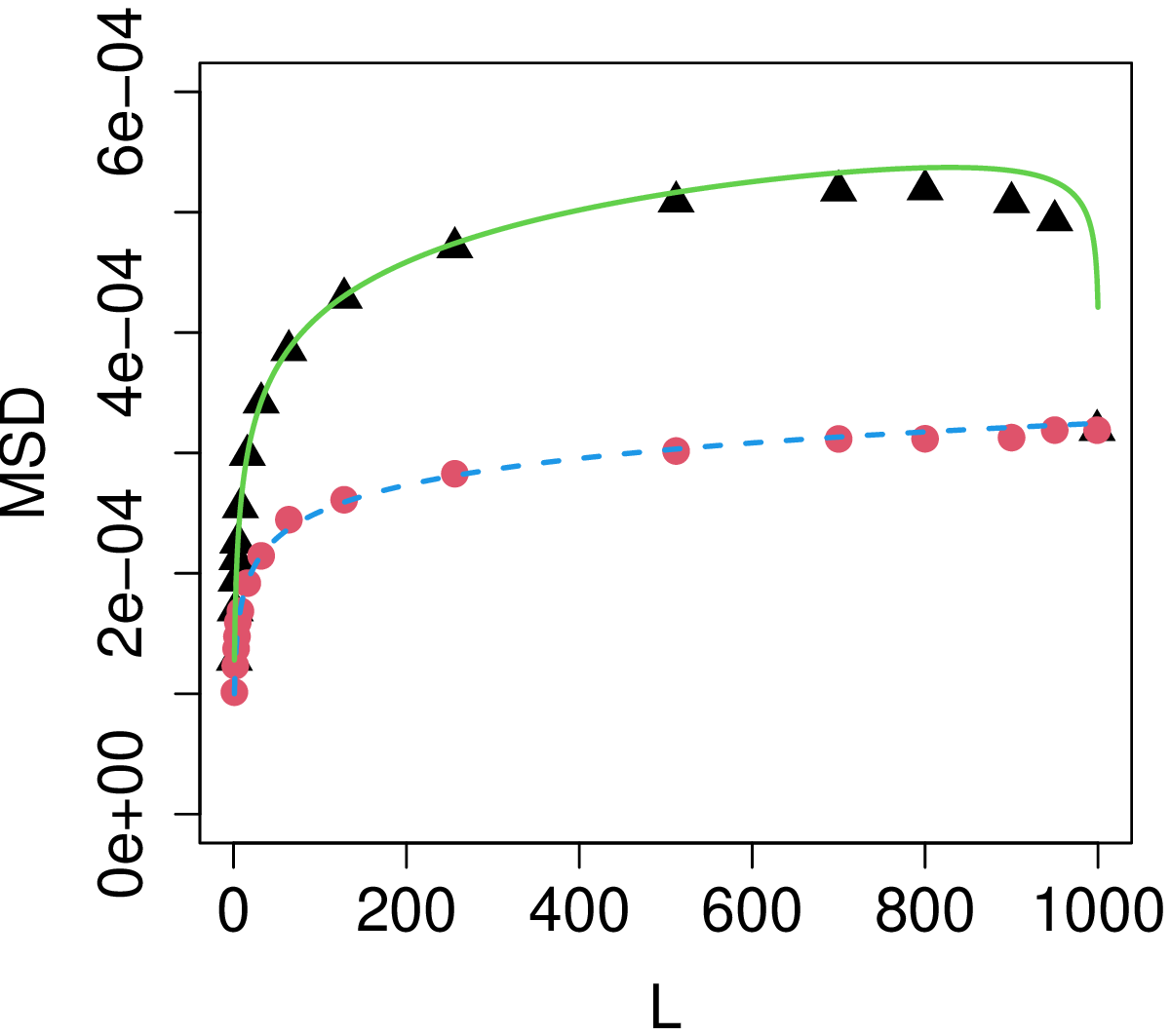}
\put(23,72){\Large(a)}
\end{overpic}
\end{minipage}
\begin{minipage}{0.98\hsize}
\centering
\begin{overpic}[width=6cm,clip]{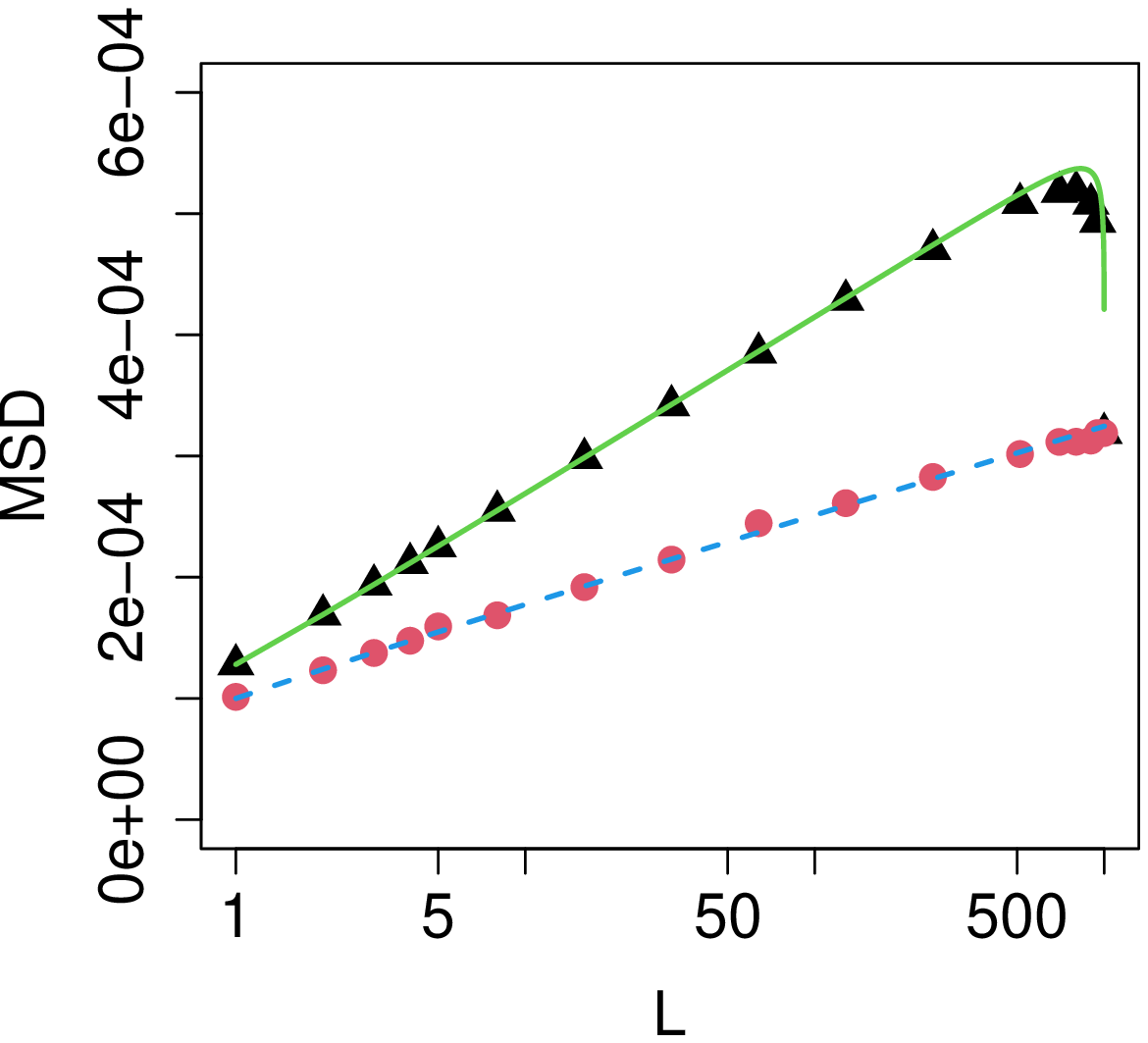}
\put(23,72){\Large(b)}
\end{overpic}
\end{minipage}
\caption{
(a) Comparison of time-averaged MSD and ensemble MSD of the numerical simulation  of model Eq. \ref{app_emsd}. The black triangles are the time-averaged MSD $<\overline{\delta^2(L)}>_j=1/W \sum_{i=1}^{W} \sum_{t=1}^{T-L}((r_j(t+L)-r_j(t))^2)/(T-L)$, the red circle is the ensemble MSD $<r_j'(L)>_{j}=1/W \sum^{W}_{j=1}r_j'(L)^2$, where $W$ is the number of ensembles and $T$ is a observation period. The green solid line is the theoretical curve for the time-averaged MSD given by Eq. \ref{app_tmsd} and the blue dashed line is for the ensemble MSD given by Eq. \ref{app_emsd}. Here, $T=1000$, $W=10000$ and $\eta_j(t)$ are sampled from the normal distribution with the mean $0$ and the standard deviation $\check{\eta}=0.01$. 
(b) Corresponding figure in a semi-log plot.  From these figures, it can be confirmed that the time-averaged MSD is different from the ensemble MSD. It can also be confirmed that the numerical simulations agree well with the theoretical curves.
}
\label{efs}
\end{figure}

 
Fig. \ref{efs} shows the results of numerical calculations comparing the time-averaged MSD and the ensemble MSD. From these figures, it can be confirmed that the time-averaged MSD is different from the ensemble MSD.  In addition, we can also confirm that the numerical simulations agree well with the theoretical curves.
Since the current analysis is rough, more detailed analysis of the differences between the time-averaged MSD and the ensemble MSD, including analysis of real data, is needed in the future.


\bibliographystyle{spphys}       
%
%

\end{document}